\documentclass[10pt,twoside,letter,amsmath,amssymb,aps,showkeys,showpacs,nofootinbib]{revtex4}
\usepackage{amsfonts}
\usepackage{graphics}
\usepackage{epsfig}
\usepackage{color}
\usepackage{amstext}
\usepackage{amssymb}
\usepackage{graphicx}

\newcommand{\KK}{$\mathcal{KK}$}
\begin{document}
\large

%% \DeclareGraphicsExtensions{.jpg,.pdf,.mps,.png}
\thispagestyle{myheadings}

\title{NLO radiative corrections for Forward-Backward and Left-Right Asymmetries at a B-Factory}

\author{Aleksandrs Aleksejevs}
\email{aaleksejevs@grenfell.mun.ca}
%\affiliation{Grenfell Campus of Memorial University, Corner Brook, Canada}

\author{Svetlana Barkanova}
\email{sbarkanova@grenfell.mun.ca}
\affiliation{Grenfell Campus of Memorial University, Corner Brook, Canada}

\author{Caleb Miller}
\email{calebmiller@uvic.ca}
\affiliation{University of Victoria, Victoria, Canada}

\author{J. Michael Roney}
\email{mroney@uvic.ca}
\affiliation{University of Victoria, Victoria, Canada}

\author{Vladimir Zykunov}
\email{vladimir.zykunov@cern.ch}
\affiliation{Joint Institute for Nuclear Research, Dubna, Russia }
\affiliation{Francisk Skorina Gomel State University,  Gomel, Belarus}

\begin{abstract}
This paper presents the first
 calculations of the parity-violating polarization asymmetry and forward-backward asymmetry of the
$e^- e^+ \rightarrow \mu^+ \mu^- (\gamma)$ process at a center-of-mass energy of 10.58~GeV
with up to one-loop electroweak radiative corrections.
%The paper discusses the one-loop (NLO) electroweak radiative corrections to the 
%$e^- e^+ \rightarrow f {\bar f} (\gamma)$ process with longitudinally polarized electrons, where electrons are excluded as final state fermions.
%The focus of this paper is an investigation of the parity-violating polarization asymmetry and forward-backward asymmetry.
%The calculations are relevant for precision electroweak measurements at the Belle~II experiment, which is being installed on the SuperKEKB $e^- e^+$ collider; designed for a center-of-mass energy at the mass of the $\Upsilon(4S)$ resonance.
The calculations are relevant for future precision electroweak measurements at the Belle~II experiment, which is now collecting data at
 the SuperKEKB $e^- e^+$ collider with a center-of-mass energy at the mass of the $\Upsilon(4S)$ resonance.
%% (aw well as experiments at future high-energy colliders ILC, CLIC etc.).
In this paper we take under full control the bremsstrahlung process at the conditions of Belle II/SuperKEKB, and the possibilities for a soft photon approach are discussed.
The scale of the obtained relative corrections to the parity-violating and forward-backward asymmetries is significant and the scattering angle dependencies of the asymmetries is non-trivial. 
As an additional validation cross-check using an independent formulation, the calculated asymmetries are compared to results from the \KK~Monte Carlo generator.
%In addition a comparison is also made with the \KK~Monte Carlo generator for the asymmetries.

\end{abstract}

 % PACS, the Physics and Astronomy Classification Scheme.


\maketitle

\section{Introduction}

Electroweak measurements can be made at a  high luminosity electron-positron collider   B-factory, such as Belle~II/SuperKEKB~\cite{BelleIITDR} operating at a center-of-mass (CM) energy of  $E_{cm}=\sqrt{s}=10.579$~GeV (the mass  of the  $\Upsilon(4S)$ meson), via $\gamma-Z$ interference in the process $e^- e^+ \rightarrow f {\bar f}$. In the Standard Model this interference term is parameterized in terms of the axial vector coupling of the fermion $f$, equal to its third component of weak isospin, $g_a(f) = I_3(f)$, and its vector coupling, $g_v(f) = I^3_f - 2 Q_f \sin^2\theta_W$ ($ I^3_{e,\mu,\tau}=-1/2, I^3_{\nu}=+1/2 $, and $\cos \theta_W = m_W/m_Z$), where $Q_f$ is its electric charge and $\theta_W$ is the weak mixing angle. The precision on the measurement of the effective weak mixing angle, and hence the effective vector couplings of the neutral current, would be comparable to those measured on the $Z^0$ pole at LEP and SLC, but at a much lower energy, if the  electron beam of the B-factory has at least a 70\% spin polarization~\cite{SuperBTDR,Roney-ICHEP2012} in a left-right asymmetry measurement. Currently, SuperKEKB does not have a polarized beam and the work presented here is a necessary component of the physics justification for installing polarization in that machine in a potential upgrade. Without polarized beam, Belle~II/SuperKEKB could still measure the forward-backward asymmetry but with a significantly lower precision on $\sin^2 \theta_W^{eff}$, as shown in this paper. A forward-backward asymmetry measurement would, however, still provide a useful measurement of the axial vector coupling  constant for the final-state fermion, $f$.

With a polarized beam,  the vector current couplings to electrons, muons, taus, $s$-quarks, $c$-quarks, and $b$-quarks can be measured and would enable a precision comparison with the Standard Model predictions of their running from 10.579~GeV to the $Z$-pole. Deviations of the
 running would signal the presence of new physics. On the other hand, assuming the
 running holds, these measurements can be used to significantly reduce the uncertainties
 on the $Z$-pole values of the couplings.
The electroweak fits that now include the measured Higgs boson parameters~\cite{LEPEWFit}
 show reasonable internal consistency, but there is a  $2.5\sigma$ deviation
associated with the determination of the $Z b {\bar b}$ couplings and  $\sin^2\theta_W^{eff}$ from the
forward-backward asymmetries for $b$-quarks at LEP. The tension is even greater, 3.2$\sigma$, between this $b{\bar b}$
 determination of  $\sin^2 \theta_W^{eff}$ and that
  from SLD, which provides the single most precise determination of  $\sin^2 \theta_W^{eff}$  using a left-right asymmetry measurement. 
 Therefore, it would be interesting to have additional precision measurements of the $Z b {\bar b}$ vertex.
Because SuperKEKB produces B mesons just above threshold it would have
 a unique ability to measure the neutral current vector coupling
of $b$-quarks in a manner that is  free from fragmentation uncertainties~\cite{SuperBTDR,Roney-ICHEP2012} and would provide
a significant decrease in its uncertainty compared to the value measured at LEP, 
where the dominant systematic error came from fragmentation uncertainties.

In order to  extract reliable information from the experimental data, it is necessary
to take into account higher order effects of electroweak theory, i.e. electroweak radiative corrections (EWC). 
The procedure for the inclusion of EWC is an indispensable part of any modern experiment, 
but will be of  paramount importance for precision electroweak measurements of Belle~II/SuperKEKB.
Consequently, theoretical predictions for the observables 
must include not only full treatment of one-loop radiative corrections (NLO) 
but also leading two-loop corrections (NNLO).

Significant theoretical effort already has been dedicated to NLO EWC to electron-positron annihilation
starting with \cite{BH82}, where EWC for this process with arbitrary polarization are calculated for center-of-mass (CM) energies between 40 and 140~GeV.
For the LEP and SLC colliders the process  $e^- e^+ \rightarrow f\bar f$
demanded consideration of the EWC at $Z$-boson pole with new precision.
The following collaborations have performed this task:
BHM and WOH  \cite{hollik,BH84},
LEPTOP \cite{LEPTOP},
TOPAZ0 \cite{TOPAZ96},
and ZFITTER \cite{ZF91,grup-bar2}.
More recent results for EWC in ``after LEP/SLC'' era are provided by  \KK~\cite{KK} and SANC~\cite{sanc-eeff} codes.

%%%% Replaced by SB for paragraph that follows
%One of the goals of this paper is to calculate the full set of
%one-loop (NLO) EWC, both numerically with no simplifications using
%semi-automatic approach (SAA) by 
%FeynArts \cite{int3}, FormCalc \cite{Hahn}, LoopTools \cite{Hahn} and Form \cite{int7}, 
%as well as analytically in a compact asymptotic form, and to compare the results.
%%%Providing a significantly simplified form of the results is very useful for comparison, verification and tuning 
%%%of an exact SAA calculation.
%The contribution of  additional virtual particles is considered in Sect. II.
%Then we take under full control the bremsstrahlung process at the lower energies of Belle~II/SuperKEKB (Sect. III). 
%The analysis of the analytical and numerical results is given in  Sect. IV. 
%Then we compare the soft photon approximation and hard photon approximation.
%Sensitivity study of polarization and forward-backward asymmetries is described in Sect. V.
%Our conclusions and future plans are discussed in Sect. VI.

The main goal of this %paper
work is to calculate the full set of one-loop (NLO) EWC with the highest precision possible.
% To avoid human error,
 In order to avoid technical errors and to provide a validation cross-check,
 we do the same calculations in two independent and different ways and compare the results – first,
 with a semi-automatic approach (“computer algebra") employing
FeynArts \cite{int3}, FormCalc \cite{Hahn}, LoopTools \cite{Hahn} and Form \cite{int7}, 
 with no simplifications, and then analytically (“by hand”), in a compact asymptotic form.
 Sect. II details the calculated differential cross sections up to one-loop. The bremsstrahlung process at the lower energies of Belle~II/SuperKEKB is fully accounted for in Sect. III, with both a soft photon approximation (SPA) and a more exact hard photon approach (HPA). The analysis of the results obtained through the semi-automatic and asymptotic methods is given in Sect. IV, as well as the comparison of the soft-photon and hard-photon approaches. 
In addition, a comparison is made with results from the \KK~Monte Carlo generator. 
 The sensitivity studies of left-right polarization and forward-backward asymmetries are described in Sect. V.
 Our conclusions and future plans are discussed in Sect. VI.

%JMR Should we be clear hear about naming/labelling: SAA = SPA; asymptotic form = HPA (?)

\section{NLO electroweak corrections at simplest case: general notations and matrix elements}

In our calculations we will start with the simplest case of $e^- e^+ \rightarrow f^-f^+ (\gamma)$ scattering, where $f=\mu$.
First we will disregarded the electron mass $m$ and final-state fermion mass $m_f$ (valid for $f=\mu$)
wherever possible, and second we treat energy in the CM system  of $e^- e^+$ 
as a small parameter, in comparison to the masses of $W/Z$ bosons:
\begin{equation}
m, m_f \ll E \ll m_{W,Z}.
\label{appr}
\end{equation}
For this case  we can obtain the total NLO EWC in compact and relatively simple form, free from unphysical parameters
and suitable for an analysis of the kinematic behavior for a given reaction.

\begin{figure}
\vspace{20mm}
\hspace{50mm}
\begin{tabular}{c}
\begin{picture}(60,60)
\put(-150,-60){
\epsfxsize=9cm
\epsfysize=9cm
\epsfbox{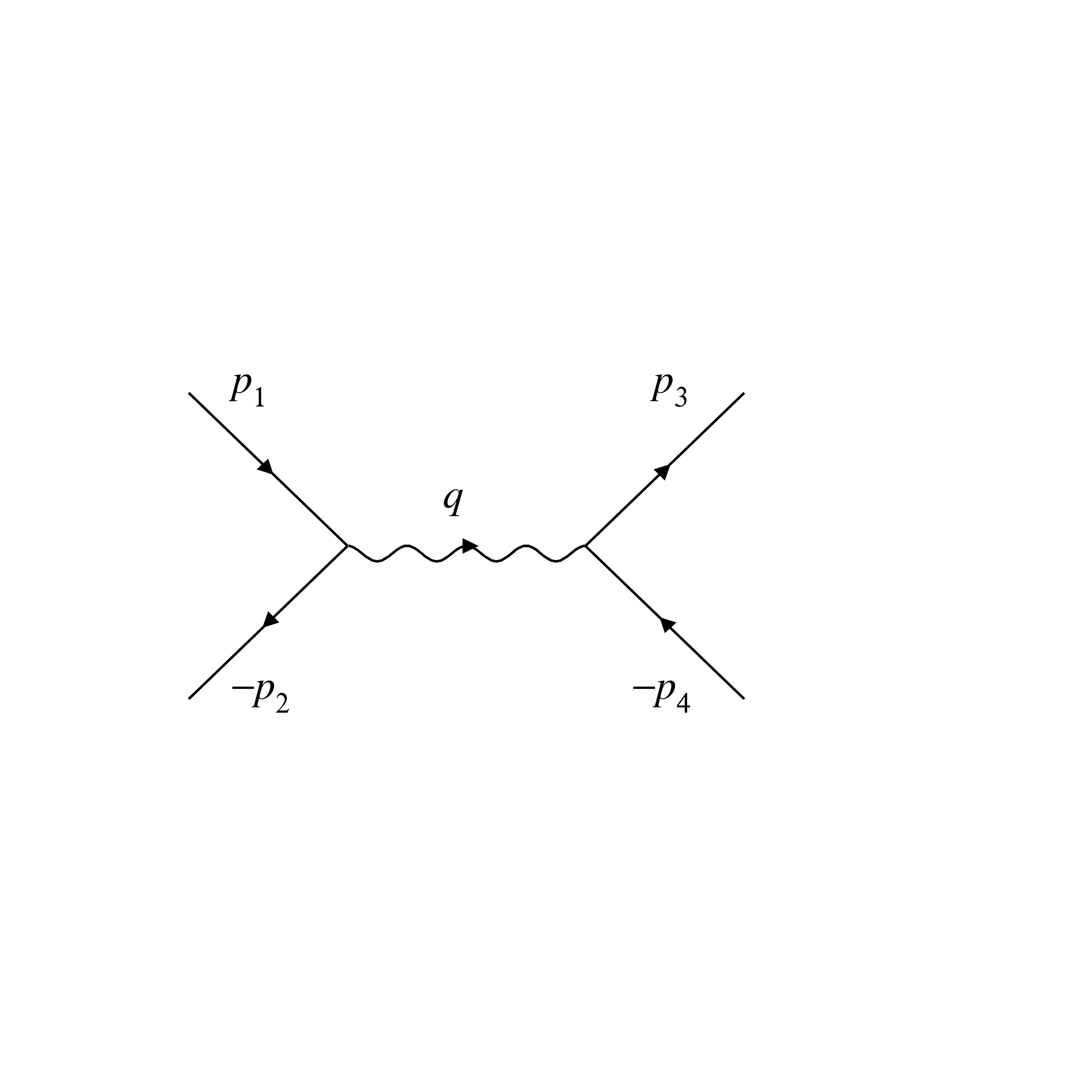} }
\end{picture}
\end{tabular}
\vspace{0mm}
\caption{\protect\it
Feynman diagram describing the process
$e^-(p_1) + e^+(p_2) \rightarrow f^-(p_3)+f^+(p_4)$ in the s-channel at tree level.
}
\label{born}
\vspace{5mm}
\end{figure}

Let us start by writing the cross section for the scattering of polarized electrons on unpolarized positrons,
\begin{equation}
e^-(p_1) + e^+(p_2) \rightarrow f^-(p_3)+f^+(p_4),
\label{0}
\end{equation}
using the Born approximation shown in Fig.~\ref{born}, we find:
\begin{equation}
\sigma \approx \frac{\pi^3}{2s} |M_0|^2.%\approx \frac{\pi^3}{2s} (M_0M_0^+ + 2 {\Re} M_1M_0^+ ).
\label{01}
\end{equation}
Here $\sigma$ is a short notation for the differential cross section
$$\sigma \equiv {d\sigma}/{d(\cos\theta)},$$
$\theta$  is the scattering angle of the detected muon
with 4-momentum $p_3$ in the CM system of the initial electron and positron. 
The  4-momenta of initial ($p_1$ and $p_2$) and final
($p_3$ and $p_4$) fermions generate a standard
set of Mandelstam variables:
\begin{equation}
s=(p_1+p_2)^2,\ t=(p_1-p_3)^2,\ u=(p_1-p_4)^2.
\label{stu}
\end{equation}
Defining $M_0$ as the Born (${\cal O}(\alpha)$) %and  one-loop (${\cal O}(\alpha^2)$) 
amplitude (matrix element),
we describe the structure of $M_0$:
\begin{eqnarray}
M_{0} = \sum_{j=\gamma,Z} M^j,\
M^j = i\frac{\alpha}{\pi} I_\mu^j  D^{j} J^{\mu,j},
\end{eqnarray}
where the electron and muon currents are
\begin{eqnarray}
I_\mu^j = \bar u(-p_2)\gamma_\mu (v_e^j-a_e^j\gamma_5) u(p_1),\
J_\mu^j = \bar u(p_3)\gamma_\mu (v_f^j-a_f^j\gamma_5) u(-p_4)
\end{eqnarray}
and $D^j$ is represented by:
\begin{equation}
D^{j}=\frac{1}{s-m_j^2+im_j\Gamma_j}\ \ (j=\gamma,Z),
\label{structure}
\end{equation}
which depends on the $Z$-boson mass ($m_Z$) and width ($\Gamma_Z$), or
on the photon mass $m_\gamma \equiv \lambda $. 
The photon mass  is set to zero everywhere with the exception of specially indicated
cases where it is taken to be an infinitesimal parameter that regularizes the infrared divergence (IRD).

The squared amplitude $M_0$  forms the Born cross section:
\begin{equation}
%\sigma^0 =\frac{\pi^3}{2s} M_0M_0^+ 
\sigma^0 =\frac{\pi^3}{2s} |M_0|^2 
= \frac{\pi \alpha^2}{s}
\sum_{i,k=\gamma,Z} D^{i}{D^{k}}^* \mu^{ikik},
\label{cs0}
\end{equation}
where
\begin{equation}
\mu^{ikjl} = T_+ \lambda_+^{ikjl} - T_- \lambda_-^{ikjl}, \ \
T_\pm = t^2 \pm u^2,\
\label{MU}
\end{equation}
and
\begin{equation}
\lambda_+^{ikjl} = \lambda_{1}^{ik} \lambda_{fV}^{jl},\ 
\lambda_-^{ikjl} = \lambda_{2}^{ik} \lambda_{fA}^{jl},\ \ \
\lambda_{1}^{ik} = \lambda_{eV}^{ik}-p_B \lambda_{eA}^{ik},\
\lambda_{2}^{ik} = \lambda_{eA}^{ik}-p_B \lambda_{eV}^{ik},
\label{Lpm}
\end{equation}
with $p_B$ representing the degree of electron polarization.
The $\lambda$-type functions have the following structures (here $g=e,f$):
\begin{equation}
 \lambda_{gV}^{ij}=v_g^iv_g^j + a_g^ia_g^j,\
 \lambda_{gA}^{ij}=v_g^ia_g^j + a_g^iv_g^j,
\label{lVA}
\end{equation}
where the vector and axial coupling constants are
\begin{equation}
 v_g^{\gamma}=-Q_g,\ a_g^{\gamma}=0,\
 v_g^Z=(I_g^3-2Q_gs_{W}^2)/(2s_{W}c_{W}),
\ a_g^Z=I_g^3/(2s_{W}c_{W}),
\end{equation}
$Q_g$ is the electric charge of particle $g$ in units of the proton's charge.
Let us recall that $ I_{g}^3=-1/2,\ I_{\nu}^3=+1/2 $ etc. and
$s_{W}\  (c_{W})$
is the sine (cosine) of the Weinberg mixing angle expressed in terms of the $Z$- and $W$-boson
masses according to the on-shell definition in the Standard Model:
\begin{equation}
c_{W}=m_{W}/m_{Z},\
s_{W}=\sqrt{1-c_{W}^2}.
\label{eqn:swDef}
\end{equation}

%At the Next-to-Leading-Order (NLO)  (${\cal O}(\alpha^2)$), we can present NLO differential cross section through an interference term.
%It is given by the second term of the following expansion:
%At the Next-to-Leading-Order (NLO)  (${\cal O}(\alpha^2)$), we can represent the NLO component of the
% differential cross section as an additional term, $M_1$, with which the Born ampliture, $M_0$, interfers:
At the Next-to-Leading-Order (NLO), we can introduce the NLO 
 differential cross section(${\cal O}(\alpha^3)$) via an interference term given by the second term of the following expansion:

\begin{equation}
%\sigma = \frac{\pi^3}{2s} |M_0+M_1|^2\approx \frac{\pi^3}{2s} (M_0M_0^+ + 2 {\Re} [M_1M_0^+] ).
\sigma = \frac{\pi^3}{2s} |M_0+M_1|^2\approx \frac{\pi^3}{2s} (M_0M_0^{\dagger} + 2 {\Re} [M_1M_0^{\dagger}] ).
\label{01a}
\end{equation}

Here, the one-loop amplitude $M_1$ has structure of the sum of boson self-energy (BSE),
vertex (Ver) and box diagrams (see  Fig. \ref{2f}):
\begin{figure}
\vspace{0mm}
\begin{tabular}{ccccc}
\begin{picture}(60,60)
\put(-100,0){
\epsfxsize=5.5cm
\epsfysize=5.5cm
\epsfbox{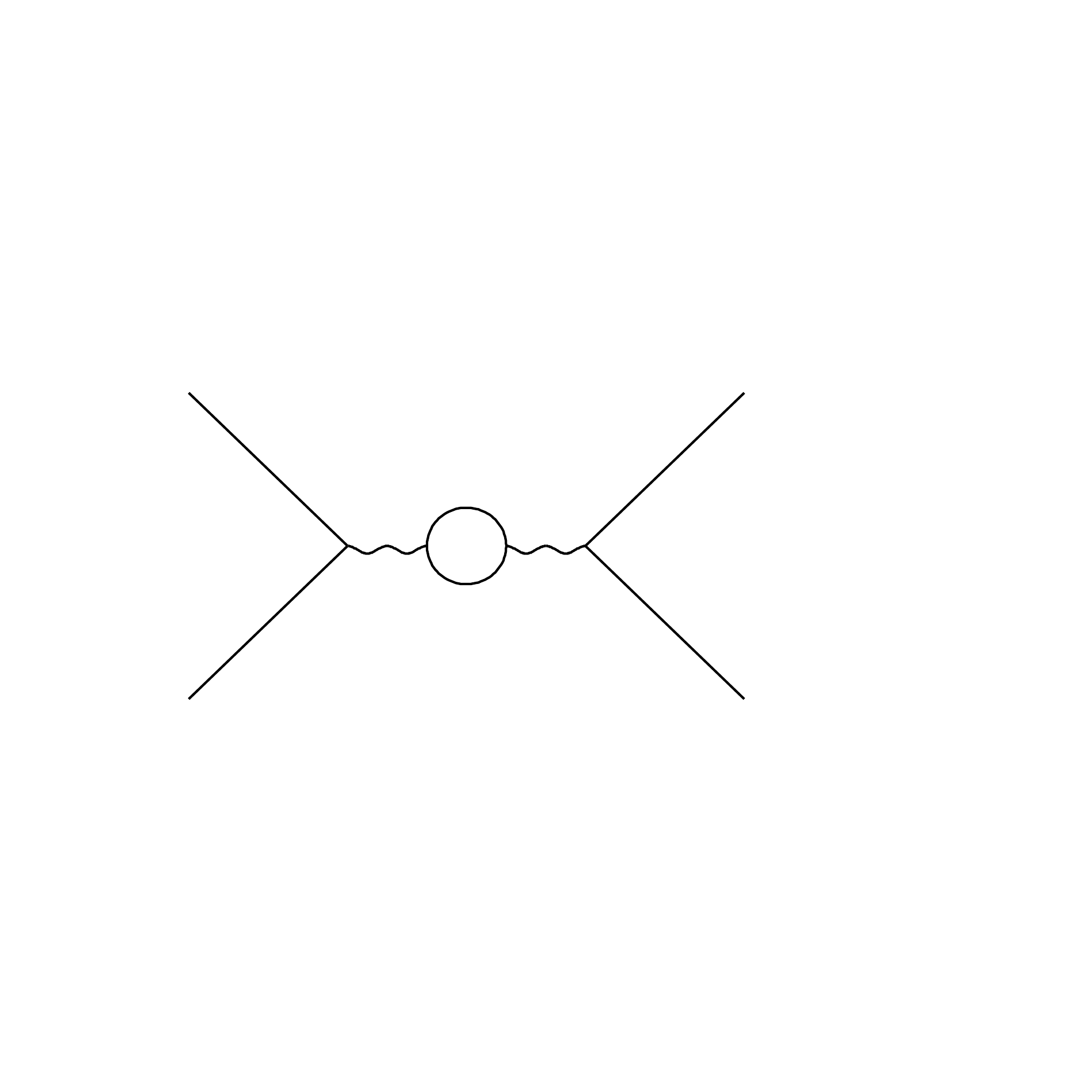} }
\end{picture}
&
\begin{picture}(60,100)
\put(-75,0){
\epsfxsize=5.5cm
\epsfysize=5.5cm
\epsfbox{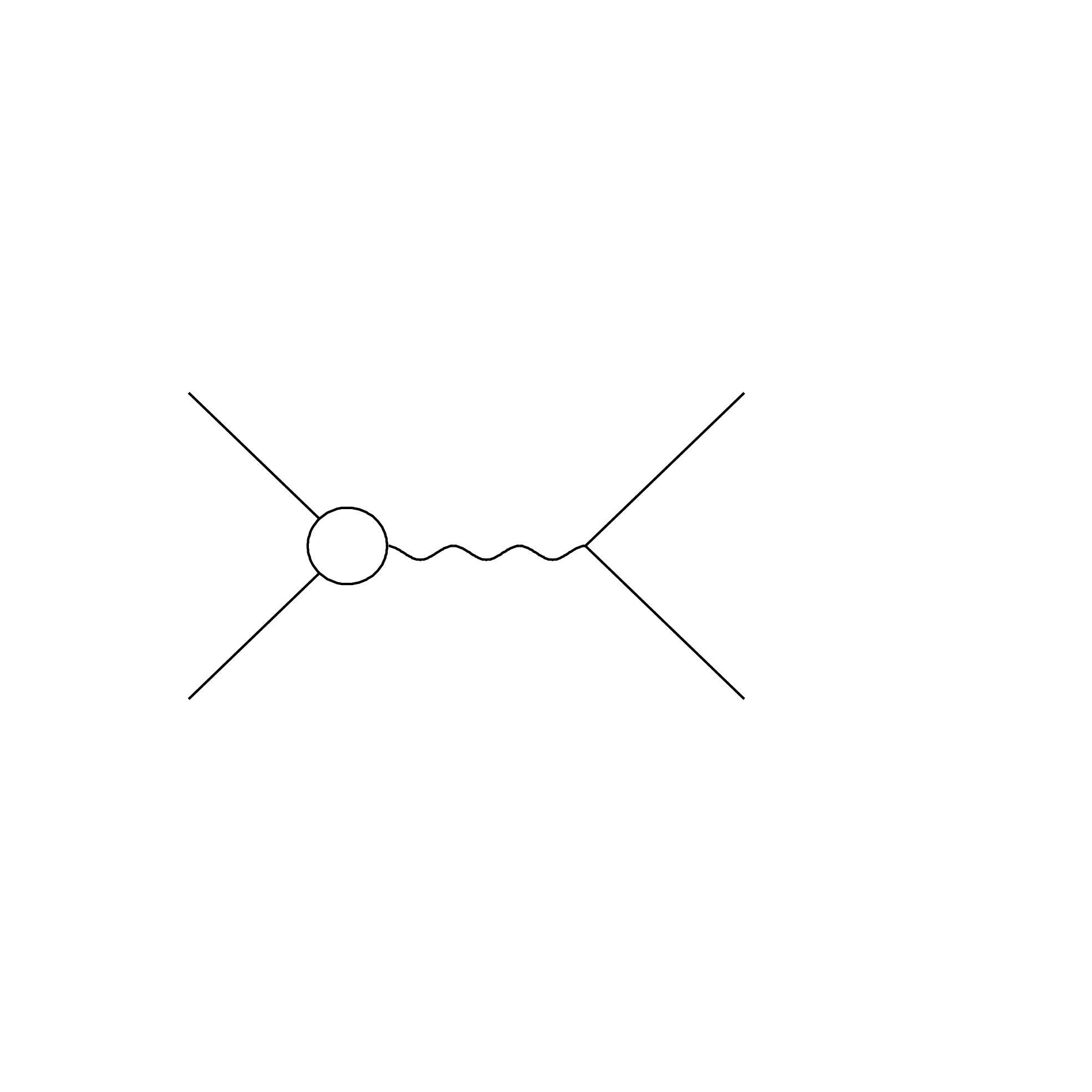} }
\end{picture}
&
\begin{picture}(60,100)
\put(-50,0){
\epsfxsize=5.5cm
\epsfysize=5.5cm
\epsfbox{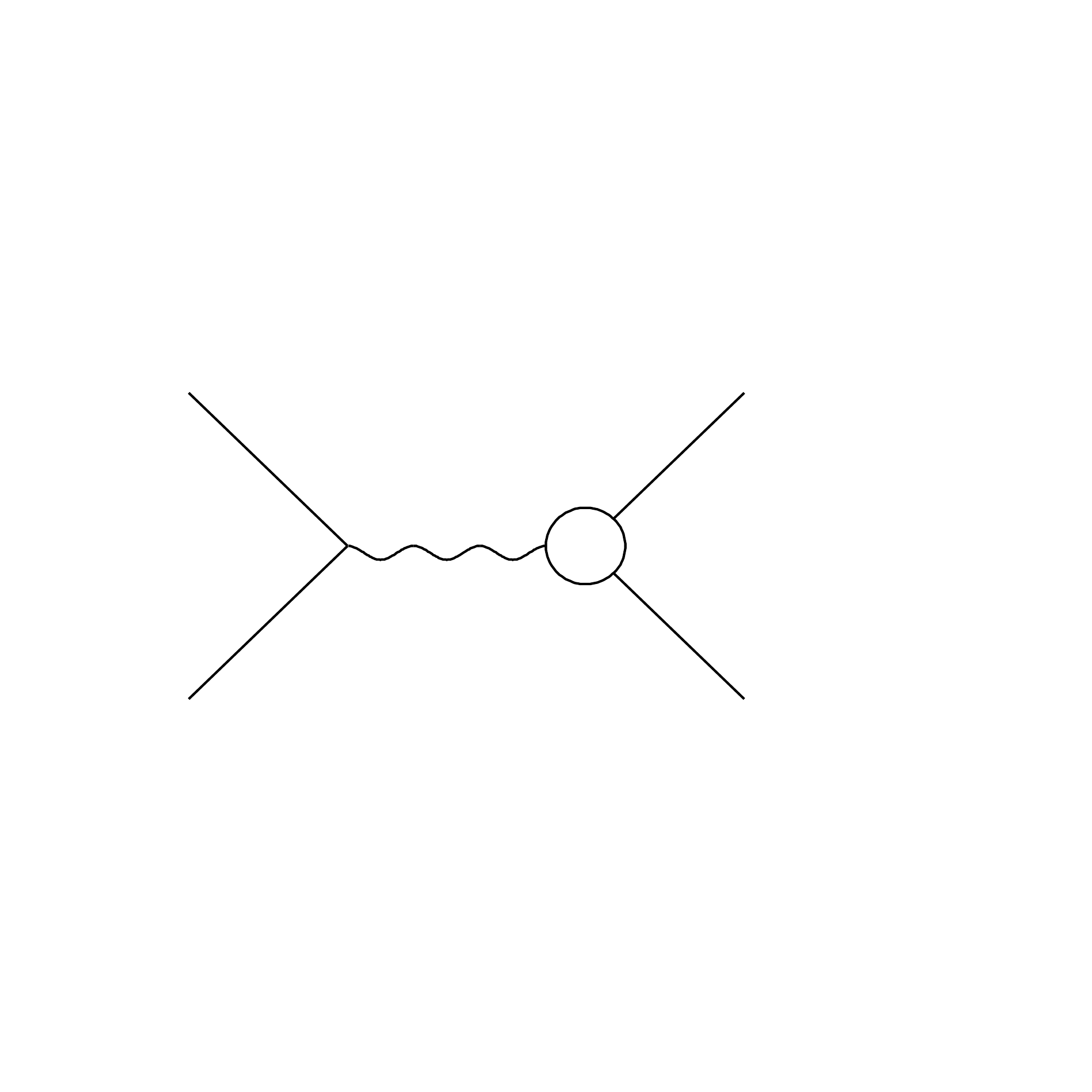} }
\end{picture}
&
\begin{picture}(60,60)
\put(-25,10){
\epsfxsize=4.5cm
\epsfysize=4.5cm
\epsfbox{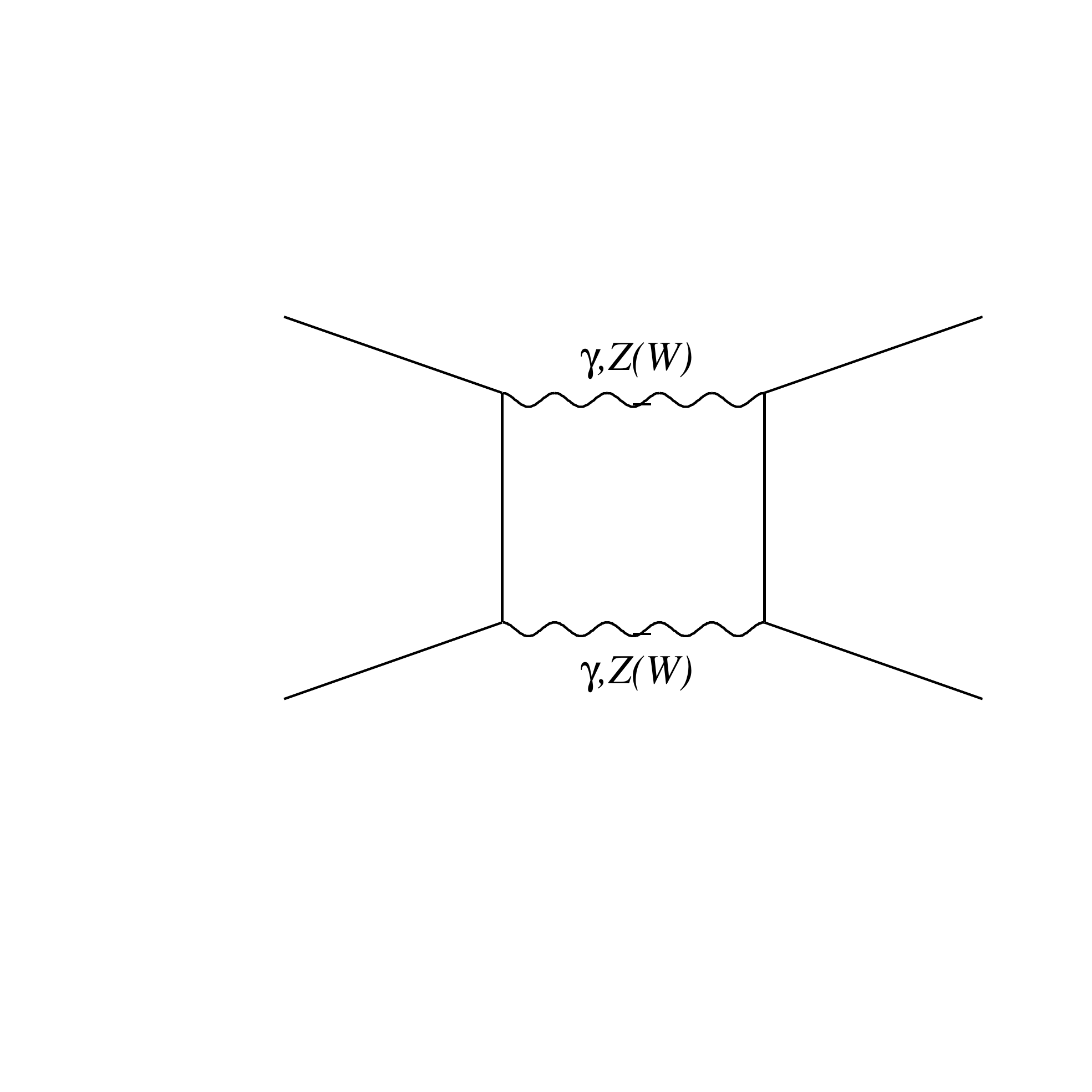} }
\end{picture}
&
\begin{picture}(60,100)
\put(10,10){
\epsfxsize=4.5cm
\epsfysize=4.5cm
\epsfbox{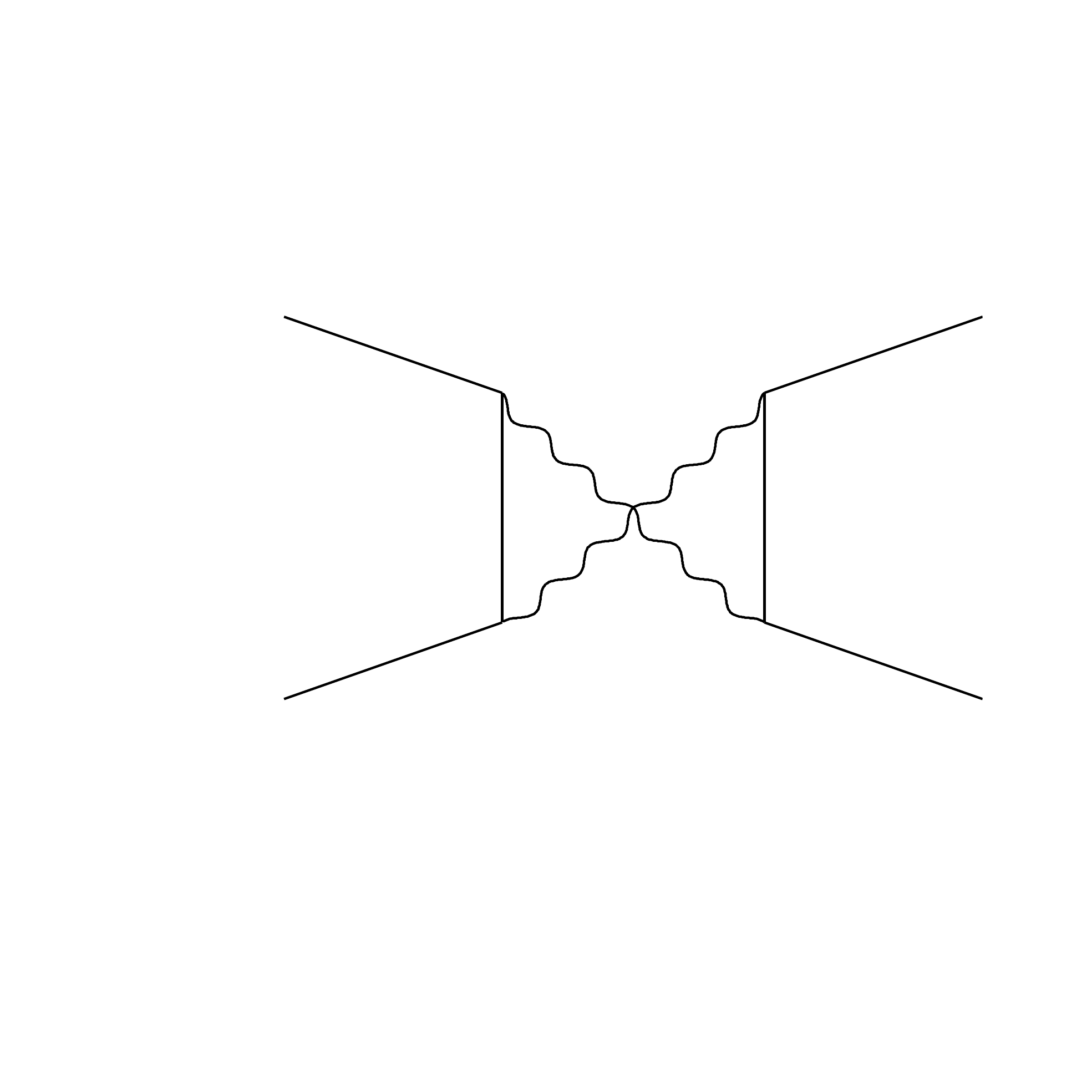} }
\end{picture}
\end{tabular}
\vspace*{-10mm}
\caption{\protect\it
One-loop diagrams:
the circles represent the contributions of self-energies and vertex functions.
Unsigned curly lines represent photon or $Z$-boson.
}
\label{2f}
\vspace{5mm}
\end{figure}
\begin{eqnarray}
M_{1}=M_{{\rm BSE}}+M_{{\rm Ver}}+M_{{\rm Box}}.
\end{eqnarray}
We use the on-shell renormalization scheme from \cite{BSH86, Denner},
so there are no contributions from the electron self-energies.
The infrared-finite BSE term can easily be expressed as:
\begin{equation}
M_{{\rm BSE}}= i\frac{\alpha}{\pi}  \sum_{i,j=\gamma,Z} I_\mu^i  D^{ij}_S J^{\mu,j},
\end{equation}
with 
\begin{equation}
D_S^{ij}=-D^{i} {\hat{\Sigma}}_T^{ij}(s) D^{j},
\label{D_S}
\end{equation}
where ${\hat{\Sigma}}_T^{ij}(s)$
is the transverse part of the renormalized photon, $Z$-boson and $\gamma Z$ self-energies.
The longitudinal parts of the boson self-energy make contributions
that are proportional to  $m^2/r$ ($r=s,t,u$);
therefore they are very small and are not considered here.

For the Belle~II experiment, the CM energy of the electron and positron is $\sqrt{s}=10.579$~GeV. 
Specifically for the Hollik renormalization conditions \cite{hollik}, we have the following numerical results for 
the truncated and renormalized self energies ($\hat{\Sigma}^{ij}_T$):
$$
\Re [{\hat{\Sigma}}_T^{\gamma\gamma}(s)] D^{\gamma s} = -0.0361,\
\Re [{\hat{\Sigma}}_T^{\gamma Z}(s)] D^{\gamma s}     = -0.0301,\
\Re [{\hat{\Sigma}}_T^{ZZ}(s)] D^{Z s}                = -0.0317,
$$
$$
\Im [{\hat{\Sigma}}_T^{\gamma\gamma}(s)] D^{\gamma s} =  0.0159,\
\Im [{\hat{\Sigma}}_T^{\gamma Z}(s)] D^{\gamma s}     = -0.0056,\
\Im [{\hat{\Sigma}}_T^{ZZ}(s)] D^{Z s}                = -0.0003.
$$

In order to derive the vertex amplitude (2nd and 3rd diagrams in Fig.~\ref{2f}),
we use the form factors notation in the manner similar to the work of \cite{BSH86}. 
Here, we will replace the coupling constants $v_g^j,\ a_g^j$ with the form factors
 $v_g^{\gamma (Z)} \rightarrow v_g^{F_{\gamma (Z)}}$,\
 $a_g^{\gamma (Z)} \rightarrow a_g^{F_{\gamma (Z)}}$,
where for the photon
\begin{eqnarray}
%\Gamma F_{\rm V}^{\gamma } & = &
v_g^{F_\gamma} & = &
  \frac{\alpha }{4\pi}
  \Bigl[ \Lambda_1^{\gamma}
   + \bigl({(v_g^Z)}^2 + {(a_g^Z)}^2 \bigr) \Lambda_2^Z
   + \frac{3}{4s_{W}^2} \Lambda_3^W \Bigr],
\label{HV1}
\end{eqnarray}
\begin{eqnarray}
%\Gamma F_{\rm A}^{\gamma } & = &
a_g^{F_\gamma} & = &
  \frac{\alpha }{4\pi}
  \Bigl[ 2v_g^Za_g^Z \Lambda_2^Z
   + \frac{3}{4s_{W}^2} \Lambda_3^W \Bigr],
\label{HV2}
\end{eqnarray}
and for  $Z$-boson
\begin{eqnarray}
%\Gamma F_{\rm V}^{Z } & = &
v_g^{F_Z}  & = &
  \frac{\alpha }{4\pi}
  \Bigl[ v_g^Z \Lambda_1^{\gamma}
   + v_g^Z \bigl({(v_g^Z)}^2 + 3{(a_g^Z)}^2 \bigr) \Lambda_2^Z +
%\nonumber
%\\[0.3cm] \displaystyle
%  &&    + 
 \frac{1}{8s_{W}^3c_{W}} \Lambda_2^W  -
 \frac{3c_{W}}{4s_{W}^3} \Lambda_3^W \Bigr],
\label{HV3}
\end{eqnarray}
\begin{eqnarray}
%\Gamma F_{\rm A}^{Z } & = &
a_g^{F_Z} & = &
  \frac{\alpha }{4\pi}
  \Bigl[ a_g^Z \Lambda_1^{\gamma}
   + a_g^Z \bigl( 3{(v_g^Z)}^2 + {(a_g^Z)}^2 \bigl) \Lambda_2^Z +
%\nonumber 
%\\[0.3cm] \displaystyle
% &&    + 
\frac{1}{8s_{W}^3c_{W}} \Lambda_2^W  -
   \frac{3c_{W}}{4s_{W}^3} \Lambda_3^W \Bigr].
\label{HV4}
\end{eqnarray}
The function $\Lambda_1^\gamma$ corresponds to the contribution of triangle diagrams with the photon in the loop,
 $\Lambda_2$ corresponds to  the triangle diagrams with the massive boson -- $Z$ or $W$, and $\Lambda_3$ corresponds to the  the triangle diagrams with 3-boson vertices  -- $WW\gamma$ or $WWZ$. These complex functions have been studied in detail
 and presented, for example, in  \cite{hollik}.
%
% We denote the vertices  with an additional photon as light vertices (LV)
% and the vertices  with an additional massive boson as heavy vertices (HV).
% The LV terms are proportional to the function $\Lambda_1$,
% and the HV terms are proportional to the combinations of functions $\Lambda_2$ and $\Lambda_3$
% as it is evident from Eqs. (\ref{HV1})-(\ref{HV4}).
%
Hence,
\begin{equation}
M_{{\rm Ver}} = i\frac{\alpha}{\pi} \sum_{j=\gamma,Z} 
\Bigl( 
 I_\mu^{F_j}  D^{js} J^{\mu,j} + I_\mu^j  D^{js} J^{\mu,{F_j}}
\Bigr).
\end{equation}
The infrared singularity is regularized by giving the photon a small mass $\lambda$
and in the vertex amplitude can be extracted in the form:
\begin{equation}
M_{{\rm Ver}}^\lambda =
-\frac{\alpha}{\pi} \Bigl( \ln\frac{s}{mm_f} -1 \Bigr)  \ln\frac{s}{\lambda^2} M_{0}.
\end{equation}
The remaining (infrared-finite) part of the vertex amplitude has a simple form
convenient for further analysis:
\begin{equation}
M_{{\rm Ver}}^f =
M_{{\rm Ver}} - M_{{\rm Ver}}^\lambda =
M_{{\rm Ver}}(\lambda^2 \rightarrow s).
\label{ver-sub}
\end{equation}

The box amplitude can be presented as a sum of all two-boson exchange contributions:
\begin{equation}
M_{{\rm Box}}= M_{\gamma\gamma} +M_{\gamma Z} + M_{ZZ} +M_{WW}.
\end{equation}
We need to account for both direct and crossed $\gamma\gamma$, $\gamma Z$ and $ZZ$-boxes:
\begin{equation}
M_{ij}= M_{ij}^D + M_{ij}^C\ \ (i,j=\gamma,Z),
\end{equation}
but, obviously, for $WW$-boxes we only need the direct expression.
The infrared parts of the $\gamma\gamma$- and $\gamma Z$-boxes are similarly given by
\begin{eqnarray}
M_{\gamma\gamma (\gamma Z)}^\lambda =
\frac{\alpha}{2\pi} \ln\frac{u}{t} \ln\frac{tu}{\lambda^4}  M_0.
\label{box-la}
\end{eqnarray}
The finite part of the $\gamma\gamma$-box can be found in \cite{13-A}.
Using asymptotic methods, we can significantly simplify the box amplitudes
containing at least one heavy boson (see, for example, \cite{ABIZ-prd}, where simplifications were done
on the cross-section level).
Finally, we provide the expressions for $M_{ii}^{D,C}$ in the low energy approximation:
\begin{eqnarray}
M_{ii}^D
=  - i \Bigl(\frac{\alpha}{\pi} \Bigr)^2   \frac{1}{16 m_i^2} 
 \bar u(-p_2) \gamma^\mu \gamma^\alpha \gamma^\nu (v_e^B-a_e^B\gamma_5) u(p_1) 
 \cdot
 \bar u(p_3) \gamma_\nu \gamma_\alpha \gamma_\mu (v_f^B-a_f^B\gamma_5) u(-p_4),
\end{eqnarray}
\begin{eqnarray}
M_{ii}^C
=  i \Bigl(\frac{\alpha}{\pi} \Bigr)^2   \frac{1}{16 m_i^2} 
 \bar u(-p_2) \gamma^\mu \gamma^\alpha \gamma^\nu (v_e^B-a_e^B\gamma_5) u(p_1) 
 \cdot
 \bar u(p_3) \gamma_\mu \gamma_\alpha \gamma_\nu (v_f^B-a_f^B\gamma_5) u(-p_4),
\end{eqnarray}
with the coupling-constants combinations for $ZZ$- and $WW$-boxes ($B=ZZ, WW$)
\begin{equation}
v^{ZZ}={(v^Z_g)}^2+{(a^Z_g)}^2,\ a^{ZZ}=2v^Z_ga^Z_g,\ v^{WW}=a^{WW}=1/(4s_W^2).
\end{equation}

Now we are ready to present the one-loop  amplitude as the sum of 
IR-divergent (index $\lambda$) and IR-finite (index $f$) parts: $M_1 = M_1^\lambda + M_1^f$,
where
\begin{equation}
M_1^\lambda = \frac{\alpha}{2\pi} {\Gamma_1^{\lambda}} M_0,\
\Gamma_1^{\lambda} = 4 B \ln\frac{\lambda}{\sqrt{s}}.
\label{mmm1}
\end{equation}
and the value $B$ can be presented in the form
\begin{equation}
B= \ln\frac{st}{m~m_f~u} - 1.
\end{equation}
%In the text below we can forget about the imaginary part of infrared cross section:
%$ \Gamma_1^{\lambda} \rightarrow {\rm Re} \Gamma_1^{\lambda} $ and 
%$ B \rightarrow {\rm Re} B $.
Using (\ref{mmm1}), it is straightforward to write the expression for the NLO cross section:
\begin{equation}
%\sigma_1^V =\frac{\pi^3}{s} {\rm Re} [M_1M_0^+] =  \sigma^{\lambda}_1 + \sigma^{f}_1,
\sigma_1^V =\frac{\pi^3}{s} {\Re} [M_1M_0^{\dagger}] =  \sigma^{\lambda}_1 + \sigma^{f}_1,
\end{equation}
where IR-divergent and regularized NLO cross section is given by
\begin{equation}
\sigma^{\lambda}_{1} = \frac{\alpha}{\pi} \Gamma_1^{\lambda}  \sigma^0.
\end{equation}
The IR-finite part can be represented using the notation of the relative correction ($\Gamma_1^f$)
\begin{equation}
\sigma^{f}_{1} = \frac{\alpha}{\pi}   \Gamma_1^f  \sigma^0 
= \sigma_{\rm BSE} + \sigma^{f}_{\rm Ver} + \sigma^{f}_{\rm Box},
\label{def}
\end{equation}
where at one-loop level the cross sections are written as follows:
\begin{eqnarray}
\sigma_{\rm BSE} 
&=& \frac{2 \pi \alpha^2}{s} {\Re}
\sum_{i,j,k=\gamma,Z} D_S^{ij}{D^{k}}^* \mu^{ikjk},
\label{cs-BSE}
\\ 
\sigma^f_{\rm Ver} 
&=& \frac{2\pi \alpha^2}{s} {\Re}
\sum_{i,k=\gamma,Z} D^{i}{D^{k}}^* 
[ \mu^{F_ikik} + \mu^{ikF_ik} ],
\label{cs-Ver}
\\
%\sigma^{f}_{\rm Box} &=&  \frac{\pi^3}{s} {\rm Re} (M^f_{\gamma\gamma} +M^f_{\gamma Z} + M_{ZZ} +M_{WW})  M_0^+.
\sigma^{f}_{\rm Box} &=&  \frac{\pi^3}{s} {\Re} (M^f_{\gamma\gamma} +M^f_{\gamma Z} + M_{ZZ} +M_{WW})  M_0^{\dagger}.
\label{cs-Box}
\end{eqnarray}
In (\ref{cs-Ver}), the IR-finite part of vertex form factors was used according (\ref{ver-sub}).

\section{Bremsstrahlung:  Cancellation of infrared divergence }

\begin{figure}
\vspace{0mm}
\begin{tabular}{cccc}
\begin{picture}(60,60)
\put(-100,0){
\epsfxsize=5.5cm
\epsfysize=5.5cm
\epsfbox{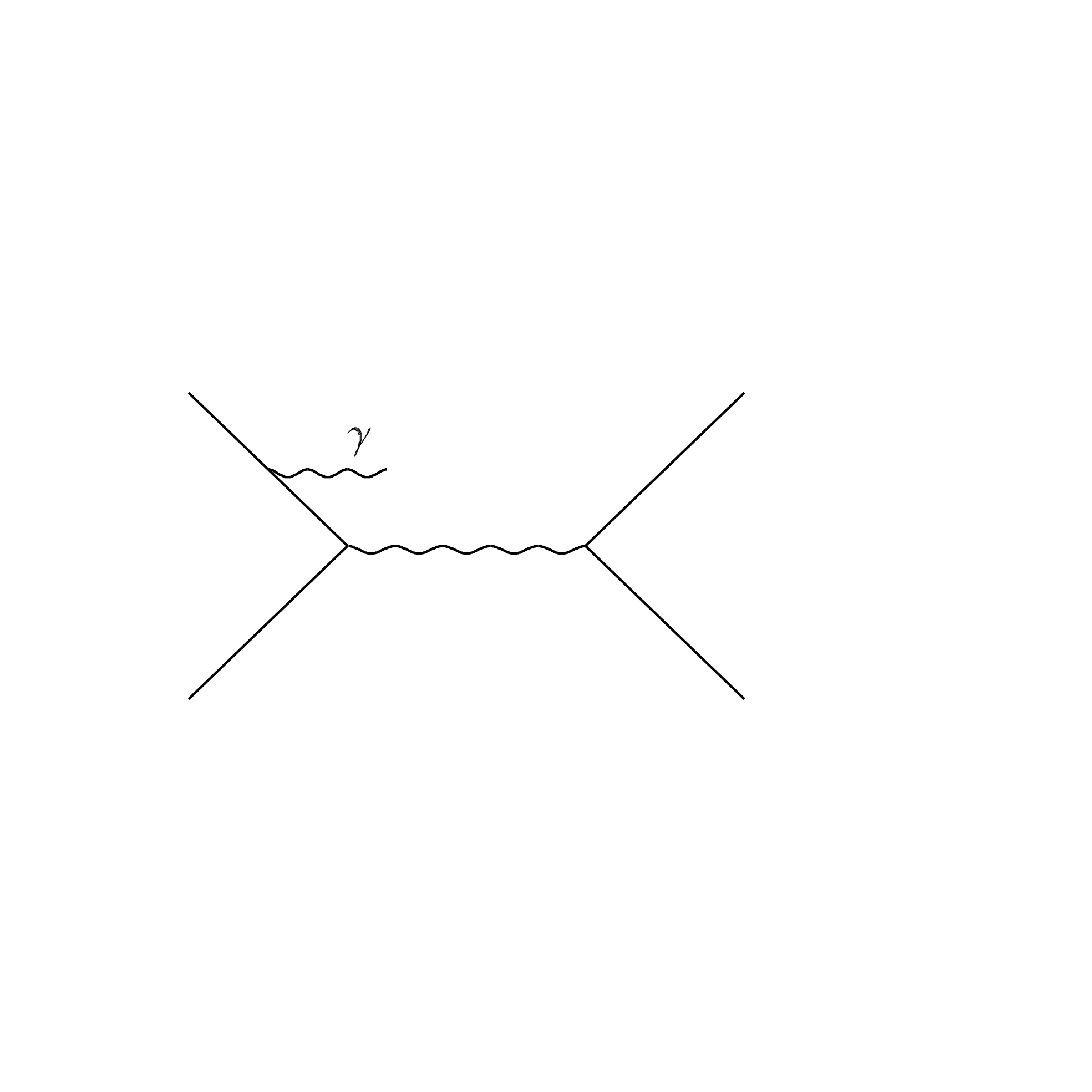} }
\end{picture}
&
\begin{picture}(60,100)
\put(-70,0){
\epsfxsize=5.5cm
\epsfysize=5.5cm
\epsfbox{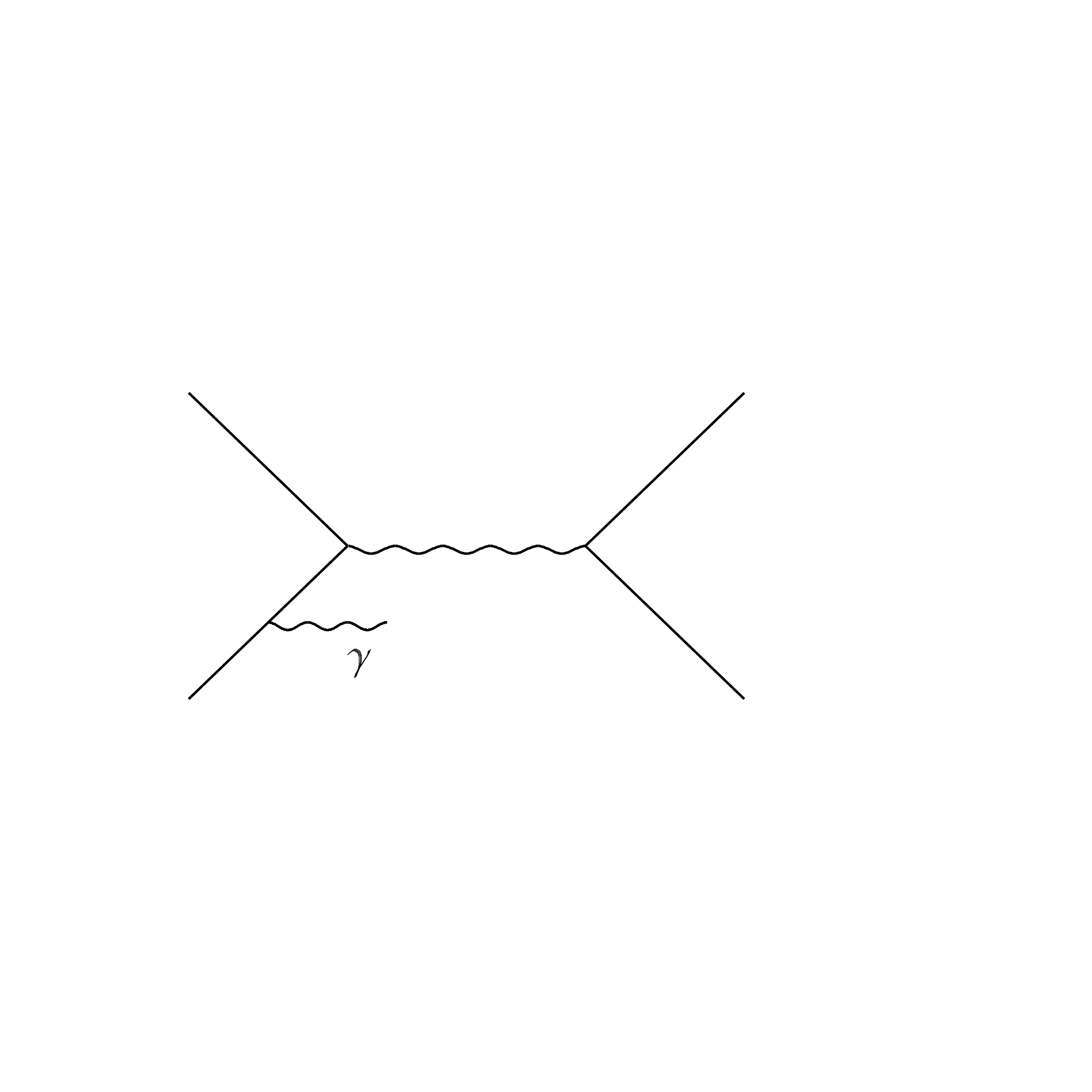} }
\end{picture}
&
\begin{picture}(60,100)
\put(-40,0){
\epsfxsize=5.5cm
\epsfysize=5.5cm
\epsfbox{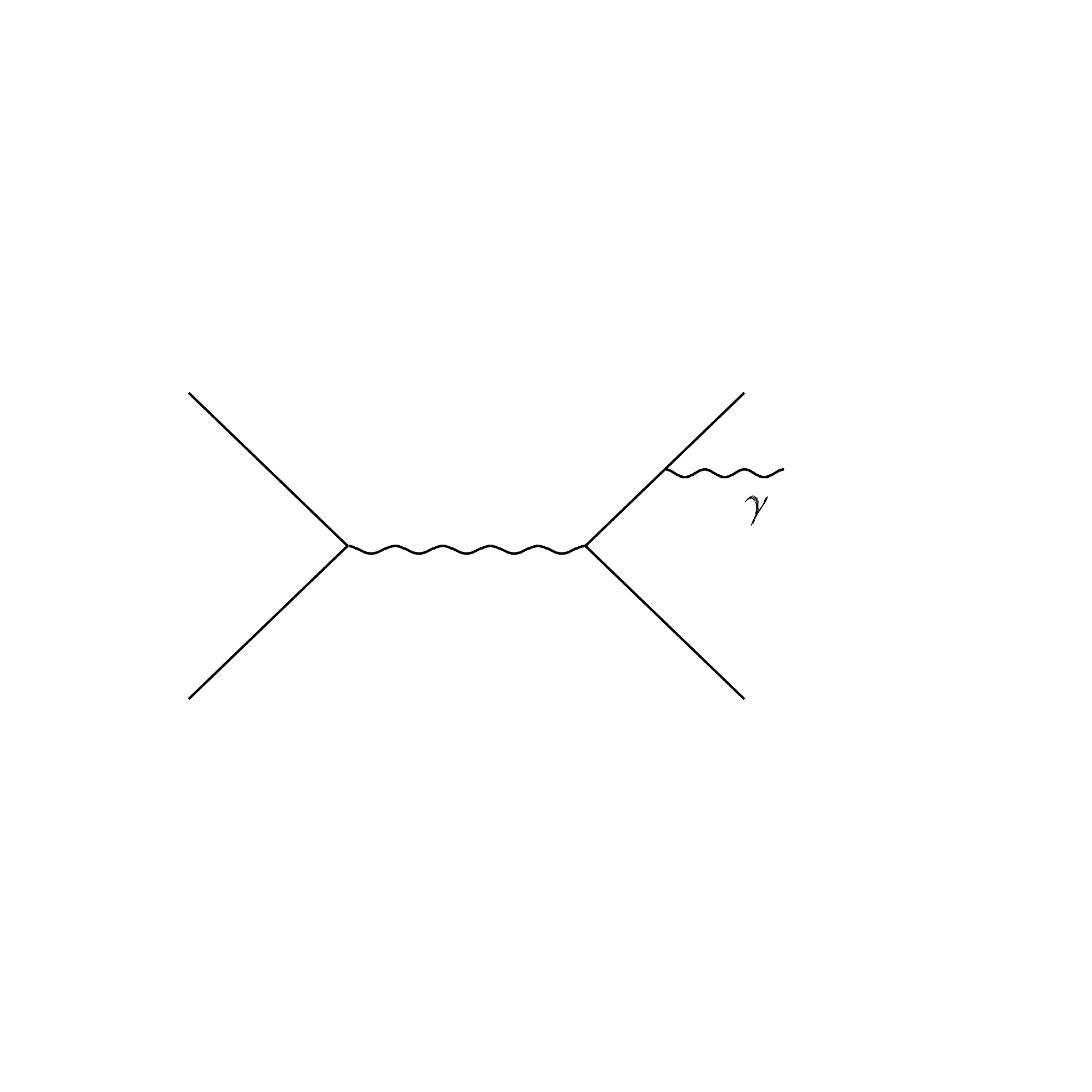} }
\end{picture}
&
\begin{picture}(60,60)
\put(-10,0){
\epsfxsize=5.5cm
\epsfysize=5.5cm
\epsfbox{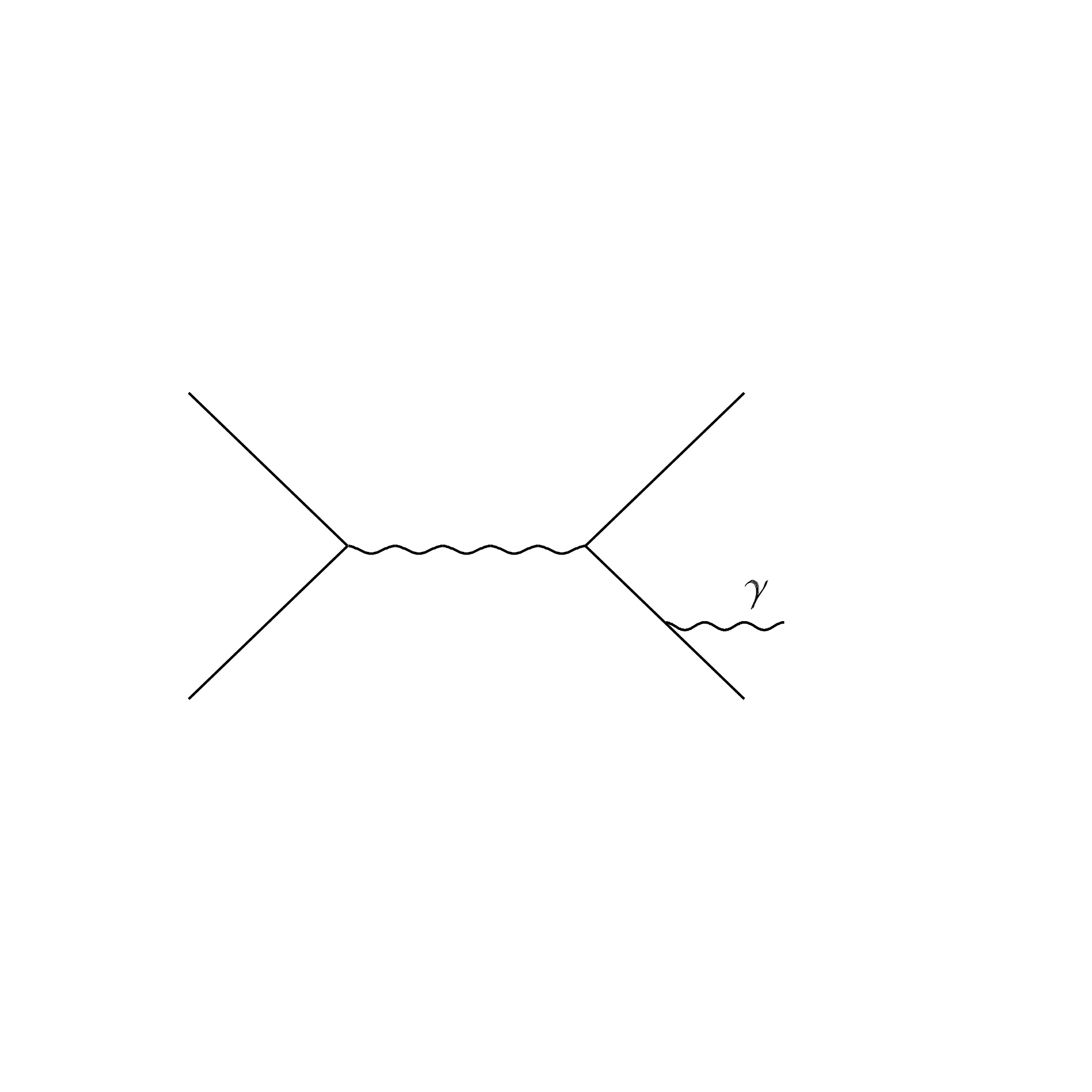} }
\end{picture}
\end{tabular}
\vspace*{-10mm}
\caption{\protect\it
Diagrams with photon emission.
}
\label{bre}
\vspace{5mm}
\end{figure}

The bremsstrahlung diagrams are illustrated in Fig.~\ref{bre},
where the first two diagrams correspond to initial state radiation (ISR),
whereas the last two correspond to final state radiation (FSR).

%.Fully differential cross section for the process
We express the full differential cross section for the process
\begin{equation}
e^-(p_1) + e^+(p_2) \rightarrow f^-(p_3)+f^+(p_4)+\gamma(p),
\label{BREM}
\end{equation}
%has the form
as
\begin{equation}
d\sigma_R = \frac{\alpha^3}{\pi^2 s} \sum |R|^2 d\Gamma_3,
\label{sig-R}
\end{equation}
where phase space is defined as
\begin{equation}
d\Gamma_3 = \Gamma (p_1+p_2-p_3-p_4-p) \frac{d^3p_3}{2{p_3}_0} \frac{d^3p_4}{2{p_4}_0} \frac{d^3p}{2{p}_0}
\label{G3}
\end{equation}
%The squared by absolute value amplitude is expanded and given by the sum of ISR, interference and FSR parts, respectively:
and
\begin{equation}
\sum |R|^2 = \sum\limits_{i,j=\gamma,Z} \bigl( Q_e^2R_e^{ij}+Q_eQ_f R_i^{ij} + Q_f^2R_f^{ij} \bigr).
\end{equation}
where the three terms in the sum are the ISR, interference and FSR parts, respectively.

The ISR part can be written as
\begin{eqnarray}
R_e^{ij} = -\Pi^i {\Pi^j}^* 
{\rm Tr} 
\Bigl[ \bigl(  \gamma^\mu \frac{-2p_1^\rho+\hat p \gamma^\rho}{z_1} 
              + \frac{2p_2^\rho-\gamma^\rho \hat p }{v_1}\gamma^\mu \bigr)
           \frac 1 2 (\lambda_1^{ij} - \lambda_2^{ij} \gamma_5 ) \hat p_1    
\nonumber \\
        \bigl(  \frac{-2p_1^\rho+\gamma^\rho \hat p }{z_1}\gamma^\nu
              + \gamma^\nu \frac{2p_2^\rho-\hat p \gamma^\rho}{v_1} \bigr)
           \frac 1 2  \hat p_2  \Bigr]
{\rm Tr} 
\bigl[ \gamma_\mu ({\lambda_{f V}^{ij}} - {\lambda_{f A}^{ij}}\gamma_5) \hat p_4 \gamma_\nu \hat p_3 \bigr ].
\label{ISR}
\end{eqnarray}
The FSR part can be found by substitution 
$ R_f^{ij} = R_e^{ij} (\Pi^j \rightarrow D^j, p_{1,2} \leftrightarrow - p_{4,3}, 
\lambda_{1,2} \leftrightarrow \lambda_{fV,fA})$,
and for the interference term we have
\begin{eqnarray}
R_i^{ij} = -\Pi^i {D^j}^* 
{\rm Tr} 
\Bigl[ \bigl(  \gamma^\mu \frac{-2p_1^\rho+\hat p \gamma^\rho}{z_1} 
              + \frac{2p_2^\rho-\gamma^\rho \hat p }{v_1}\gamma^\mu \bigr)
           \frac 1 2 (\lambda_1^{ij} - \lambda_2^{ij} \gamma_5 ) \hat p_1   
        \gamma_\nu \frac 1 2 \hat p_2  \Bigr]
\nonumber \\
{\rm Tr} 
\bigl[ \gamma_\mu ({\lambda_{f V}^{ij}} - {\lambda_{f A}^{ij}}\gamma_5) \hat p_4
        \bigl(  \frac{-2p_4^\rho-\gamma^\rho \hat p }{v}\gamma^\nu
              + \gamma^\nu \frac{2p_3^\rho+\hat p \gamma^\rho}{z} \bigr)
      \hat p_3  \bigr] -
\nonumber \\
 -D^i {\Pi^j}^* 
{\rm Tr} 
\Bigl[ \gamma_\mu \frac 1 2 (\lambda_1^{ij} - \lambda_2^{ij} \gamma_5 ) \hat p_1
\bigl(   \frac{-2p_1^\rho+\hat p \gamma^\rho}{z_1} \gamma^\nu
              + \gamma^\nu \frac{2p_2^\rho-\gamma^\rho \hat p }{v_1} \bigr)
        \frac 1 2 \hat p_2  \Bigr]
\nonumber \\
{\rm Tr}
\bigl[ 
        \bigl( \gamma^\mu \frac{-2p_4^\rho-\gamma^\rho \hat p }{v}
              + \frac{2p_3^\rho+\hat p \gamma^\rho}{z} \gamma^\mu  \bigr)
({\lambda_{f V}^{ij}} - {\lambda_{f A}^{ij}}\gamma_5) \hat p_4
\gamma_\nu  \hat p_3  \bigr].
\label{INT}
\end{eqnarray}
For the radiative case, the truncated propagator has the following form
\begin{equation}
\Pi^{j}=\frac{1}{s-z-v-m_j^2+im_j\Gamma_j}\ \ (j=\gamma,Z).
\label{rad-structure}
\end{equation}

In the last three equations we have used four radiative invariants (they tend to zero at $p \rightarrow 0 $):
\begin{equation}
 z_1=2p_1p,\ v_1=2p_2p,\ z=2p_3p,\ v=2p_4p,
\end{equation}
together with three invariants $s,\ t,\ u$, and taking into account the 
momentum conservation, we can write the following identities
\begin{equation}
z-z_1=v_1-v,\ \ s+t+u=v+2m^2+2m_f^2.
\end{equation}
Here we have five ($4+3-2=5$) independent variables in the description of bremsstrahlung process.
Phase space of the emitted photon $d\Gamma_3$ can be expressed in the basis of these invariants
\begin{equation}
d\Gamma_3 = \frac{\pi}{16 s} \frac{dt~dv~dz~dv_1}{\sqrt{-\Gamma_4}},
\label{G3a}
\end{equation}
and  $-\Gamma_4$  is a usual Gram determinant.

Next we divide the bremsstrahlung cross section into soft and hard parts using a separator $\omega$.
The soft part $\sigma^{\gamma}(\omega)$ is  integrated under the condition that the photon energy 
(all energies are in the CM system  of $e^- e^+$) is less than $\omega$.
The hard part of bremsstrahlung cross section $\sigma^{\gamma}(\omega,\Omega)$  corresponds 
to the photon energy  greater than $\omega$ and less than $\Omega$.
To evaluate the cross section induced by the emission of a single soft photon,
we follow the methods of Berends {\it et al.}  \cite{BGG} (see also  \cite{HooftVeltman}, \cite{KT1}).
To obtain the result, we must calculate the 3-dimensional
integral over the phase space of the emitted real soft photon:
\begin{eqnarray}
  L(\lambda,\omega) =
-\frac{1}{4\pi}
\int\limits_{p_0<\omega}  \frac{d^3 p}{p_0}
T^{\alpha}(p) T_{\alpha}(p)   =  -\Gamma_1^{\lambda} + R_1,
\label{1}
\end{eqnarray}
where
\begin{eqnarray}
T^{\alpha}(p)
=\frac{p_1^{\alpha}}{p_1p} - \frac{p_2^{\alpha}}{p_2p} +\frac{p_3^{\alpha}}{p_3p} - \frac{p_4^{\alpha}}{p_2p},
\end{eqnarray}
and
\begin{eqnarray}
R_1=-4B \ln\frac{\sqrt{s}}{2\omega} 
- \Bigl( \ln\frac{m^2}{s} +\frac{1}{2}\ln^2\frac{m^2}{s} + \frac{\pi^2}{3} \Bigr)
\nonumber\\
 		- \Bigl( \ln\frac{m_f^2}{s} +\frac{1}{2}\ln^2\frac{m_f^2}{s} + \frac{\pi^2}{3} \Bigr)
		+2 \mbox{Li}_2\frac{-t}{u} -2 \mbox{Li}_2\frac{-u}{t}.
\end{eqnarray}
%As a results the soft cross section can be factorized by Born cross section in soft-photon bremsstrahlung approximation:
As a result the soft cross section can be factorized in terms of the Born cross section in this soft-photon bremsstrahlung approximation:
\begin{eqnarray}
\sigma^{\gamma}(\omega)= \frac{\alpha}{\pi} \bigl[ -\Gamma_1^{\lambda} +R_1 \bigr] \sigma^0.
\label{SPA}
\end{eqnarray}
In the rest of the article we will refer to it as the Soft Photon Approximation (SPA). 
The contribution due to soft photons is evaluated in with our semi-automatic approach, with no further simplifications. 
%The SPA has a $\mu^+\mu^-$ final state, with no on-shell photon.

The hard photon approach (HPA) fully accounts for the photon in the final state, where the HPA emission cross section is calculated with a Monte Carlo integration technique
using the VEGAS routine \cite{VEGAS} in the region $\omega \leq p_0 \leq \Omega$. 
The hard photon bremsstrahlung cross section can be expressed as
\begin{eqnarray}
\sigma^{\gamma}(\omega,\Omega)= 
\frac{\alpha^3}{8\pi s} 
\int\limits_{\omega \leq p_0 \leq \Omega} 
\frac{dv~ dz~ dv_1}{\sqrt{-\Gamma_4}} \frac{s-v}{s} \sum |R|^2 \theta(-\Gamma_4).
\label{HPA}
\end{eqnarray}
Here we have used the ultra-relativistic form of the Jacobian $(s-v)/s$, which originates in the transition from radiative $t$ invariant
\begin{eqnarray}
t= \frac{1}{2} \biggl( 2m^2+2m^2_f-s+v+\cos\theta\sqrt{\frac{s-4m^2}{s}}\sqrt{(s-v)^2-4m^2_fs} \biggr)
\label{trad}
\end{eqnarray}
to the cosine of the scattering angle: $\cos\theta$. The integral in (\ref{HPA}) can be evaluated first analytically  over the 
variables $v_1$ and $z$ (explicit details are given in \cite{w}), and then numerically.

%To provide better convergence, first we can integrate analytically over two of the radiative invariants.
%We choose for analytical integration $v_1$ and $z$. But previously we made the change of variables:
%$(z,v)\rightarrow (z,x)$, where $x=z+v$.
%As a result the last variable of integration $x$ will acquire the following limits:
%$ 2\omega\sqrt{s} \leq x \leq 2\Omega\sqrt{s}$.
%% since in the c.m.s the energy $p_0=(v+z)/\sqrt{4s}$ ...
%For analytical integration  we use  the integral expressions calculated early one of us in \cite{w}.

Putting it all together at one-loop, we get:
\begin{equation}
\sigma^{1} 
= \sigma_1^V + \sigma^{\gamma}(\omega) + \sigma^{\gamma}(\omega,\Omega).
\label{1IR}
\end{equation}
Obviously $\sigma^{1}$ does not depend on either $\lambda $ or $\omega$.
The independence on the mass of the photon can be justified by direct analytical cancellations of $\lambda$, 
and as a result we get
\begin{equation}
\sigma_1^{\lambda} + \sigma^{\gamma}(\omega) 
=   \frac{\alpha}{\pi} R_1  \sigma^0.
\end{equation}
Independence on $\omega$ is obvious by definition.
But since the hard photon bremsstrahlung integration was performed numerically, we verify that and
observe $\omega$ independence with a relative numerical uncertainty not exceeding order of $10^{-4}$.

\section{Numerical Results}
\label{sec:NumRes}
Electroweak input parameters of the on-shell renormalization scheme
($m_{W},$ $m_{Z}$ and $\alpha$) are naturally defined as measurable
quantities with fixed values at all orders of perturbation theory.
As a result, the $s^2_W=1-\frac{m_W^2}{m_Z^2}$ definition of the weak mixing angle 
is also fixed at all orders of perturbation theory.
%\begin{eqnarray}
%\sin^{2}\theta_{W} & = & 1-\frac{m_{W}^{2}}{m_{Z}^{2}}.\label{eq:a1}
%\end{eqnarray}
From muon decay one can establish the relationship between the most
precisely measured quantity, Fermi constant $G_{\mu}=1.1663787(6)\cdot10^{-5}\ \mbox{GeV}^{-2}$,
and the $m_{W}$. This can be achieved by comparing muon
lifetimes calculated in Fermi four-fermion interaction theory and
the Standard Model calculations at one-loop level. This gives the following
relationship:
\begin{eqnarray}
m_{W}^{2} & = & \frac{\pi\alpha}{\sqrt{2}G_{\mu}s^{2}_{W}(1-\Delta r)}.\label{eq:a2a}
\end{eqnarray}
Here $\Delta r$ is a radiative correction which is calculated in
the on-shell renormalization scheme~\cite{Hollik} and has the following structure:
\begin{eqnarray}
\Delta r & = & \frac{\Re\hat{\Sigma}_{WW}(0)}{m_{W}^{2}}+\frac{\alpha}{4\pi s_{W}^{2}}\left(6+\frac{7-4s_{W}^{2}}{2s_{W}^{2}}\ln c_{W}^{2}\right)+\frac{c_{W}^{2}}{m_{Z}^{2}s_{W}^{2}}\Re\left[\frac{\hat{\Sigma}_{\gamma Z}^{2}(m_{Z}^{2})}{m_{Z}^{2}+\hat{\Sigma}_{\gamma\gamma}(m_{Z}^{2})}\right].\label{eq:a3}
\end{eqnarray}
Here, $\hat{\Sigma}_{V_1 V_2}$ is defined as truncated and renormalized self-energy graph for $V_1\rightarrow V_2$ mixing.

The formulae (\ref{eq:a2a}) and (\ref{eq:a3}) gives the effective
$m_{W}$ value of $80.4628 \ \mbox{GeV}$, which we use in our calculations. For the numerical calculations we have used $\alpha=1/137.035 999$, $m_Z=91.1876\ \mbox{GeV}$,\ and $m_H=125\ \mbox{GeV}$  as input parameters  according to \cite{PDG16}. 
%\cite{PDG08}
%We use the effective value of  $m_W=80.4628\ \mbox{GeV}$.  
The electron, muon, and $\tau$-lepton
masses are taken as $m_e=0.510 998 910\ \mbox{MeV}$,  $m_\mu=0.105 658 367\ \mbox{GeV}$, $m_\tau=1.776 84\ \mbox{GeV}$
and the quark masses for loop contributions as
$m_u=0.069 83\ \mbox{GeV}$,\ $m_c=1.2\ \mbox{GeV}$,\ $m_t=174\ \mbox{GeV}$,
$m_d=0.069 84\ \mbox{GeV}$,\ $m_s=0.15\ \mbox{GeV}$, and $m_b=4.6\ \mbox{GeV}$.
The light quark masses provide a shift in the fine structure constant due to hadronic
vacuum polarization $\Delta \alpha_{had}^{(5)}(m_Z^2)$=0.02757 \cite{jeger},
where
\begin{equation}
\Delta \alpha_{had}^{(5)}(s)=\frac{\alpha}{3\pi} \sum_{q=u,d,s,c,b} Q_q^2 \biggl(\ln\frac{s}{m_q^2}-\frac{5}{3}\biggr).
\end{equation}
Here, we choose to use the light quark masses
as parameters regulated by the hadronic vacuum polarization.
%Finally, for the mass of the Higgs boson, we take $m_H=125\ \mbox{GeV}$.
%Although this mass is still to be determined experimentally with sufficient accuracy, 
%the dependence of EWC on $m_H$ is rather weak. 

Let us introduce superscript $C$ which corresponds to the specific type of contribution to a cross section or asymmetry.
$C$ can be 0 (Born contribution), 1 (one-loop EWC contribution), or 0+1 (both these types):
$C=\{0, 1, 0\!\!+\!\!1\}$.
The relative correction to the unpolarized differential cross section (denoted by subscript $00$) is
\begin{eqnarray}
\delta_{00}= 
 \frac{\sigma^{1}_{L}+\sigma^{1}_{R}}
      {\sigma^{0}_{L}+\sigma^{0}_{R}}=
\frac{\sigma^{1}_{00}}{\sigma^0_{00}},
\label{rc1}
\end{eqnarray}            
where the subscripts $L$ and $R$ on the cross sections correspond
to the degree of polarization for electron $p_{B}$ = $-1$ and $p_{B}$ = $+1$, respectively.
The relative correction to the unpolarized total cross section is
\begin{eqnarray}
\delta_{T}= 
 \frac{\Sigma^{1}_{F}+\Sigma^{1}_{B}}
      {\Sigma^{0}_{F}+\Sigma^{0}_{B}}=
\frac{\Sigma^{1}_{T}}{\Sigma^0_{T}},
\label{rc2}
\end{eqnarray}            
where forward and backward cross sections are defined as
$$ \Sigma^C_{F} =\int\limits_0^{\cos a} \sigma^{C}_{00} \cdot d(\cos\theta),\ \ \
   \Sigma^C_{B} =\int\limits_{-\cos a}^0 \sigma^{C}_{00} \cdot d(\cos\theta).
$$
The relative correction to integrated cross section is
\begin{eqnarray}
\delta_{\Sigma}= 
 \frac{\Sigma^{1}_{L}+\Sigma^{1}_{R}}
      {\Sigma^{0}_{L}+\Sigma^{0}_{R}},
%=
%\frac{\Sigma^{1}_{00}}{\Sigma^0_{00}},
\label{rc3}
\end{eqnarray}            
%where  left and right integrated (over $a\le \cos \theta \le b$)  cross sections are given by
where the left and right integrated  cross sections are given by
$$ \Sigma^C_{L} =\int\limits_{\cos b}^{\cos a} \sigma^{C}_{L} \cdot d(\cos\theta),\ \ \
   \Sigma^C_{R} =\int\limits_{\cos b}^{\cos a} \sigma^{C}_{R} \cdot d(\cos\theta),
$$
and the integration is over the cosine of the polar angle of the out-going negative fermion.

The parity-violating (left-right) asymmetry is defined in a traditional way
\begin{equation}
A^C_{LR} =
 \frac{\sigma^C_{L}-\sigma^C_{R}}
      {\sigma^C_{L}+\sigma^C_{R}},
\label{A}
\end{equation}
which is at the Born level has the following structure

 \begin{eqnarray}
 A^0_{LR}&=&-\frac{s}{4m_W^2}\frac{(y-1)^2}{2(y-1)y+1}\frac{1-4s_W^2}{s^2_W}
\nonumber \\
\nonumber \\
&=&-\frac{2s}{m^2_Z}\Bigg{[}a_e v_\mu+a_\mu v_e \frac{(1-2y)}{2(y-1)y+1}\Bigg{]},
 \label{ALRtree}
 \end{eqnarray}
with $y=-t/s$. The left-right integrated asymmetry is constructed from integrated cross sections 

 \begin{equation}
 A^C_{LR\Sigma} =
  \frac{\Sigma^C_{L}-\Sigma^C_{R}}
       {\Sigma^C_{L}+\Sigma^C_{R}}.
 \label{ALRS}
 \end{equation}
Born results for the integrated asymmetry can be written in the following form

 \begin{eqnarray}
 A^0_{LR\Sigma}&=&-\frac{s}{8m_W^2}\frac{1-4s_W^2}{s^2_W}\frac{2\cos{a}\cos{b}+6(\cos{a}+\cos{b})+\cos{2a}+\cos{2b}+8}
{2\cos{a}\cos{b}+\cos{2a}+\cos{2b}+8}
\nonumber \\
\nonumber \\
&=&-\frac{2s}{m^2_Z}\Bigg{[}a_e v_\mu+ a_\mu v_e \frac{6(\cos{a}+\cos{b})}{2\cos{a}\cos{b}+\cos{2a}+\cos{2b}+8}\Bigg{]}.
 \label{ALRtreeSigma}
 \end{eqnarray}

In the case, when we consider full acceptance ($a=0^\circ$ and $b=180^\circ$), expressions for the integrated asymmetry simplify considerably:

 \begin{eqnarray}
 A^0_{LR\Sigma}\vert_{0^\circ}^{180^\circ}=-\frac{s}{8m_W^2}\frac{1-4s_W^2}{s^2_W}=-\frac{2s}{m^2_Z}a_e v_\mu=-\frac{\sqrt{2}G_\mu s}{\pi \alpha}s_W^2 c_W^2 a_e v_\mu =-\frac{1}{\sqrt{2}}\frac{G_\mu s}{\pi \alpha} g_a(e) g_v(\mu).
 \label{ALRtreeSigmaTotal}
 \end{eqnarray}

The choice of the polarization asymmetry (or integrated asymmetry) as one of the observables is driven by its high sensitivity
to Weinberg mixing angle. In case the physics beyond the Standard Model has a parity violating contributor (as for a  $Z^{\prime}$ boson), it would be best to use
$A^C_{LR}$ and $A^C_{LR\Sigma}$ in the study of the properties of new physics particles.
By analogy, the forward-backward asymmetry is defined as
 \begin{equation}
 A^C_{FB} =
  \frac{\Sigma^C_{F}-\Sigma^C_{B}}
       {\Sigma^C_{F}+\Sigma^C_{B}},
 \label{AFB}
 \end{equation}

At the Born level $A^0_{FB}$ is found to be:

\begin{equation}
 A^0_{FB} =a_{e}a_{\mu}\frac{6s  \cos a}{3+\cos^2 a} \ \frac{s(1+2v_{e}v_{\mu})-m_Z^2}{(s-m_Z^2)^2+2s\ v_{e}v_{\mu} (s-m_Z^2)+s^2(v_{e}v_{\mu}+a_{e}a_{\mu})},
 \label{AFBtree}
 \end{equation}
here, and in the above formulas, $\{v_{f},a_{f}\}\equiv \{v^Z_{f},a^Z_{f}\}$. Since $A^0_{FB}$ is directly proportional to the product $a_{e}a_{\mu}$, it is a very useful observable if we would like to search for the candidates beyond the SM, with an axial part of the coupling only.

Finally, we would like to define the NLO absolute corrections to the Born asymmetries: 
\begin{eqnarray}
\Delta_{{LR}}
= {A_{ LR}^{0+1}-A_{ LR}^0},\ \
 \Delta_{FB}
 = {A_{ FB}^{0+1}-A_{ FB}^0},\ \
 \Delta_{LR\Sigma}
 = {A_{ LR\Sigma}^{0+1}-A_{ LR\Sigma}^0}.
 \label{abs-cor-A}
 \end{eqnarray}

In our analysis we start with comparison between the asymptotic and full semi-automatic calculations.
The results for the relative correction $\delta_{00}$ using the SPA approach can be found in Table \ref{tab1}
 for different $\mu^-$ scattering angles in the CM of the $e^-e^+$ system.
Table~\ref{tab1} shows the asymptotic and full semi-automatic results, respectively.
For the cut on the maximum energy of emitted soft photon,  we take 
$\gamma_1=\omega/\sqrt{s}$.
Here we used $\gamma_1=0.05$; this corresponds to the maximum photon energy 
$0.05\cdot \sqrt{s} = 0.52885\ \mbox{(GeV)}$ for Belle~II conditions.
We also found very good agreement between the two approaches for any reasonable choice of $\gamma_1$.

\begin{table}[ht]
\caption{ 
SPA relative corrections to unpolarized differential cross sections,  $\delta_{00}$,  
 at the Belle~II/SuperKEKB CM energy for the $e^- e^+ \rightarrow \mu^- \mu^+(\gamma)$ process at $\gamma_1=0.05$ comparing asymptotic (2nd row) and semi-automatic (3rd row) calculations at  different 
$\mu^-$ polar angles, $\theta$, in the $e^-e^+$ CM system.}

\label{tab1}
\begin{center}
{\vspace*{5mm}
\begin{tabular}{|c|c|c|c|c|c|c|c|c|c|}
\hline
  \multicolumn{1}{|c|}{ $\theta^\circ$   } 
& \multicolumn{1}{ c|}{ 10 } 
& \multicolumn{1}{ c|}{ 30 } 
& \multicolumn{1}{ c|}{ 50 } 
& \multicolumn{1}{ c|}{ 70 } 
& \multicolumn{1}{ c|}{ 90 }
& \multicolumn{1}{ c|}{110 } 
& \multicolumn{1}{ c|}{130 } 
& \multicolumn{1}{ c|}{150 } 
& \multicolumn{1}{ c|}{170 } \\
\hline 
  asymptotic approximation &$  0.0180 $&$ -0.0456 $&$ -0.0738 $&$ -0.0935 $&$ -0.1099 $&$ -0.1264 $&$ -0.1460 $&$ -0.1743 $&$ -0.2378 $ \\
  semi-automatic approach    &$  0.0179 $&$  -0.0455 $&$ -0.0738 $&$ -0.0934 $&$ -0.1099 $&$ -0.1263 $&$ -0.1459 $&$ -0.1742 $&$ -0.2372 $ \\
\hline
\end{tabular}
}
\end{center}
\end{table}

Various numerical results for asymmetries and radiative corrections are presented on Figs.~\ref{4}--\ref{12}.
Here, for the cut on energy of the emitted hard photon,  
in the center-of-mass system of $e^-$ and $e^+$, we used $\Omega = 2.0 \  \mbox{GeV}$.

%As we can see on Fig.\ref{4} correction to the unpolarized cross section in the forward/backward  kinematics is not negligible. 
As we can see on Fig.\ref{4}, the correction to the unpolarized cross section related to the forward/backward  kinematics is not negligible. 
The correction in the region $50^{\circ} \le \theta \le 130^{\circ}$ is linearly decreasing with its central value at  $\sim5.0\%$. 
It is important to note that our comparison between asymptotic and full semi-automatic results (see Tbl.\ref{tab1}) has used only the soft-photon contribution to the unpolarized cross-section and that obviously disagrees with the values of the correction on Fig.\ref{4} (left plot), where the hard photon bremsstrahlung contribution was also included. For the L-R polarization asymmetry on Fig.\ref{5}, we observe a
 standard dependence of the asymmetry on scattering angle. 
Here, as expected, the asymmetry reaches its maximum value at forward angles, which is explained by the short range interaction regime, 
where the parity violating $Z$-boson exchange dominates the contribution to the numerator of the asymmetry term. At backward angles we observe that the asymmetry
 is trending towards a zero value due to the large range interaction regime, where short range 
$Z$-boson exchange has a negligible contribution, and hence the entire L-R asymmetry goes to zero. 

The total cross section and NLO correction, as a function of detector acceptance, are shown on Fig.\ref{6}.
%The correction to the total cross section reaches the value of $\sim46.8\%$, for $2\pi$ geometry of the detector, and is 
The correction to the total cross section reaches the value of $\sim46.8\%$, for full geometrical acceptance, and is relatively constant.% at $\sim47\%$
%, and drops to the value of $\sim 22\%$ for $40^{\circ} \le \theta \le 140^{\circ}$ acceptance. 

The integrated L-R asymmetry $A^{0+1}_{LR\Sigma}$ and its NLO correction $\Delta_{LR\Sigma}$ are shown on Fig.\ref{9}. The maximum value of $A^{0+1}_{LR\Sigma}$ (for $a=10^{\circ}$ and $b=170^{\circ}$) is approximately equal to the average value of differential
 L-R asymmetry, which also corresponds to $A_{LR}^{0+1}$ at $\theta=90^{\circ}$.
 Results for the calculated $A^{0+1}_{FB}$ asymmetry are shown in Fig.\ref{7}.
 %The absolute correction on the Fig.\ref{7} shows roughly linear behavior and reaches its maximum value for $2\pi$ acceptance.
 
Figs.\ref{7}-\ref{12} are dedicated to the sensitivity study of calculated observables to the cuts on the energy of emitted soft photons. In these plots we show dependencies of the observables on the photon's energy cut  $\Omega$, where the dashed line was 
obtained using the soft-photon approach only, and the solid line corresponds to the calculation with hard-photon emission. 

As it can be seen, for the asymmetries, either $A_{LR}^{0+1}$, $A_{FB}^{0+1}$ or $A_{LR\Sigma}^{0+1}$, the two approaches start 
to deviate significantly at $\Omega\approx 0.5\ \mbox{GeV}$. 
This justifies the importance of inclusion of hard-photon emission calculations when it is required to provide analysis for 
observables such as asymmetries. 
%On the contrary, for the various cross sections such as $d\sigma^{0+1}$, $\Sigma^{0+1}_T$ or $\Sigma^{0+1}_{00}$ 
However, for the various cross sections such as $d\sigma^{0+1}$, $\Sigma^{0+1}_T$ or $\Sigma^{0+1}_{00}$ 
the discrepancy between two approaches start to become visible only at $\Omega\approx 4.5\ \mbox{GeV}$, 
which is rather close to the maximum energy of emitted photons, $\Omega =  5.2885\ \mbox{GeV}$. 
Since the calculations in the soft-photon approach are considerably simpler, we can rely on SPA when dealing with cross section calculations.\\

%Caleb Text \input{kk2f.tex}
{\bf Comparisons with \KK~Monte Carlo:}\\
The $\mathcal{KK}$\cite{KK} Monte Carlo code is used by a number of particle physics experiments, including BaBar, Belle and Belle~II, to simulate
$e^+e^-\rightarrow\mu^+\mu^- (n\gamma)$ and $e^+e^-\rightarrow\tau^+\tau^- (n\gamma)$ events. 
In \KK,  photon emission effects from the initial beams as well as outgoing fermions are calculated up to second order,
 including interference effects, using Coherent Exclusive Exponentiation (CEEX)~\cite{CEEX} and
electroweak corrections using the {\em DIZET} library, which is based on the on-shell renormalization scheme~\cite{DIZET}.
The calculations of this work are compared to those provided by \KK~version 4.19, which uses {\em DIZET} version 6.05.
 In order to carry out these comparisons the particle masses used in \KK~were changed to match those in Sect.~\ref{sec:NumRes} and
 the Weinberg mixing angle, which is also an input to \KK,  was set to the value corresponding to the on-shell value of
$\sin^2\theta_W = 0.221392$,  as described by (\ref{eqn:swDef}).   %.Equation \ref{eqn:swDef}.

  Two billion $e^+e^-\rightarrow\mu^+\mu^-$ events were generated with \KK~for both a left-handed polarized e$^-$ beam and a right-handed polarized e$^-$ beam. Each simulated event was required to produce both muons within an angular acceptance
 of $a=10^\circ$ and $b=170^\circ$. From the simulated events comparisons were made with each observable in Figs.\ref{4}-\ref{12}. 
 For Figs.\ref{4}-\ref{5} the \KK~results were binned in $\cos\theta$ with bins 0.125 in width.
 The mean of each bin was used to determine the $\cos\theta$ value of the points.
 In both of these figures the \KK~results are in agreement with our calculations.
 In order to obtain the differential cross section in \KK~we calculate the integrated cross section in the bin and then normalize it by the width of the bin. In Figs.\ref{6}-\ref{9} the \KK~events are binned by angular acceptance. Note that as one end of the bin is fixed and the other moved to various angular cuts, some \KK~events populate multiple bins and therefore the points are not statistically independent. 
 Using \KK~the forward-backward asymmetry was determined with two separate methods. The
 first method counts the events that fall in an angular acceptance between $\pm a$ and 90$^\circ$ with a 2~GeV $\Omega$ cut, shown in Fig.\ref{7}(Left).
 The second method counts events as a function of the $\Omega$ cut in an angular acceptance of $a=30^\circ$ and $90^\circ$, as seen in Fig.\ref{7}(Right). The forward-backward asymmetry seen in Fig.\ref{7} shows an offset of a  few percent between our calculations and the \KK~results.
  This is most likely a result of $A_{FB}$ receiving a substantial contribution  from the IR-finite part
 of the photon bremsstrahlung terms (see Fig.8, left plot). In our case we consider only one-photon emission in initial and final states,
 while \KK~accounts for higher photon multiplicity when the bremsstrahlung contribution to $A_{FB}$ is calculated. 

 In Fig.\ref{7} we switch from angular acceptances to cuts on the energy of the emitted photon. As multiple photons are produced in \KK, not just a single photon, we define $\Omega$ as $\Omega_{KK} =\frac{\sqrt{s}}{2}\left(1-\frac{s'}{s}\right)$, where $s$ is the square of the center-of-mass energy and $s^{\prime}$ is the square of the invariant mass of the muon pair. In Figs.\ref{11}-\ref{12} the \KK~events are binned according to the $\Omega$ cut value of the event. As some events populate multiple bins as the $\Omega$ cut value is varied, this again leads to statistical correlations between bins on these plots. 
 The level of agreement between the \KK~cross sections and the NLO corrected (0+1) cross sections can be seen in Fig.\ref{11}.
 Fig.\ref{12}  compares the  \KK~results with our calculations of $A_{LR}$ as a function of the  $\Omega$ cut.
 In order to compare our calculations of $A_{LR}$ at $\theta=90^{\circ}$ as a function of the $\Omega$ cut  to those of \KK~(Fig.\ref{12}(Left)), 
 the acceptance for charged muons generated by \KK~is set to $70^\circ<\theta<110^\circ$, a region over which the $A_{LR}$ dependence on $\cos\theta$ is linear
 to a good approximation  (see Fig.\ref{5}).
 Note that, again, the point-to-point correlations are large.
 From Fig.\ref{12}(Right), it is evident that in the region 1~GeV$<\Omega <$3~GeV  the \KK~results integrated over $10^\circ<\theta<170^\circ$ 
 are in good agreement with the HPA calculation, within the \KK~statistical uncertainties.
  This statistical uncertainty arises from the finite number of events generated by \KK~for each of  the two e$^-$ polarization states.
 Accounting for the numbers of \KK~events from each sample that survive the acceptance and $\Omega$ requirements,
the absolute \KK~statistical uncertainty on $A_{LR}$ is $\pm 1.9\times 10^{-5}$.

%5\% statistical uncertainty of \KK~.
 It is evident that at low values of $\Omega$ there is significant  disagreement between \KK~and the SPA and HPA treatments. This is a result of the fact that \KK~addresses infrared divergences via exponentiation whereas in the SPA and HPA treatments, the infrared divergences persist.
% However, for $\Omega$ between 1.5~GeV and 2.5~GeV the HPA and \KK calculations agree to better than 1\%
%and the sensitivity of A$_{LR\Sigma}$ to the value of $\Omega$ is less than 0.8\% in that range of $\Omega$. This motivates setting the value of
% $\Omega$ to be near 2~GeV in Belle~II in order to minimize sensitivity to this effect. 

% In addition, the cross-section calculations as a function of $\Omega$ are presented on the left side of Fig.~\ref{12} where it can be seen that
% \KK agrees with the HPA calculation to better than 1\% for $\Omega$ between 0.5~GeV and 5~GeV.

\section{Sensitivity Study}

We next study the sensitivities 
of the  observables $A^{0+1}_{LR\Sigma}$ (\ref{ALRS}) and $A^{0+1}_{FB}$ (\ref{AFB})
to the effective weak mixing angle $(\bar{s}_{W}^{2}\equiv\sin^2\theta_W^{eff})$ and vector part
of the $Z$-boson to fermion coupling ($v^{Z}_{eff}=I_3 - 2 Q_g \bar{s}_{W}^{2}$).

%Next, it would be interesting to see which of the observables in (\ref{A}), (\ref{ALRS}) and (\ref{AFB}) is most sensitive to 
%parameters such as an effective weak mixing angle $(\bar{s}_{W}^{2}\equiv\sin^2\theta_W^{eff})$ or vector part
%of the $Z$-boson to fermion coupling ($v^{Z}_{eff}=I_3 - 2 Q_g \bar{s}_{W}^{2}$). 
%For that, we would like to compare sensitivities of $A^{0+1}_{LR}$, $A^{0+1}_{LR\Sigma}$ and $A^{0+1}_{FB}$.

In order to represent the $e^{+}e^{-}\rightarrow\mu^{+}\mu^{-}$ matrix
element with the simple effective Born-like amplitude, we can use 
leading order low energy one-loop oblique corrections to the Born matrix element. 
Overall we can write for the QED and electroweak parts \cite{Hollik}: 
\begin{eqnarray}
M_{\gamma} & = & \frac{\alpha(s)Q_{e}Q_{\mu}}{s}\left(\bar{v}_{e}\gamma_{\nu}u_{e}\right)\left(\bar{u}_{\mu}\gamma^{\nu}v_{\mu}\right),\nonumber \\
\label{eq:a4}\\
M_{Z} & = & \frac{G_{\mu}}{\sqrt{2}}\kappa\frac{m_{Z}^{2}}{s-m_{Z}^{2}+i\frac{s}{m_{Z}}\Gamma_{Z}}\left(\bar{v}_{e}\gamma_{\nu}
\left[I^{3}_{e}-2\bar{s}_{W}^{2}(s)Q_{e}-I^{3}_{e} \gamma_{5}\right]u_{e}\right)\left(\bar{u}_{\mu}\gamma^{\nu}
\left[I^{3}_{\mu}-2\bar{s}_{W}^{2}(s)Q_{\mu}-I^{3}_{\mu}\gamma_{5}\right]v_{\mu}\right).\nonumber 
\end{eqnarray}
Here $\alpha(s)$ represents the running value of fine structure constant,
defined as 
\begin{eqnarray*}
\alpha(s) & = & \frac{\alpha}{1+\Re\left[\hat{\Sigma}_{\gamma\gamma}(s)\right]/s},
\end{eqnarray*}
and $\bar{s}_{W}^{2}(s)$ defines
the effective running Weinberg mixing angle through the following
expression 
\begin{eqnarray}
\bar{s}_{W}^{2}\left(s\right) & = & s_{W}^{2}-s_{W}c_{W}\frac{\Re\left[\hat{\Sigma}_{\gamma Z}(s)\right]}{s+\Re\left[\hat{\Sigma}_{\gamma\gamma}(s)\right]}.
\label{eq:a5}
\end{eqnarray}
Parameter $\kappa$, is defined based on relationship to expression (\ref{eq:a3})
in the following way:
\begin{eqnarray}
\kappa & = & \frac{1-\Delta r}{1+\Re\left[\frac{\partial}{\partial s}\hat{\Sigma}_{ZZ}(s)\right]}.\label{eq:a6}
\end{eqnarray}
The effective mixing angle is frequently used as one
of the primary parameters in precision electroweak physics and here we study the dependencies of
 $A^{0+1}_{LR\Sigma}$ and $A^{0+1}_{FB}$ on $\bar{s}^2_W$.
%it is our goal to study dependencies of
%$A^{0+1}_{LR}$, $A^{0+1}_{LR\Sigma}$ and $A^{0+1}_{FB}$ on $\bar{s}^2_W$
To start with, we show on Tbl.\ref{SW2}, $\bar{s}_{W}^{2}(s)$
computed in different renormalization schemes at zero and $Z$-pole kinematics.
Our calculated on-shell values of $\bar{s}_{W}^{2}(s)$ compare favorably
%with the results obtained in $\overline{MS}$ \cite{Ferroglia-2004}, and also with PDG $\overline{MS}$ values.
with those calculated in the $\overline{MS}$ scheme, as reported in the PDG $\overline{MS}$.
\begin{table}
\begin{centering}
\begin{tabular}{|c|c|c|}
\hline 
$s$ ($\mbox{GeV}^{2})$ & $\bar{s}_{W, {\rm on-shell}}^{2}$ &  $\bar{s}_{W, \overline{MS}}^{2}$, PDG(2016)
\tabularnewline
\hline 
\hline 
$0$ & 0.23821 &  0.23857\tabularnewline
\hline 
$m_{Z}^2$ & 0.23124 & 0.2313\tabularnewline
\hline 
\end{tabular}
\par\end{centering}

\caption{Results of $\bar{s}_{W}^{2}$ in on-shell and $\overline{MS}$
renormalization schemes.}
\label{SW2}
\end{table}

%\subsection{Sensitivity of $A_{LR}$ to $\bar{s}_{W}^{2}(s)$}

For the kinematics relevant to the Belle~II experiment, $\sqrt{s}=10.58$~GeV,
the on-shell effective value of $\bar{s}_{W}^{2}(s)$ is equal to $0.23413$.
In order to study the sensitivity of the polarization asymmetry
to the variation of $\bar{s}_{W}^{2}(s)$, we can simply vary the value of $m_{W}$, then calculating
$\bar{s}_{W}^{2}(s)$ and asymmetries, we construct parametric dependencies of the asymmetry on $\bar{s}_{W}^{2}(s)$ or $v^{Z}_{eff}$.
It is important to note that in the analysis of the sensitivity of the asymmetries we took the cut on the bremsstrahlung photons at $2.0 \   \mbox{GeV}$.

%%%%% JMR ALR and AFB errors discussion
In order to evaluate the experimental asymmetry uncertainties that feed into the sensitivities, we  make the following reasonable assumptions 
regarding  pertinent experimental parameters  that potentially can be achieved at Belle~II/SuperKEKB if there is an upgrade that
introduces polarization:
\begin{itemize}
  \item the  electron beam polarization is $p_B= 0.7000 \pm 0.0035$, the positron beam is unpolarized.
  \item $p_B$ can measured with 0.5\% precision, and this dominates the systematic error on $A_{LR}$.
  \item $A_{FB}$ can be measured  with an absolute systematic uncertainty of 0.005.
  \item Belle~II collects 20~ab$^{-1}$ of data with the electron beam  polarization 
        and selects $e^- e^+ \rightarrow \mu^- \mu^+(\gamma)$   events with  50\% efficiency.
% JMR New text following referee's comments on centre-of-mass energy dependencies near the Y(4S) peak.
  \item The average $\sqrt{s}$, which has a root-mean-square (RMS) spread of 5~MeV \cite{BelleIITDR}, is known to $\pm 1.2~$MeV
 of the peak of the $\Upsilon(4S)$ resonance\footnote{SuperKEKB operations, following past practice of previous generation $e^+e^-$  B-factories, will ensure that $\sqrt{s}$ is at the peak of the $\Upsilon(4S)$ by scanning the energy of one of the beams in a manner that maximizes the rate of $e^+e^-\rightarrow$ hadrons throughout of data-taking runs. As the RMS spread in $\sqrt{s}$ is significantly smaller than the $\Upsilon(4S)$ width ($20.5\pm2.5~$MeV), the average value of $\sqrt{s}$ will be known to $\pm 1.2~$MeV, the experimental precision on the $\Upsilon(4S)$ mass \cite{PDG16}.}.
%%%%%%%%%%%%%%%%%%%%%%%%%%%%%%%%%%%%%%%%%%%%%
\end{itemize}

With such parameters we can expect an absolute statistical uncertainty on both $A_{FB}$ and $A_{LR}$ of $9.4\times 10^{-6}$.
This gives a total uncertainty on $A_{LR \Sigma}$(with $b=170^{\circ}$) of $\pm$0.0000094(stat)$\pm$0.0000030(syst)=$\pm$0.0000097(total).
The error is  dominated by the statistical uncertainty and  gives a relative uncertainty on
$A_{LR \Sigma}$ of 1.6\%.
The total uncertainty on $A_{FB}$(with $a=10^{\circ}$; $b=170^{\circ}$) is $\pm$ 0.0050 (total).
In this case, the uncertainty  is completely dominated by the systematic uncertainty and gives a 
relative error on $A_{FB}$ of 9.4\%.

The reason for this difference in relative uncertainties is that the systematic
error on $A_{LR}$ scales as the relative error because $p_B$ is a multiplicative correction
needed for the measurement and has no other large systematic error since essentially all
other detector systematic errors cancel. %This is because  only the beam polarization is changed between positive and negative helicities.
On the other hand, for $A_{FB}$  the dominant systematic errors arise in the detector and do not fully cancel: it is necessary to
measure the angles and forward and backward acceptances, the boost to transform into the CM frame,
and understand any charge asymmetries in the detector. As these are systematic uncertainties in the 
detector asymmetries,  they are absolute uncertainties on $A_{FB}$.

%%%%% JMR ALR and AFB errors discussion

\begin{figure}
\begin{centering}
\includegraphics[scale=0.32]{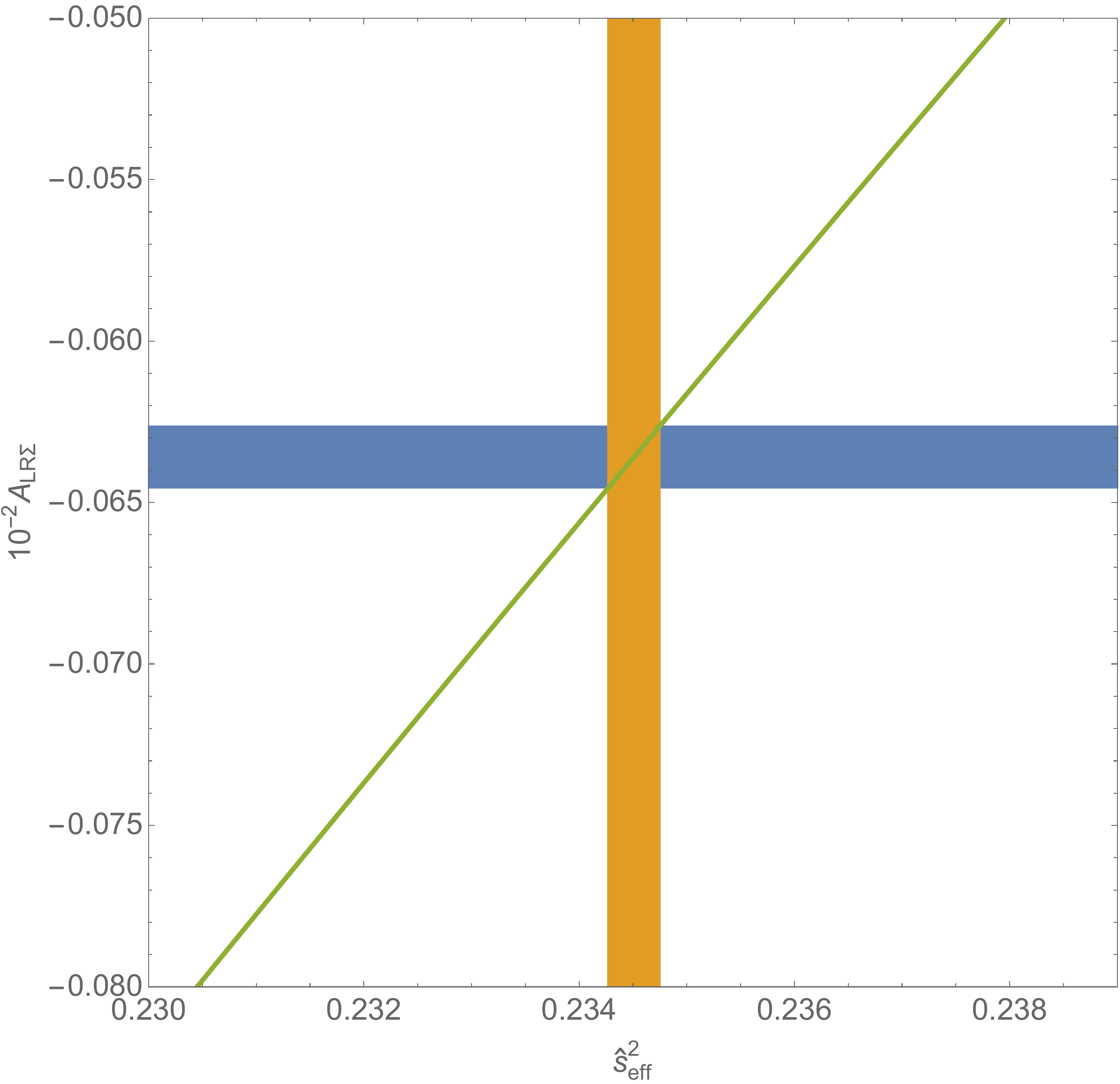}
\includegraphics[scale=0.32]{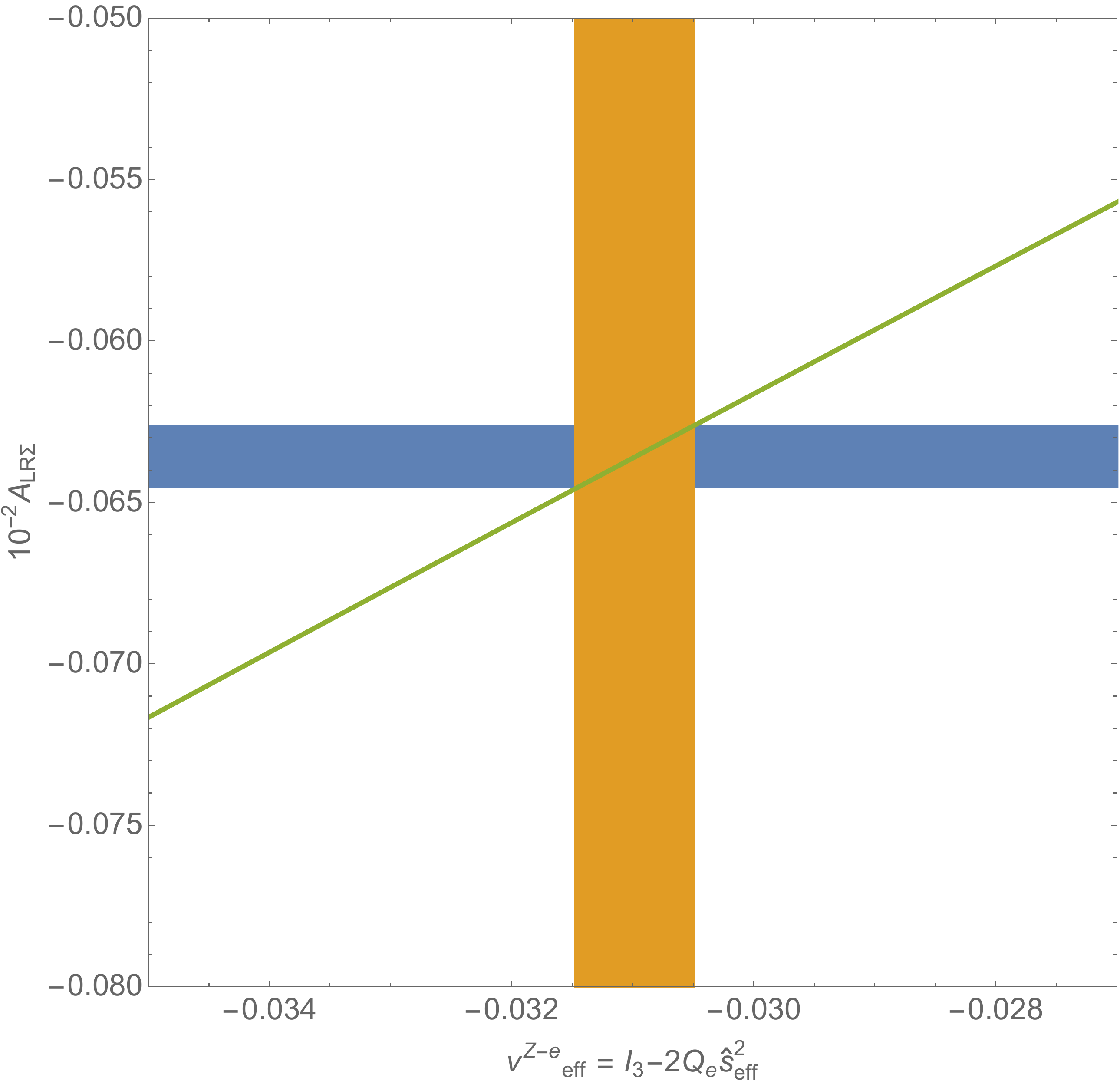}
\par\end{centering}
\caption{Dependence of the  integrated left-right  asymmetry on the effective Weinberg mixing
angle (left)  at $\sqrt{s}=10.58$~GeV and vector part of the electroweak coupling (right). Horizontal bands show the central value
 of $A^{0+1}_{LR\Sigma}=-0.00063597$ determined with the cut on soft-photons at $2.0\ \mbox{GeV}$. 
The width of the band corresponds to the $\pm  0.0000097$ uncertainty on the central value of $A^{0+1}_{LR\Sigma}$. }
%\caption{Dependence of polarization (top row) and integrated (bottom row)  asymmetries on the effective Weinberg mixing
%angle (left column) at $\sqrt{s}=10.58$~GeV and vector part of the electroweak coupling (right column). Horizontal bands show the central value
% of $A^{0+1}_{LR}(90^{\circ})= -0.0006124$  and $A^{0+1}_{LR\Sigma}=-0.00063597$ determined with the cut on soft-photons at $2.0\ \mbox{GeV}$. 
%The width of the band corresponds to the $\pm  0.0000097$ uncertainty on the central value of $A^{0+1}_{LR}(90^\circ)$. }
\label{figALRSW2}
\end{figure}

%Fig.\ref{figALRSW2} shows a linear dependence of $A^{0+1}_{LR}(90^{\circ})$ on $\bar{s}_{W}^{2}\ (s=10.58^{2}\ \mbox{GeV}^{2})$.
%That is evident from the fact that the polarization asymmetry is proportional
%to the interference term: $2\Re\left[M_{\gamma}M_{Z}^{*}\right]$,
%which is linearly proportional to $\bar{s}_{W}^{2}(s)$. 

%As it can be seen from Fig.\ref{figALRSW2}, if we take the absolute uncertainty for $A^{0+1}_{LR}(90^{\circ})$ equal to 
%$\pm 0.000097$, that translates to 0.22\% uncertainty on $\bar{s}_{W}^{2}(s)$ at $s=10.58^{2}\ \mbox{GeV}^{2}$.
%Also, if we take the same absolute uncertainty for the integrated asymmetry, we get a slightly improved sensitivity
%with the uncertainty on $\bar{s}_{W}^{2}(s)$ being at 0.21\% level.

Fig.\ref{figALRSW2} shows a linear dependence of $A^{0+1}_{LR \Sigma}$ on $\bar{s}_{W}^{2}\ (s=10.58^{2}\ \mbox{GeV}^{2})$.
That is evident from the fact that the polarization asymmetry is proportional
to the interference term: $2\Re\left[M_{\gamma}M_{Z}^{*}\right]$,
which is linearly proportional to $\bar{s}_{W}^{2}(s)$. 
As it can be seen from Fig.\ref{figALRSW2},  the absolute uncertainty for $A^{0+1}_{LR\Sigma}$ equal to 
$\pm 0.000097$, translates into an uncertainty of 0.21\% on $\bar{s}_{W}^{2}(s)$ at $s=10.58^{2}\ \mbox{GeV}^{2}$.

In general the on-shell extraction of $\bar{s}_{W}^{2}(s)$ from an experimental
polarization asymmetry could be done by determining the effective
$m_{W}$ from the measured $A^{0+1}_{LR\Sigma}$, and then determine $\bar{s}_{W}^{2}(s)$
for that specific effective $m_{W}$ from equation (\ref{eq:a5}).

% JMR New text following referee's comments on centre-of-mass energy dependencies near the Y(4S) peak.
It is known that in the time-like region, the value of $\bar{s}_{W}^{2}(s)$ changes  rapidly near resonances. Although we have not included
 the effect of hadronic resonances in our treatment, we can estimate the impact of such an effect on the precision of an asymmetry measurement made
 at the peak of the $\Upsilon(4S)$ resonance using Fig.~1 of reference\cite{Jegerlehner2017}. 
From that figure, $\sin^2\theta_W$  changes by approximately 0.003 over the 20.5~MeV full width of the $\Upsilon(4S)$
 resonance and therefore there is a sensitivity of $\Delta_{\sin^2\theta_W}/\Delta_{\sqrt{s}}=0.00015$/MeV 
 in the region of the peak of the $\Upsilon(4S)$ resonance. As $\sqrt{s}$ is known to $\pm$1.2~MeV, in the 
interpretation of the integrated $A_{LR}$ measurement in terms of  $\bar{s}_{W}^{2}(s)$,
this translates into an uncertainty on  $\bar{s}_{W}^{2}(s)$ of approximately $0.00018$, or 0.08$\%$, which contributes
a small additional uncertainty:  adding this in quadrature with the 0.21$\%$ coming from the other uncertainties yields a total uncertainty on
$\bar{s}_{W}^{2}(s)$ of 0.22$\%$. 
We note that this uncertainty will be common to measurements from each fermion species and therefore
will cancel in evaluations of fermion universality of the  weak mixing angle performed with $A_{LR\Sigma}$ at Belle~II.
%%%%%%%%%%%%%%%%%%%%%%%%%%%%%%%%%%%%%%%%%%%%%%%%%%%%%%%%%%%%%%%%%%%%%%%%%%%%%%%%%%%%%%%%%%

%\subsection{Sensitivity of $A_{FB}$ to $\bar{s}_{W}^{2}(s)$}

\begin{figure}
\begin{centering}
\includegraphics[scale=0.315]{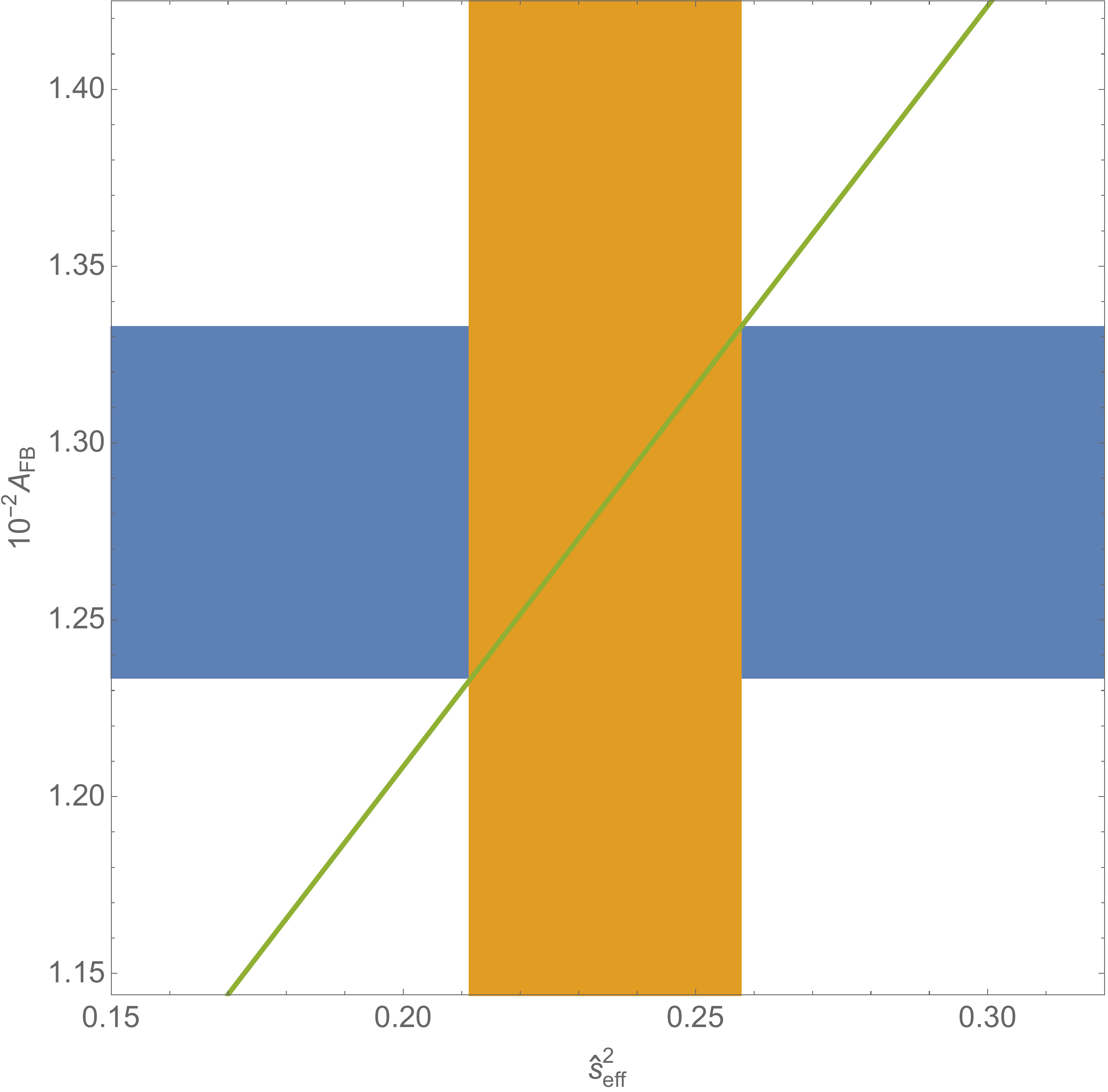}
\includegraphics[scale=0.325]{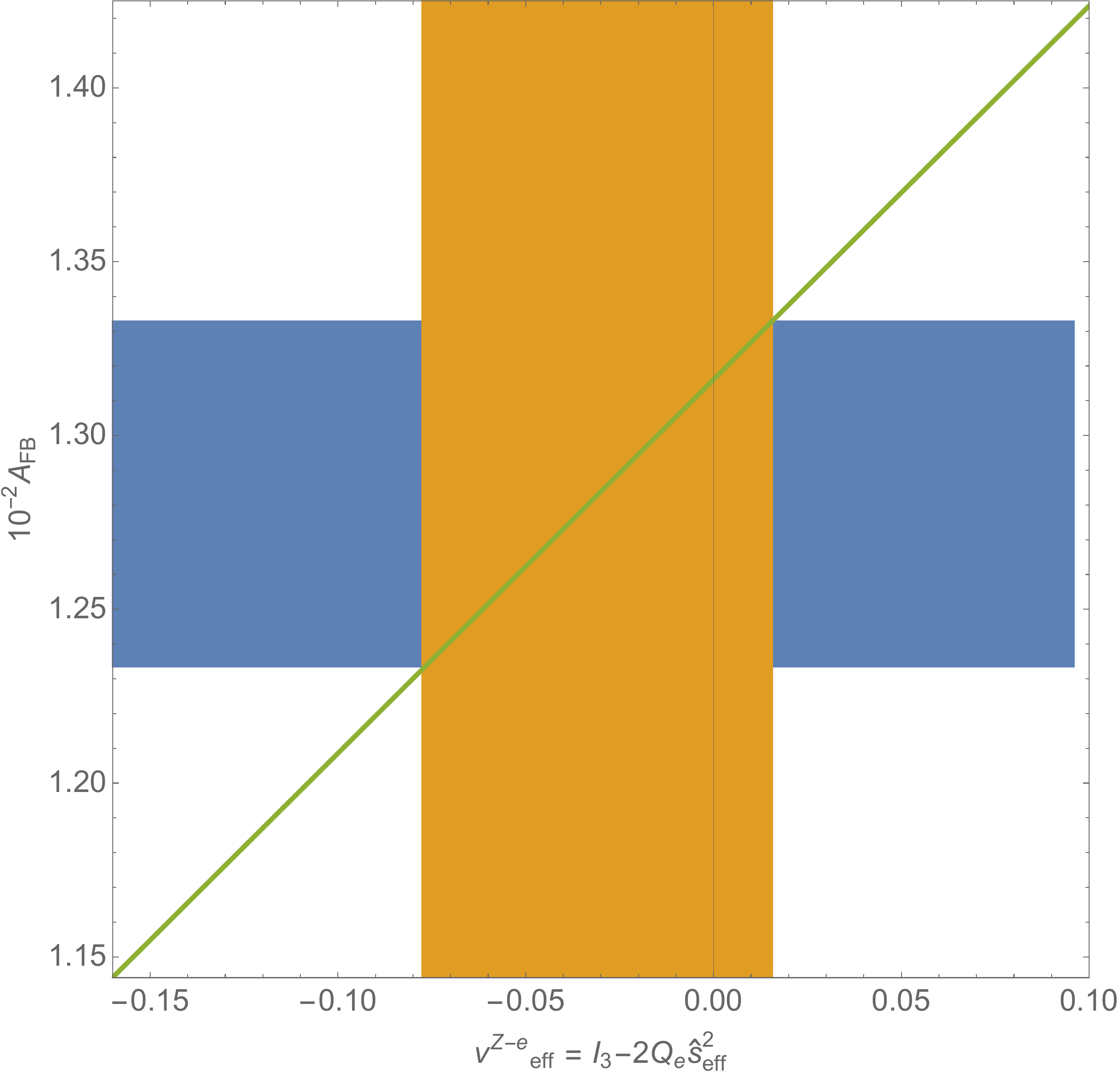}
\par\end{centering}

\caption{Forward-backward asymmetry for the $a=10^{\circ}$ and $b=170^{\circ}$ as
a function of $\bar{s}_{W}^{2}(s)$. Horizontal band shows the central
value of $A^{0+1}_{FB}=0.01283$ determined with the cut on hard-photons at $2.0\ \mbox{GeV}$. Width of the band corresponds
to the 1\% uncertainty on the central value of $A^{0+1}_{FB}$. }
\label{figAFBSW2}
\end{figure}

In a similar fashion we can study the sensitivity of the forward-backward asymmetry
to the variations of $\bar{s}_{W}^{2}(s)$. The Fig.\ref{figAFBSW2} shows the similar
dependence of $A^{0+1}_{FB}$ on $\bar{s}_{W}^{2}(s)$, but with a substantially smaller
slope when compared to Fig.\ref{figALRSW2}.
Although the numerical value of $A^{0+1}_{FB}$ is much larger than $A^{0+1}_{LR\Sigma}$,
% $A^{0+1}_{LR}(90^\circ)$ or $A^{0+1}_{LR\Sigma}$,
it's sensitivity to $\bar{s}_{W}^{2}(s)$ is rather low. This translates
to an uncertainty of 19.8\% on $\bar{s}_{W}^{2}(s)$, if we consider
$\pm 0.00050$ uncertainty in $A^{0+1}_{FB}$.

Clearly $A^{0+1}_{FB}$ contains substantial contributions, which are not sensitive to parity-violating physics. At this point we would like 
to determine the most dominant contributions to $A^{0+1}_{FB}$ and their nature.
We will start with the basic definition of various QED and Weak contributions
in the forward-backward asymmetry:
\begin{align}
A^{0+1}_{FB} & =\frac{\Sigma_{F}^{{0+1}}-\Sigma_{B}^{{0+1}}}{\Sigma_{F}^{{0+1}}+\Sigma_{B}^{{0+1}}}=\frac{\Gamma_{FB}^{{0+1}}}{\Sigma_{T}^{0+1}}\label{eq:a1}\\
\nonumber 
\end{align}
The denominator of (\ref{eq:a1}), is defined as a total integrated
unpolarized cross section including one-loop corrections. We will
keep this part of $A^{0+1}_{FB}$ unmodified. This way contributions to the asymmetry are additive.
As for the numerator of $A^{0+1}_{FB}$,
it will be divided into Born, various infrared finite NLO, and soft-bremsstrahlung
contributions. More specifically:
\begin{align}
\Gamma_{FB}^{0+1} & =\Gamma_{FB}^{0}+\Gamma_{FB}^{\gamma-SE(\gamma)}+\Gamma_{FB}^{\gamma-TR(\gamma)}+\Gamma_{FB}^{\gamma-BB(\gamma)}+\Gamma_{FB}^{\gamma-SE(Z)}+\Gamma_{FB}^{\gamma-TR(Z)}+\Gamma_{FB}^{\gamma-BB(Z)}+\nonumber \\
 & \Gamma_{FB}^{Z-SE(\gamma)}+\Gamma_{FB}^{Z-TR(\gamma)}+\Gamma_{FB}^{Z-BB(\gamma)}+\Gamma_{FB}^{Z-SE(Z)}+\Gamma_{FB}^{Z-TR(Z)}+\Gamma_{FB}^{Z-BB(Z)}+\Gamma_{FB}^{soft}\label{eq:a2}
\end{align}
Here, $\Gamma_{FB}^{0}$ is the forward-backward Born contribution to the numerator
of $A^{0+1}_{FB}$, and $\Gamma_{FB}^{\gamma-SE(\gamma)}$ (for example) and
corresponds to the interference term between Born QED and $\gamma-\gamma$
Self-Energies (SE). Furthermore TR and BB stand for Triangle and Box
type graphs, respectively. 

Our starting point is to show Born and fully corrected forward-backward asymmetries.
We do this for both renormalization conditions, based on \cite{hollik} and \cite{Denner}. 
\begin{figure}
\begin{centering}
\includegraphics[scale=0.33]{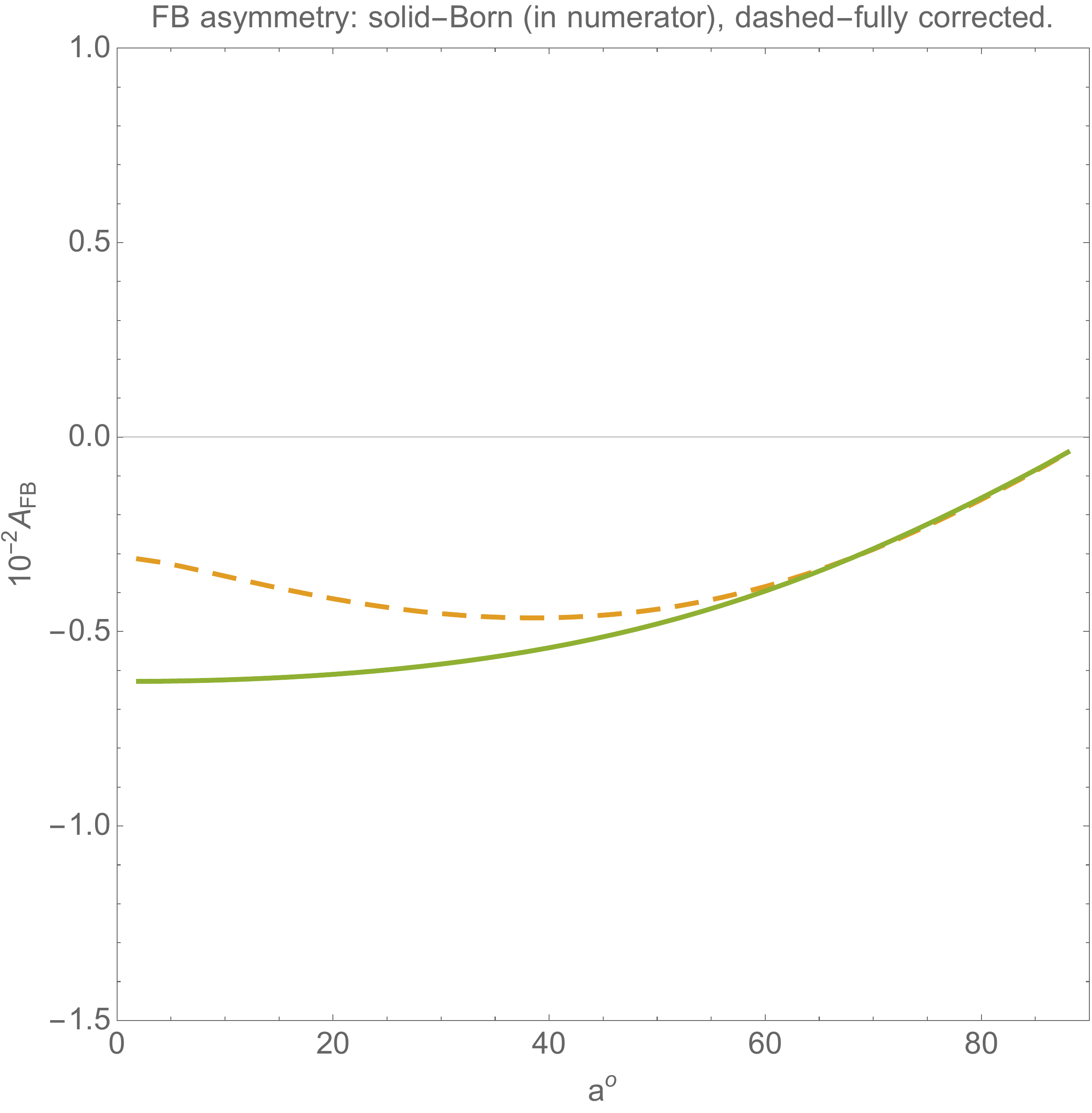}
\par\end{centering}
\caption{Born (first term in Eq.\ref{eq:a2}) and fully corrected $A^{0+1}_{FB}$.
Born is represented by solid line, and corrected $A^{0+1}_{FB}$ is shown
by dashed line. }

%JMR
\label{fig1}
\end{figure}
Results on Fig.\ref{fig1} are represented by the infrared finite parts
of virtual and soft-bremsstrahlung corrections only. That would also
be true for all partial NLO contributions appearing in (\ref{eq:a2}).
We have observed practically zero contributions coming from $\Gamma_{FB}^{\gamma-SE(\gamma)},\Gamma_{FB}^{\gamma-TR(\gamma)},\Gamma_{FB}^{\gamma-SE(Z)},\Gamma_{FB}^{\gamma-BB(Z)},\Gamma_{FB}^{Z-SE(\gamma)},\Gamma_{FB}^{Z-SE(Z)},\Gamma_{FB}^{Z-TR(Z)},\Gamma_{FB}^{Z-BB(\gamma)}$
and $\Gamma_{FB}^{Z-BB(Z)}$ terms in (\ref{eq:a2}). This implies
that contributions coming from all types of self-energies and electroweak
($\gamma-Z,\,Z-Z$ and $W-W$) boxes are negligible, and can be disregarded.
It is important to note that generally electroweak self-energies
or vertex correction graphs are not gauge invariant and hence their
independent contributions have no physical meaning. However, for the
forward-backward asymmetry this can be bypassed, since gauge dependent contributions
largely cancel out even for separate parts, such as self-energies
or vertex correction graphs. We have verified this by comparing self-energies
(or triangles) contributions in the different renormalization conditions
(Denner and Hollik) and found that the results are identical.
At this point we only show contributions which are substantial and
can not be avoided in the calculations of $A^{0+1}_{FB}$. 

\begin{figure}
\begin{centering}
\includegraphics[scale=0.33]{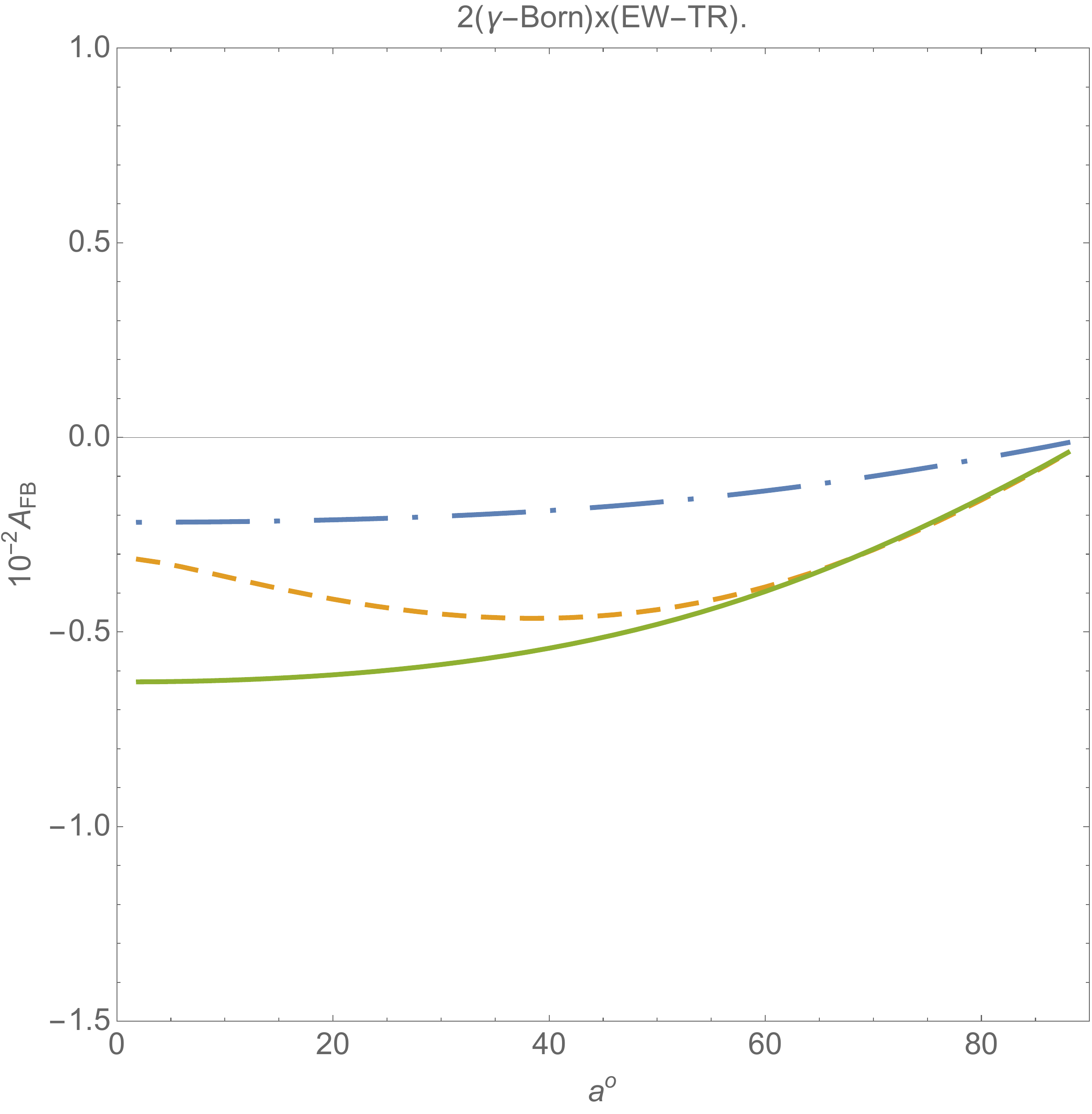} \includegraphics[scale=0.33]{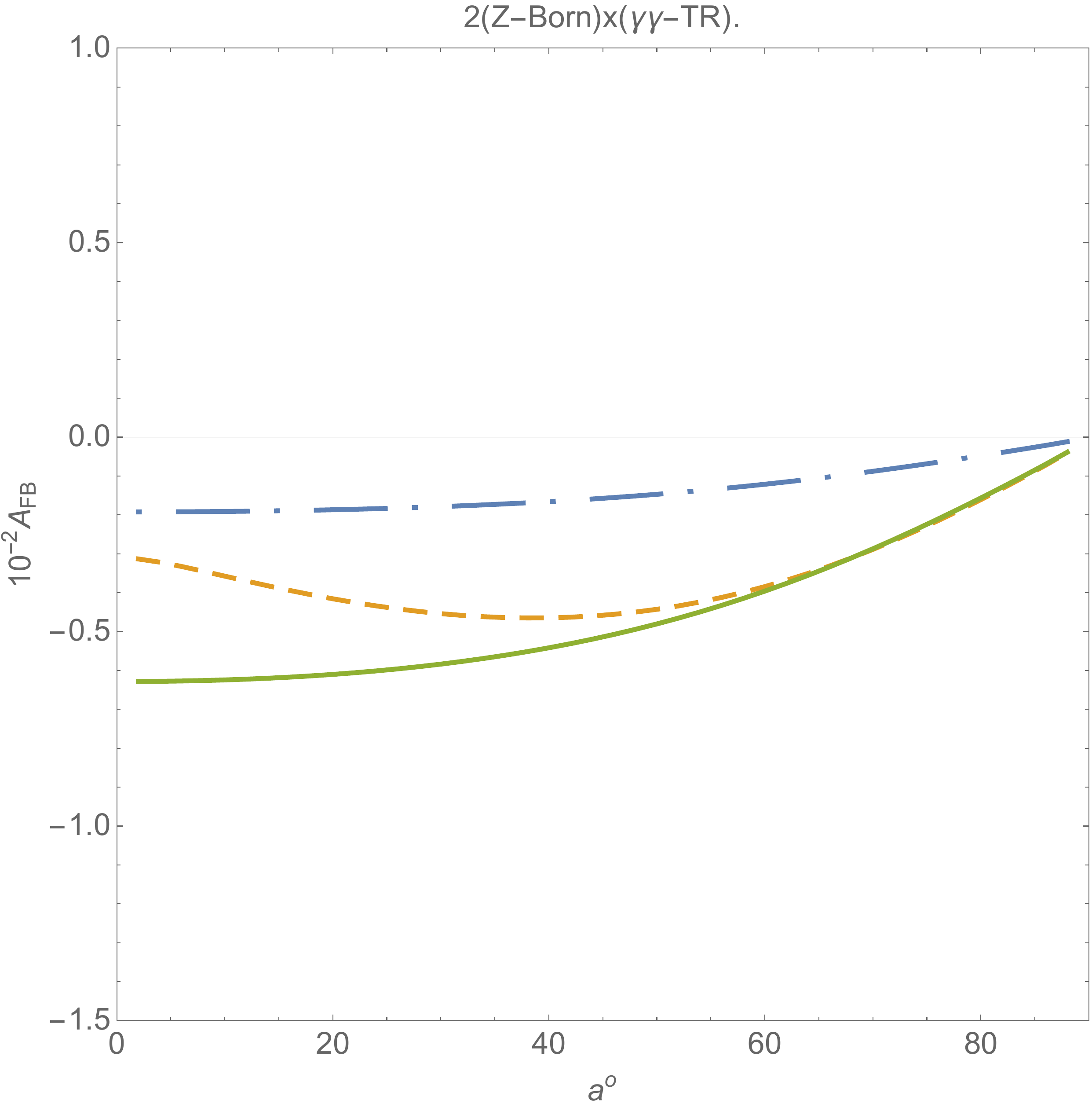} 
\par\end{centering}
\begin{centering}
\includegraphics[scale=0.33]{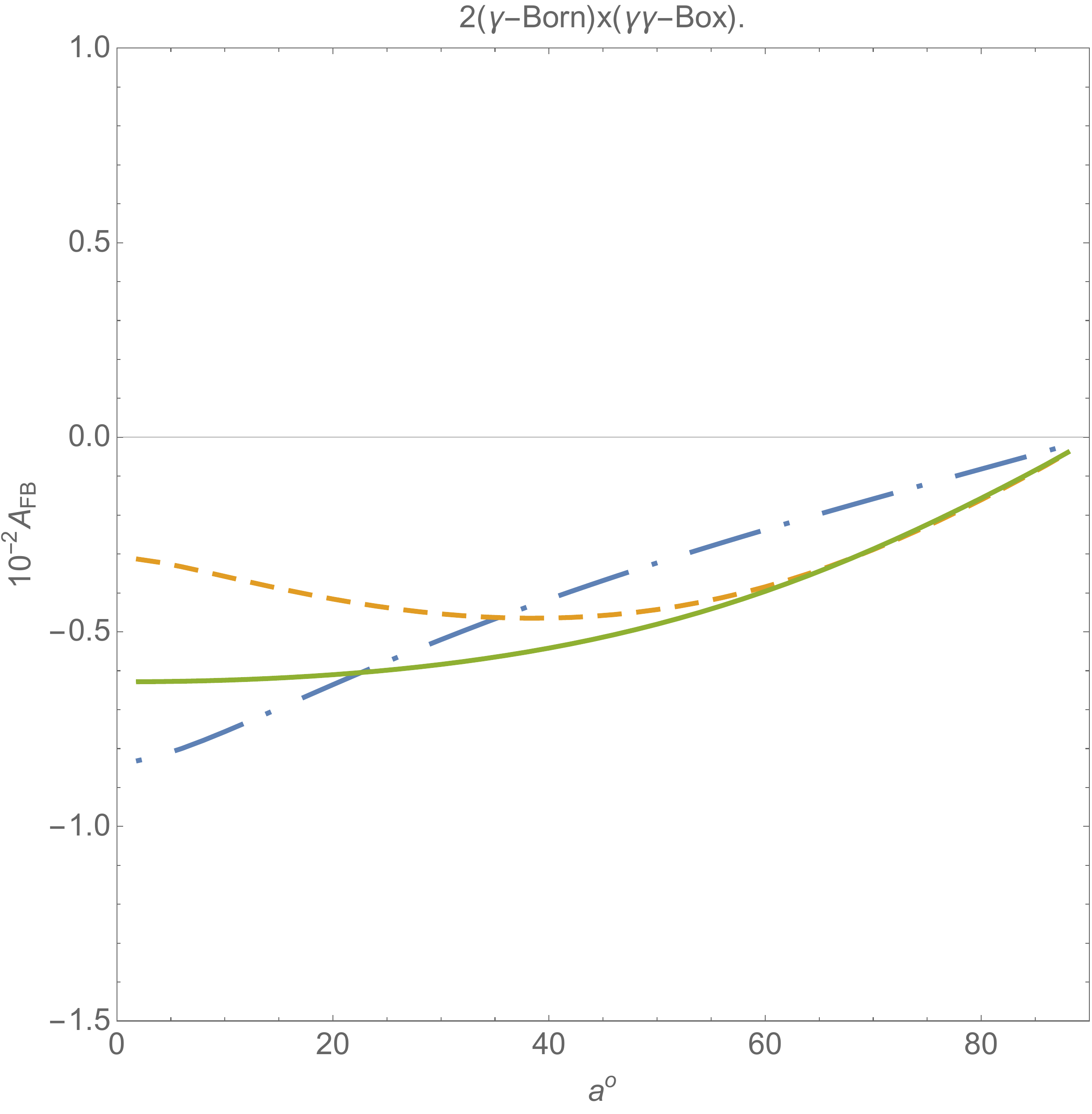}\includegraphics[scale=0.33]{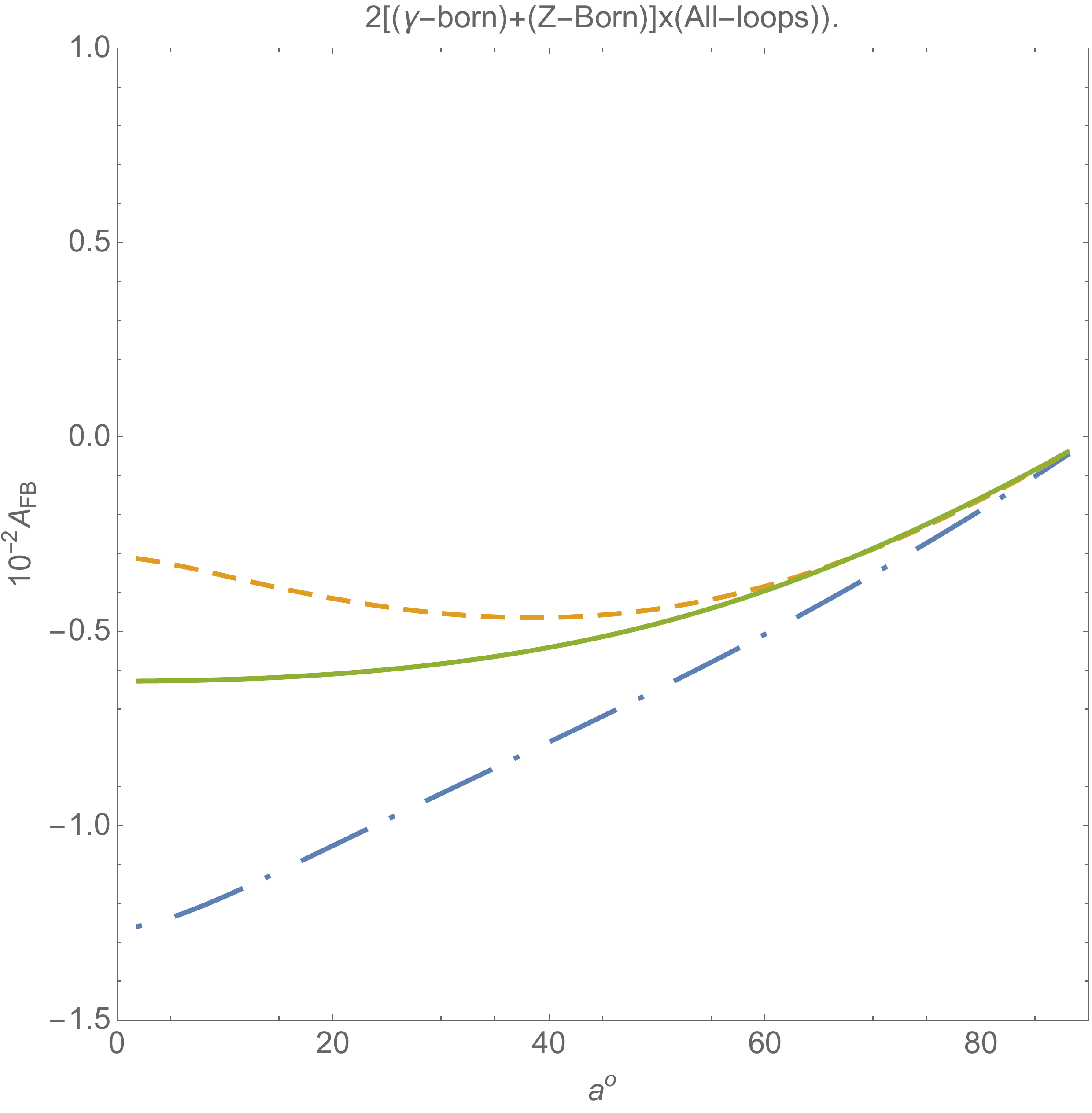}
\par\end{centering}
\caption{Various contributions to $A_{FB}$. On all graphs, solid green line
corresponds to Born contribution (in numerator of $A_{FB}$), dashed
yellow line is fully corrected asymmetry and dot-dashed blue shows
various NLO parts of $A_{FB}$. }
\label{fig2}
\end{figure}

As we can see from Fig.\ref{fig2} (two top graphs), we have identical
(symmetrical) contributions from interference terms, such as: $2\Re[M_{0}^{\gamma}M_{one-loop}^{EW-TR}]$
and $\Re[M_{0}^{Z}M_{one-loop}^{\gamma-TR}]$. The biggest contribution
comes from the interference term between $\gamma$-Born and $\gamma\gamma$-Box
(see Fig.\ref{fig2}, second row, left). Overall, all one-loop contributions
are systematically additive and the result is shown on Fig.\ref{fig2},
second row, right graph. Since it is clear now that the addition of one-loop
contributions (blue, dot-dashed curve) and Born (green, solid curve)
term would not reproduce the full result for $A^{0+1}_{FB}$, we turn our attention
to IR finite terms of soft-photon bremsstrahlung. 

\begin{figure}
\begin{centering}
\includegraphics[scale=0.33]{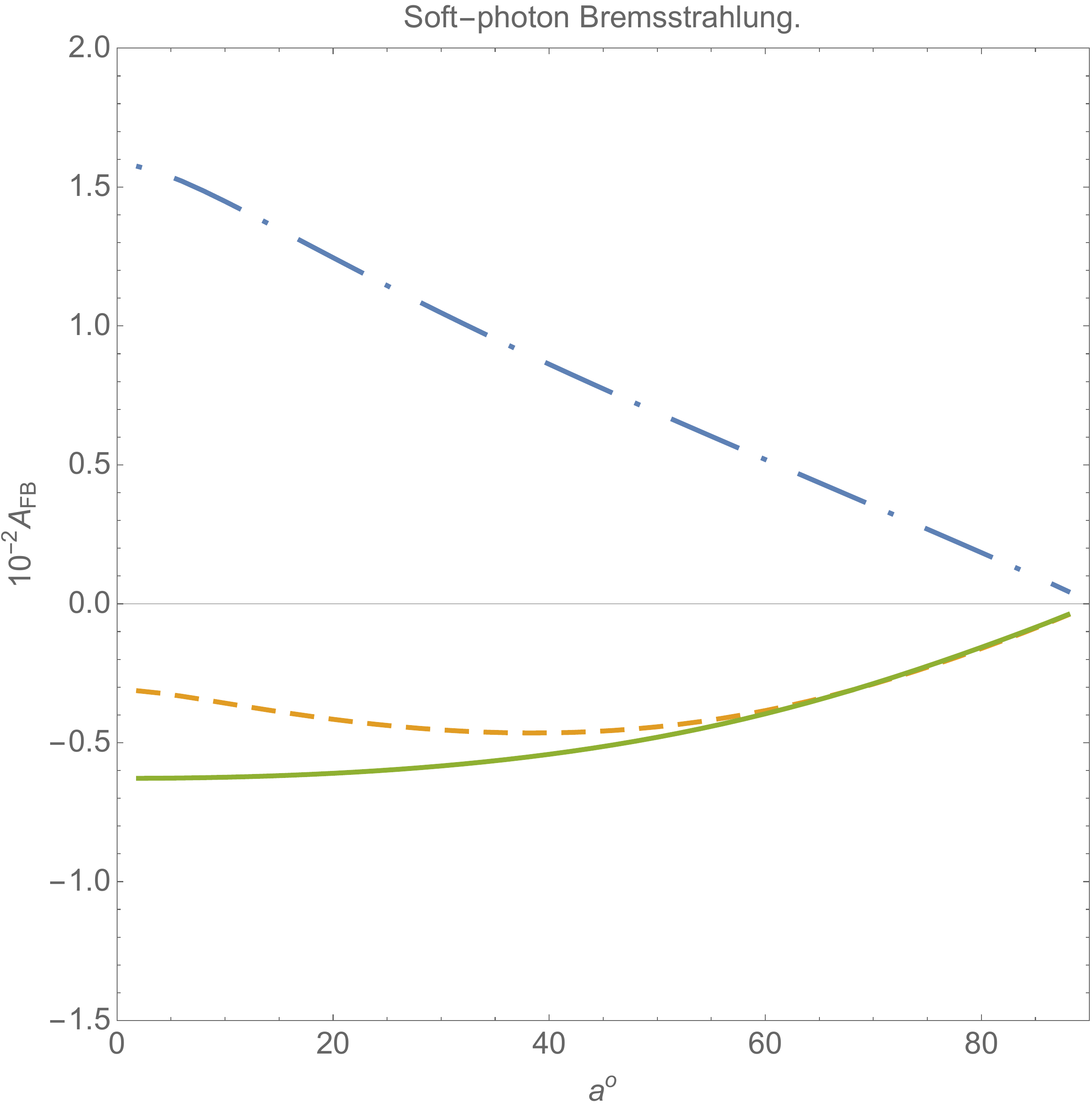} \includegraphics[scale=0.33]{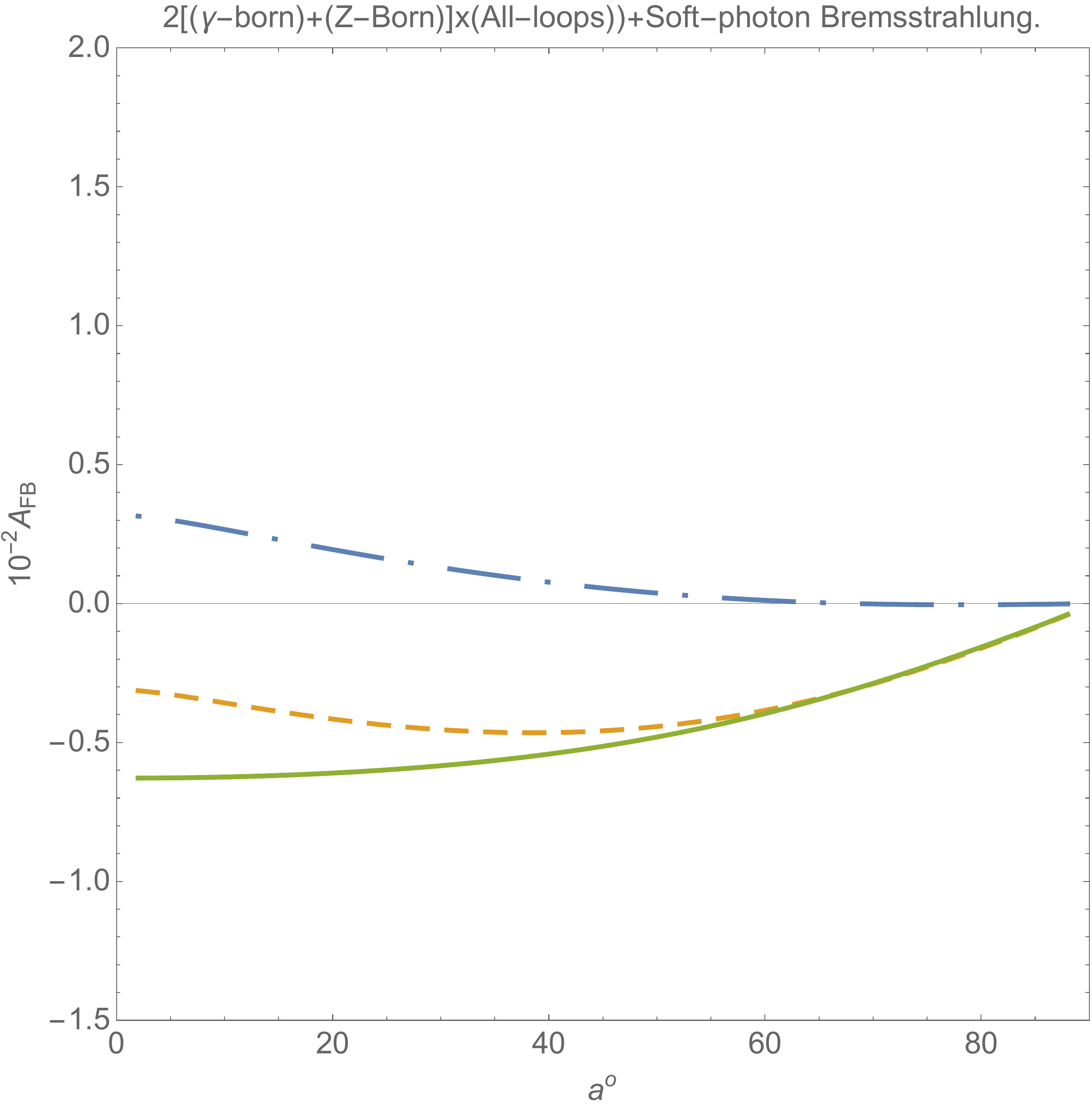}
\par\end{centering}
\caption{IR finite soft-photon bremsstrahlung (left plot) and total correction
(right plot) to the numerator of the asymmetry. Solid green line
corresponds to Born contribution (in numerator of $A_{FB}$), dashed
yellow line is fully corrected asymmetry and dot-dashed blue shows
various NLO parts of $A_{FB}$.}
\label{fig3}
\end{figure}

It is clearly visible on Fig.\ref{fig3} (left) that the bremsstrahlung
contribution largely cancels out one-loop results and produce the correction,
shown on  Fig.\ref{fig3} (right and blue dot-dashed curve). The addition of the one-loop correction (Fig. \ref{fig3}, left and blue dot-dashed curve ) 
and Born result (solid green curve on the same plot) produce the final result for $A^{0+1}_{FB}$ (dashed yellow curve).
One of the possible explanations for such a large cancellation could be found
in the fact that both of the IR finite parts of virtual one-loop correction and soft-photon
bremsstrahlung contain collinear divergent terms, which cancel out
in the final result.

%Overall we can conclude that $A^{0+1}_{LR}$ asymmetry (same could be said for $A^{0+1}_{LR\Sigma}$) is the most sensitive observable
Overall we conclude that $A^{0+1}_{LR\Sigma}$ is the observable most sensitive 
to the effective electroweak parameters. As such, in order to search for physics beyond the Standard Model at the  
 precision frontier of neutral-current measurements,
 it is crucial to have polarized electron beams in Belle~II/SuperKEKB in order to measure $A^{0+1}_{LR\Sigma}$.

\section{Conclusion}

In this paper we compare the results for the full set of
one-loop EWC to parity violating polarization and forward-backward asymmetries
at the Belle~II/SuperKEKB CM energy obtained by different methods.
The soft photon approximation using an exact semi-automatic approach is validated by an asymptotic approach with simplifications giving a compact form.
We take under full control the bremsstrahlung process and 
compare results for the soft and hard photon calculations.
We also evaluate the sensitivity to the variation of $\bar{s}_{W}^{2}$ for both polarization and forward-backward asymmetries. We find that the highest sensitivity is achieved for the measurements using $A_{LR\Sigma}$ with a polarized electron beam. 
In addition, we have analyzed various NLO contributions to the IR finite part of $A_{FB}^{0+1}$. 
As a result, we found that the large contribution arising from interference terms between $\{\gamma, \ Z\}$-Born, $\{\gamma-\gamma\}$-Box, and $\{\gamma, \ Z\}$-Triangle graphs are compensated by the IR finite part of the soft-photon bremsstrahlung contribution and that self-energies, although important for the overall cross-sections, cancel out for the
forward-backward asymmetry and therefore have an overall negligible contribution to that asymmetry. A comparison is also made with the \KK~Monte Carlo generator for the $A_{LR}$ and $A_{FB}$ asymmetries.
Where infrared divergencies are small, our current calculations are in good agreement with those of the \KK~Monte Carlo.

We plan to broaden these studies to include left-right asymmetries in $e^+e^-$ collisions for Bhabha scattering and for massive final-state fermions 
(tau leptons, charm and bottom quarks), where the negligible-mass assumption is not valid.
% and to 
% do a study of at least the leading two-loop EWC in order to meet the precision goals of future experiments.
%Also we work on some extension to new physics for Belle~II/SuperKEKB 
%in mode discussed in this paper.
In order to further reduce the theoretical uncertainties, our next step is 
 to include the two-loop EWC in the on-shell renormalization scheme, and compare these to the calculation in the $\overline{MS}$ scheme.
% to make sure that the best approach is chosen for the analysis on the experimental data.
Nonetheless, the results of this paper  demonstrate that the Standard Model predictions for $A_{LR}$ at 10.58~GeV, and consequently the weak mixing angle at that
 energy, 
are already under excellent theoretical control and provide encouragement to upgrade SuperKEKB with a polarized e$^-$ beam in order to provide a new tool in the 
 search for physics beyond the Standard Model.

\section{ACKNOWLEDGMENTS}
We are grateful to \framebox{Eduard Alekseevich Kuraev} for stimulating discussions. 
V.Z. thanks the Acadia University (Wolfville, Canada) and Memorial University (Corner Brook, Canada)  for hospitality. 
This work was supported by the Belarusian State Program of Scientific Research ``Convergence-2020''
and the Natural Sciences and Engineering Research Council of Canada.

%\newpage

\newpage

\begin{figure}
	\centering
	\begin{tabular}{cc}
		\includegraphics[trim=2cm 5cm 1cm 5cm,width=0.5\linewidth]{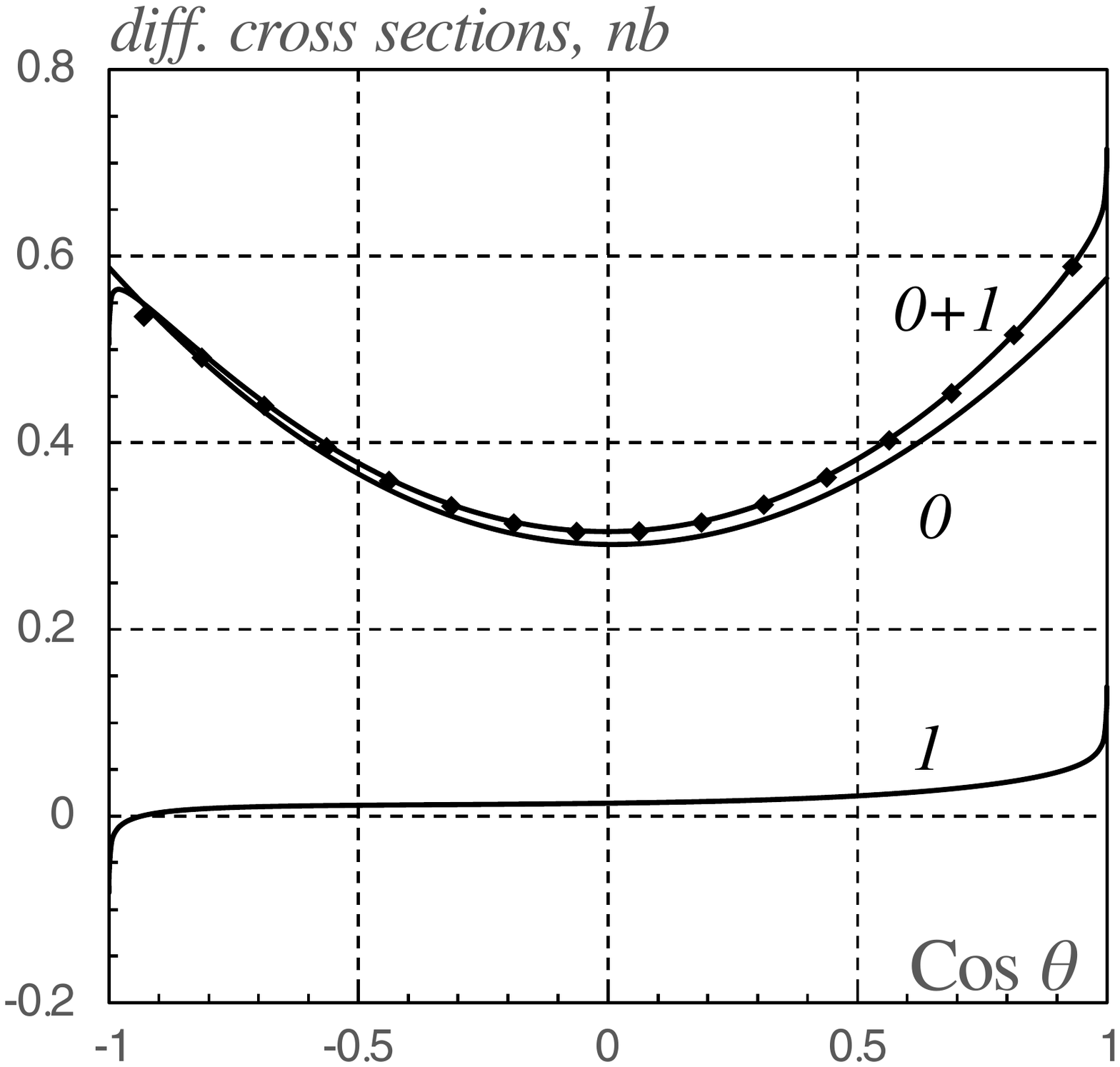}
		&
		\includegraphics[trim=2cm 5cm 1cm 5cm,width=0.5\linewidth]{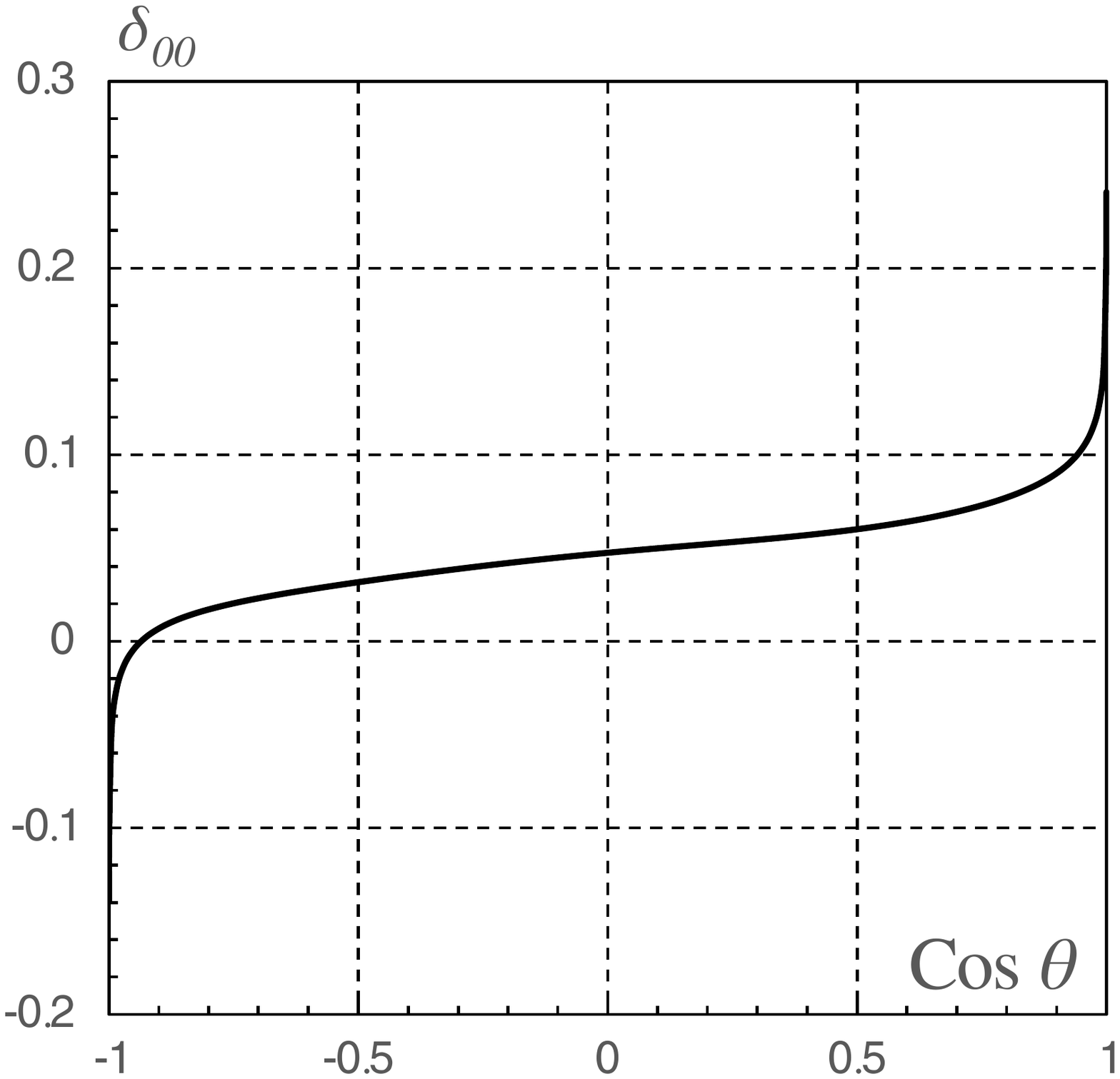}
	\end{tabular}
	%\vspace{8mm}
	\caption{\protect\it
		Left: unpolarized  
		NLO corrected (0+1), 
		Born (0), 
		and their difference (1)
		differential cross sections vs scattering angle $\theta$.
		Right: the relative NLO correction to unpolarized Born cross section vs $\theta$. Calculations are done at an $\Omega$ cut of 2~GeV. The points are the results obtained from running the \KK~Monte Carlo generator as described in the text, where the error bars
		represent the statistical errors from the number of Monte Carlo events generated.
	}
	\label{4}
	\vspace{5mm}
\end{figure}
\begin{figure}
	\centering
	\begin{tabular}{cc}
		\includegraphics[trim=2cm 5cm 1cm 5cm,width=0.5\linewidth]{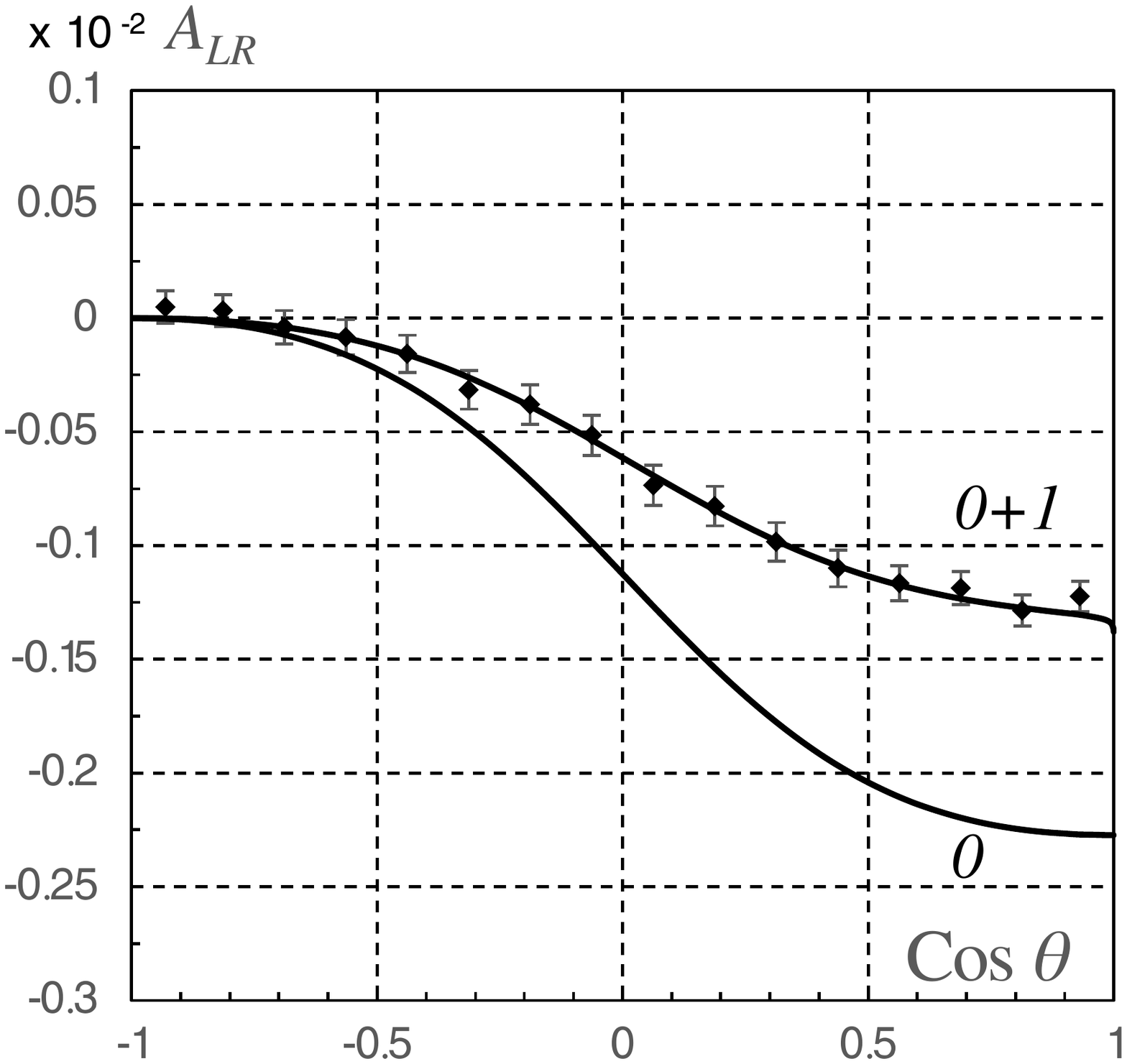}
		&
		\includegraphics[trim=2cm 5cm 1cm 5cm,width=0.5\linewidth]{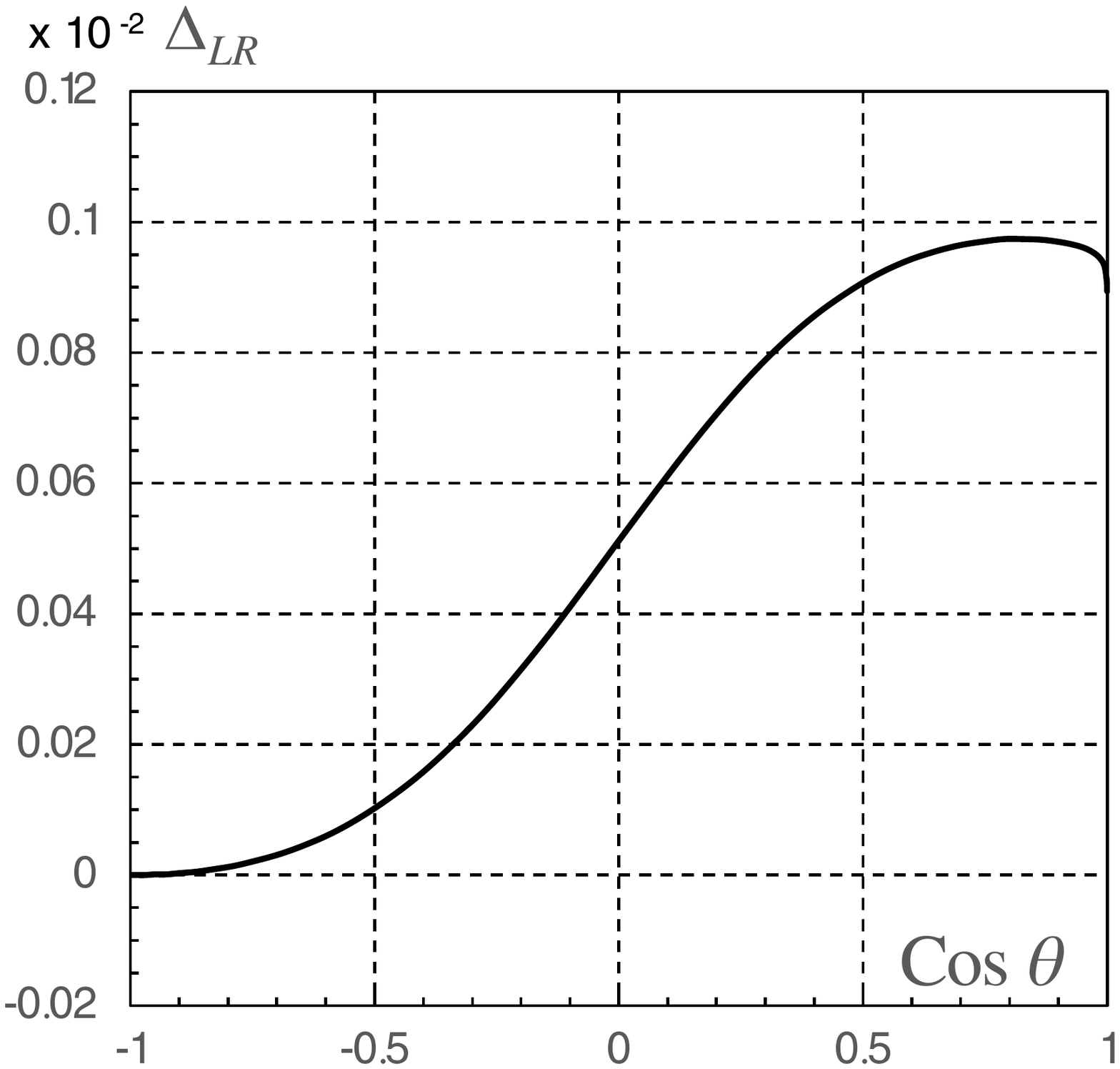}
	\end{tabular}
	%\vspace{8mm}
	\caption{\protect\it
		Left: the polarization Born asymmetry (0) 
		and asymmetry taking into account the NLO EWC (0+1) vs scattering angle $\cos\theta$, \KK~Monte Carlo points are integrated in $\cos\theta$ bins 0.125 in width. 
		Right: the absolute NLO correction to polarization Born asymmetry vs $\cos\theta$. Calculations are done at an $\Omega$ cut of 2~GeV. The points are the results obtained from running the \KK~Monte Carlo generator as described in the text, where the error bars
		represent the statistical errors from the number of Monte Carlo events generated.
	}
	\label{5}
	\vspace{5mm}
\end{figure}

\newpage

\begin{figure}
	\centering
	\begin{tabular}{cc}
		\includegraphics[trim=2cm 5cm 1cm 5cm,width=0.5\linewidth]{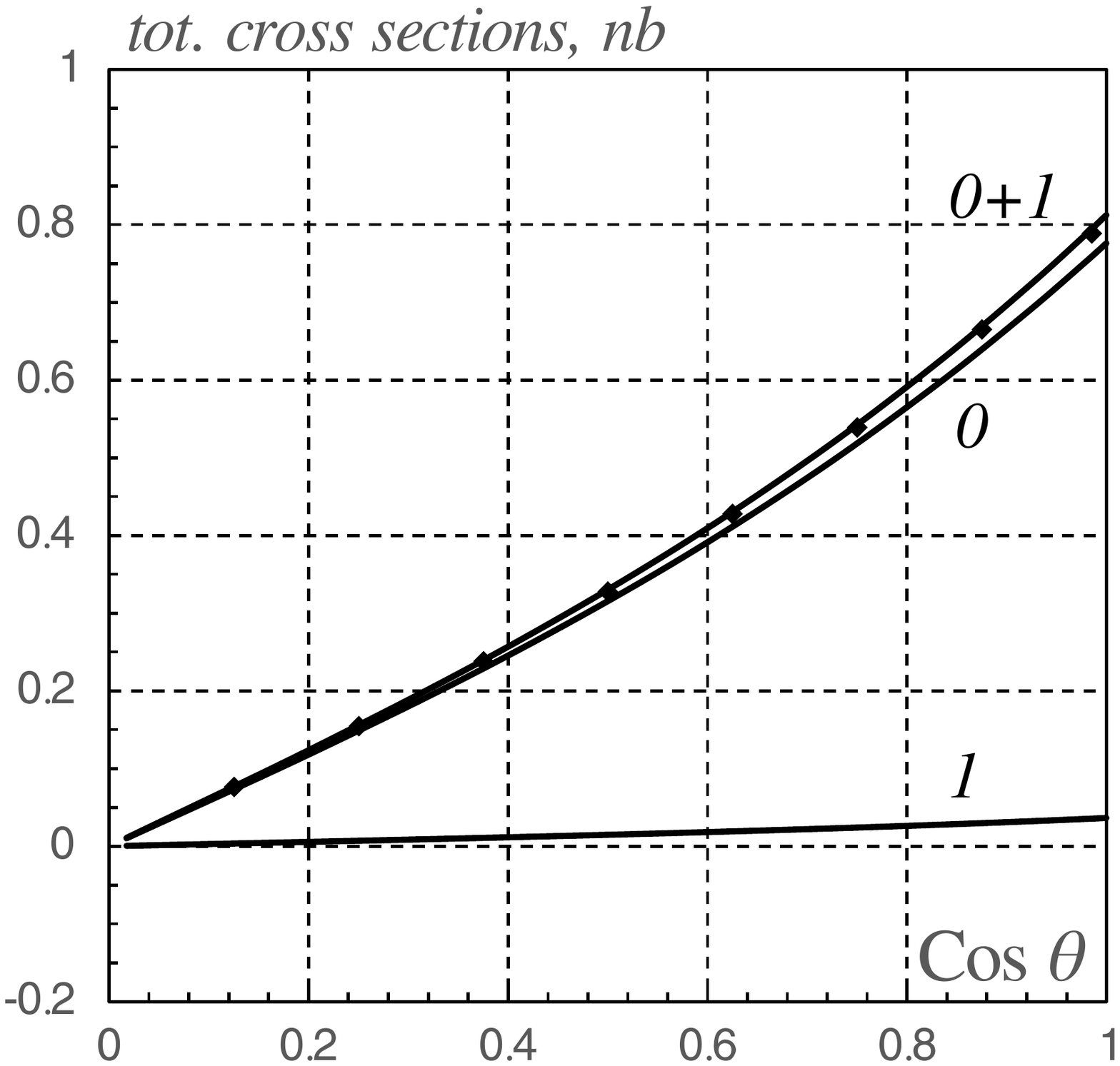}
		&
		\includegraphics[trim=2cm 5cm 1cm 5cm,width=0.5\linewidth]{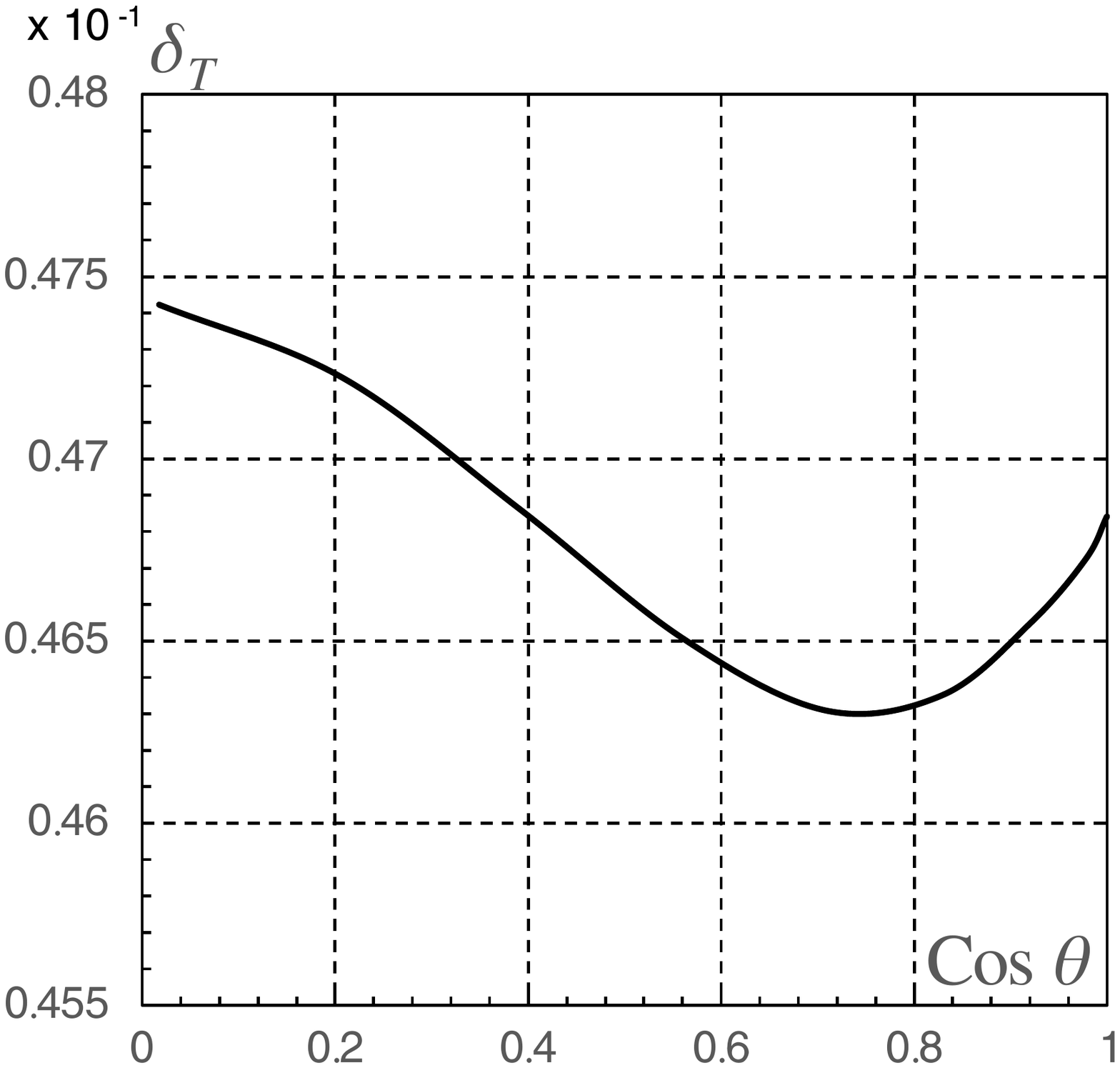}
	\end{tabular}
	%\vspace{8mm}
	\caption{\protect\it
		Left: unpolarized  
		NLO corrected (0+1), 
		Born (0), 
		and their difference (1) 
		total cross sections vs angle $a$.
		Right: the relative NLO correction to unpolarized total Born cross section vs $a$. Calculations are done at an $\Omega$ cut of 2~GeV. The points are the results obtained from running the \KK~Monte Carlo generator as described in the text, where the error bars
		represent the statistical errors from the number of Monte Carlo events generated.
	}
	\label{6}
	\vspace{5mm}
\end{figure}
\begin{figure}
	\centering
	\begin{tabular}{cc}
		\includegraphics[trim=2cm 5cm 1cm 5cm,width=0.5\linewidth]{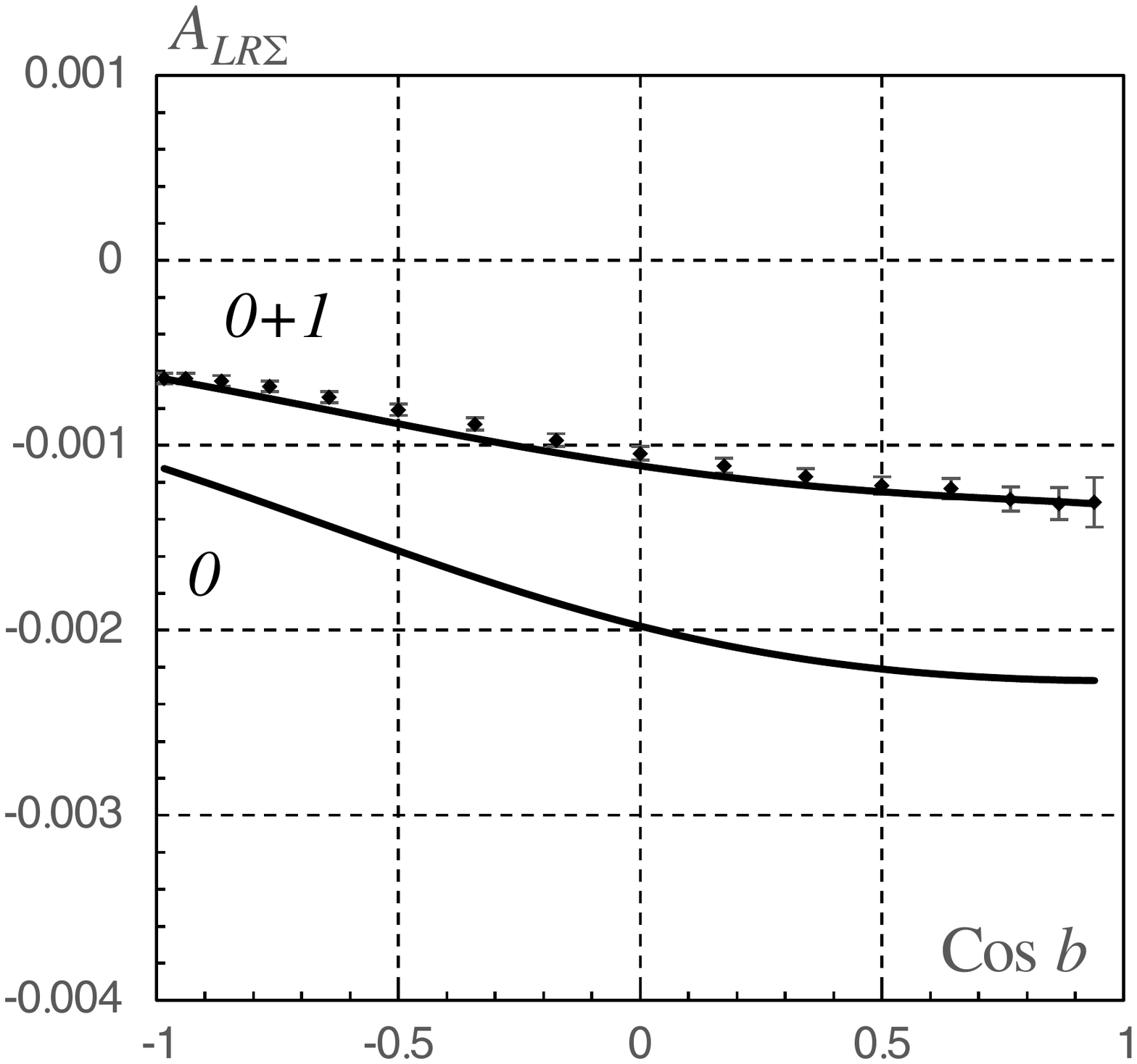}
		&
		\includegraphics[trim=2cm 5cm 1cm 5cm,width=0.5\linewidth]{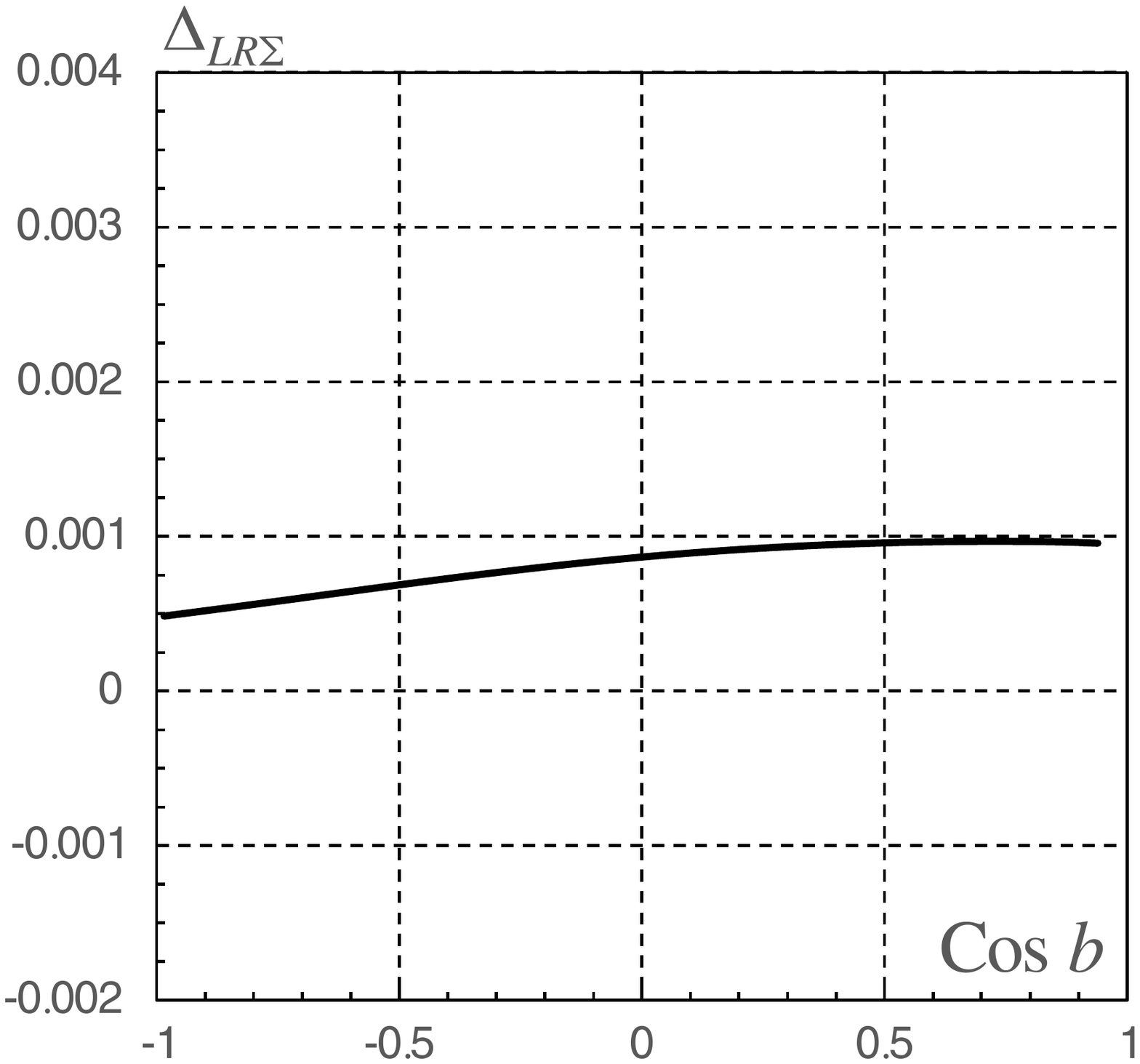}
	\end{tabular}
	%\vspace{8mm}
	\caption{\protect\it
		Left: the left-right integrated Born asymmetry (0) 
		and asymmetry taking into account the NLO EWC (0+1) vs angle $b$ at $a=10^\circ$.
		%JMR add:
		Right: the absolute NLO correction to left-right integrated Born asymmetry vs $b$ at $a=10^\circ$. Calculations are done at an $\Omega$ cut of 2~GeV. The points are the results obtained from running the \KK~Monte Carlo generator as described in the text, where the error bars
		represent the statistical errors from the number of Monte Carlo events generated.
	}
	\label{9}
	\vspace{5mm}
\end{figure}
\begin{figure}
	\centering
	\begin{tabular}{cc}
		\includegraphics[trim=2cm 5cm 1cm 5cm,width=0.5\linewidth]{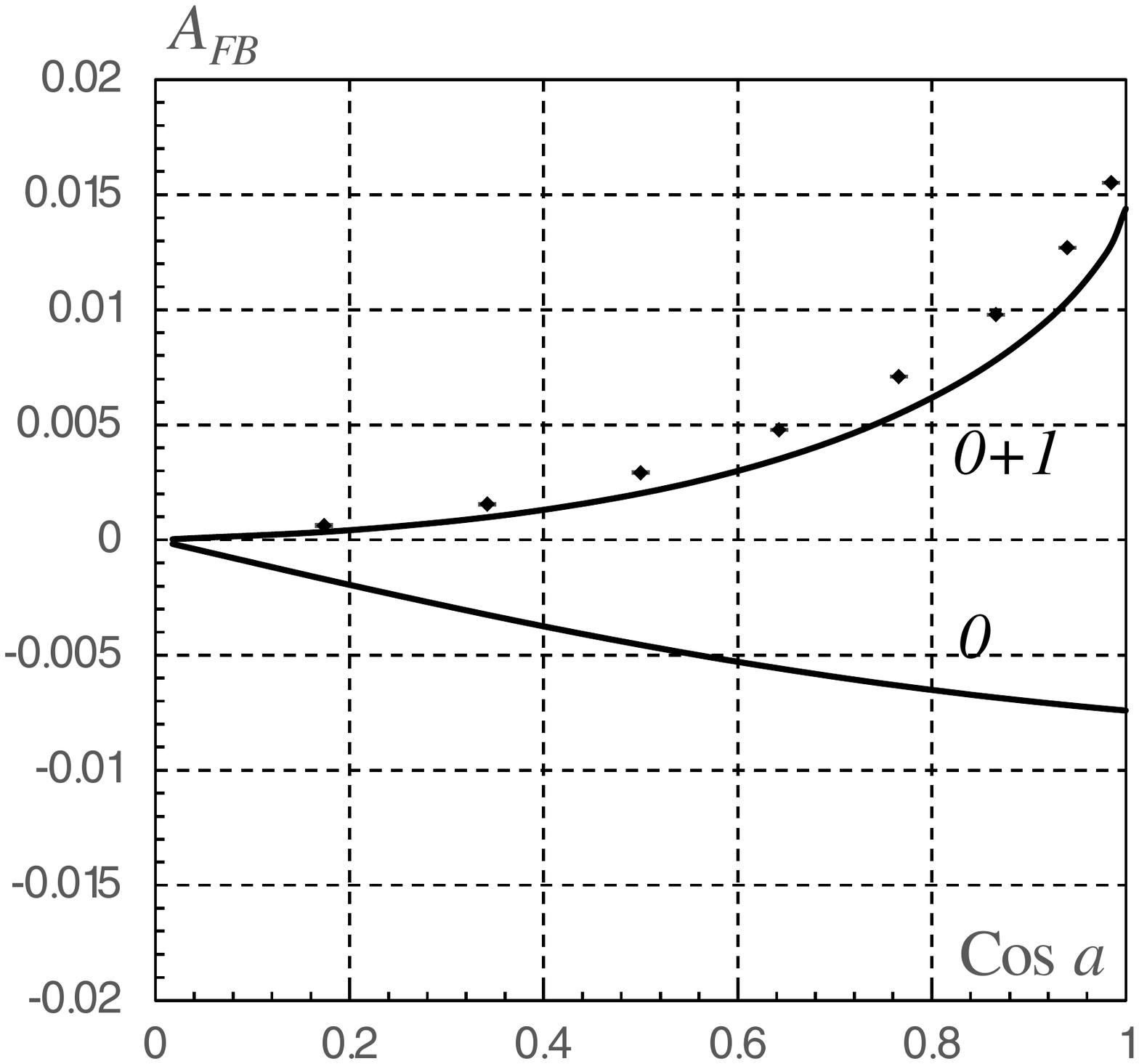}
		&
		\includegraphics[trim=2cm 5cm 1cm 5cm,width=0.5\linewidth]{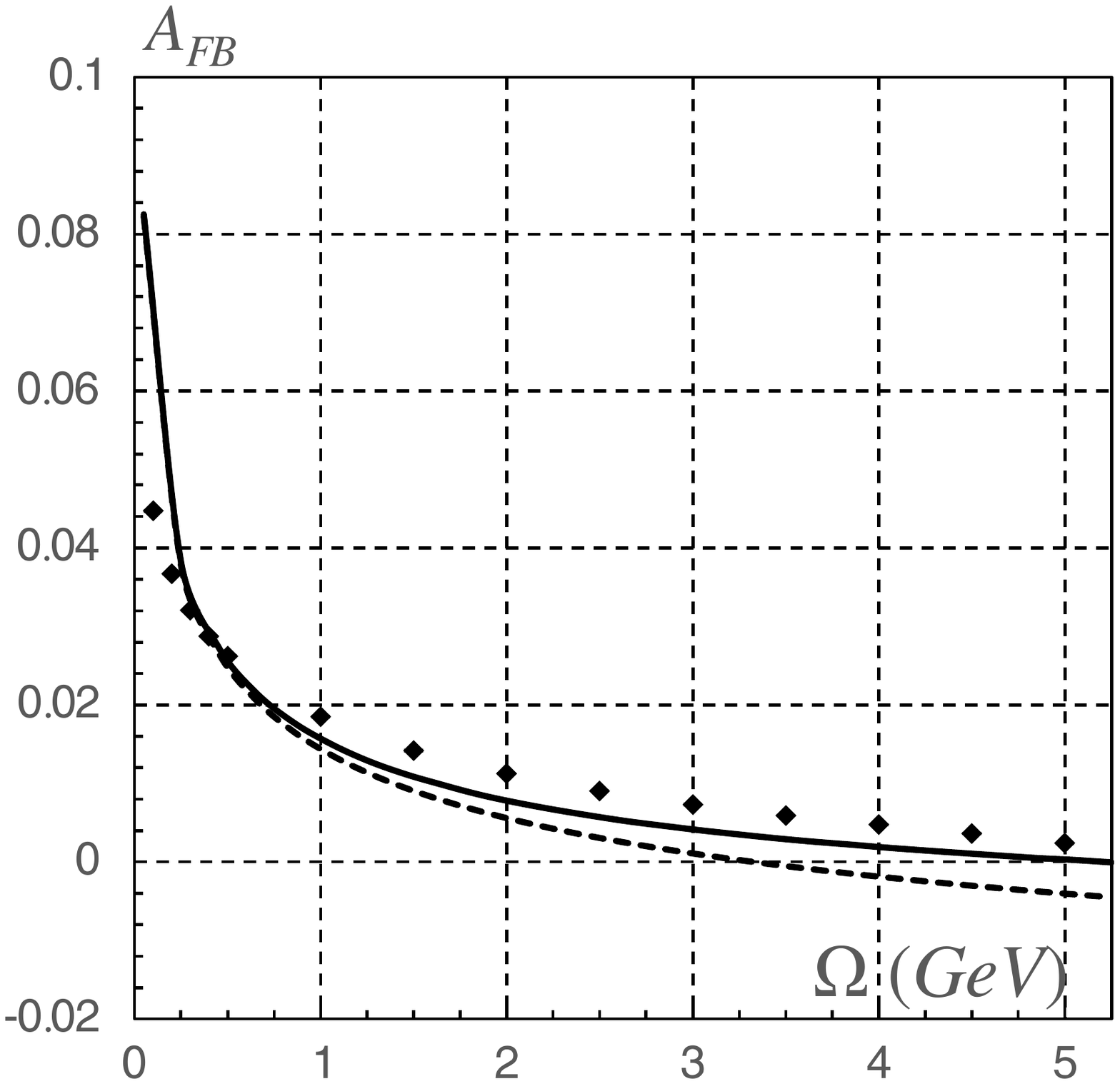}
	\end{tabular}
	%\vspace{8mm}
	\caption{\protect\it
		Left: the forward-backward Born asymmetry (0) 
		and asymmetry taking into account the NLO EWC (0+1) vs angle $a$ at an $\Omega$ cut of 2~GeV.
		Right: Calculations in two approaches: SPA (dashed line) and HPA (solid line), the NLO corrected forward-backward asymmetry at $a$=30$^\circ$. The points are the results obtained from running the \KK~Monte Carlo generator as described in the text, where the error bars
		represent the statistical errors from the number of Monte Carlo events generated.
	}
	\label{7}
	\vspace{5mm}
\end{figure}

\newpage

\begin{figure}
	\centering
	\begin{tabular}{cc}
		\includegraphics[trim=2cm 5cm 1cm 5cm,width=0.5\linewidth]{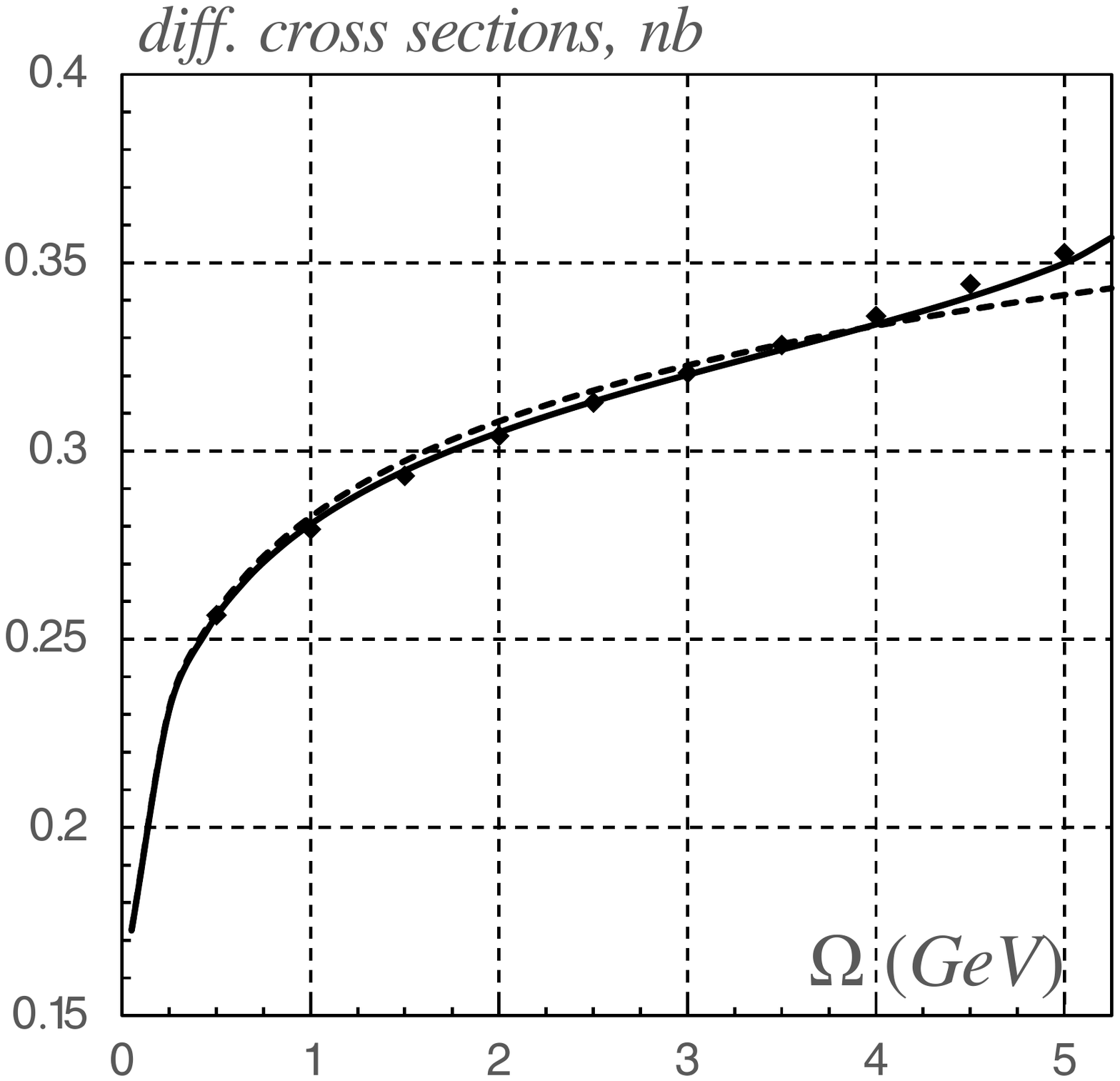}
		&
		\includegraphics[trim=2cm 5cm 1cm 5cm,width=0.5\linewidth]{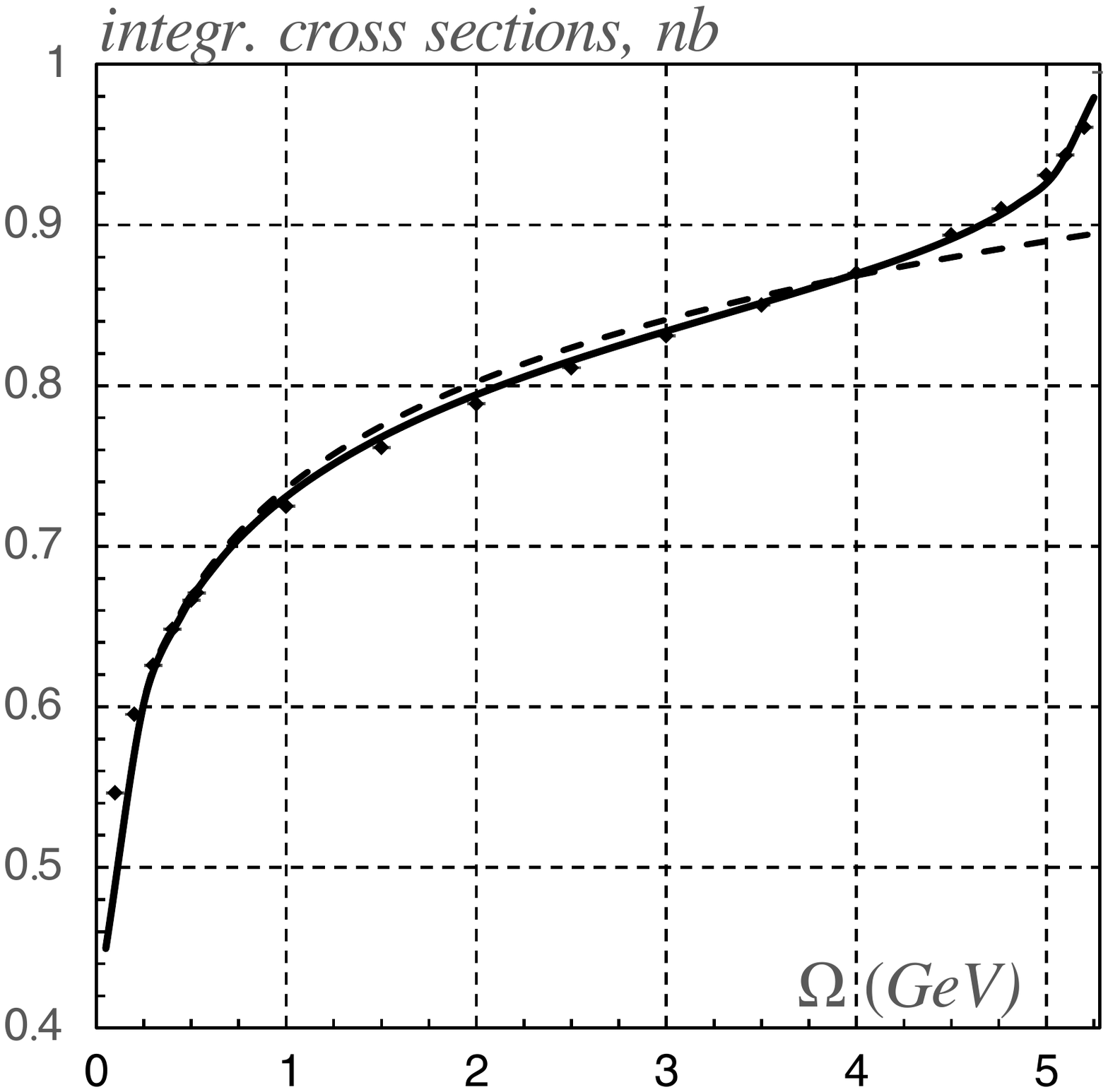}
	\end{tabular}
	%\vspace{8mm}
	\caption{\protect\it
		Calculations in two approaches: SPA (dashed line) and HPA (solid line). 
		Left: 
		The NLO corrected unpolarized differential cross section at $a$=$30^\circ$ vs $\Omega$.
		Right: 
		the NLO corrected unpolarized integrated cross section at $a$=$10^\circ$, $b$=$170^\circ$ vs $\Omega$. The points are the results obtained from running the \KK~Monte Carlo generator as described in the text, where the error bars
		represent the statistical errors from the number of Monte Carlo events generated.
	}
	\label{11}
	\vspace{5mm}
\end{figure}

\begin{figure}
	\centering
	\begin{tabular}{cc}
		\includegraphics[trim=2cm 5cm 1cm 5cm,width=0.5\linewidth]{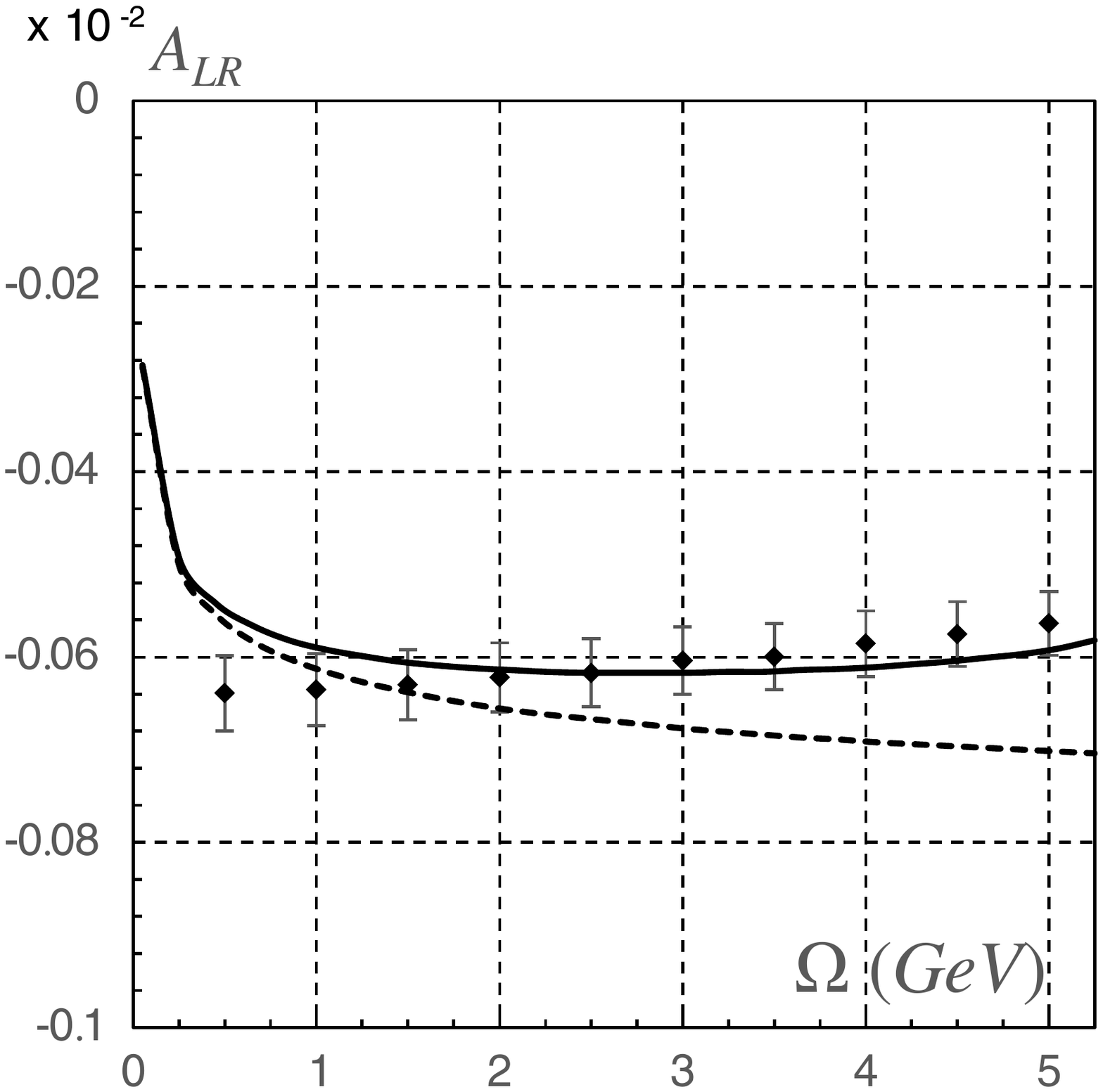}
		&
		\includegraphics[trim=2cm 5cm 1cm 5cm,width=0.5\linewidth]{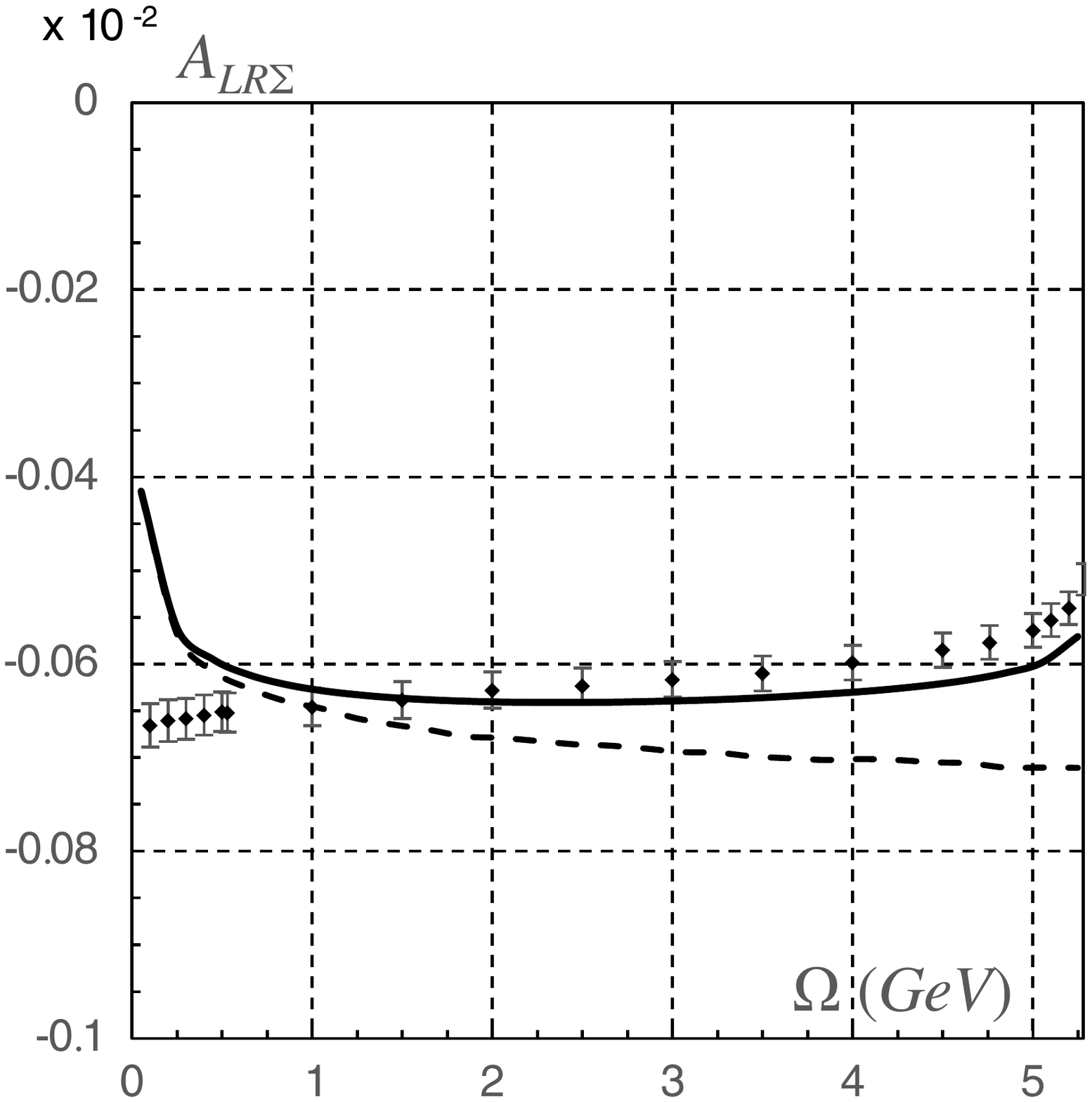}
	\end{tabular}
	%\vspace{8mm}
	\caption{\protect\it
		Calculations in two approaches: SPA (dashed line) and HPA (solid line).
		Left: the NLO corrected polarization asymmetry at $\theta$=90$^\circ$, \KK~Monte Carlo integrated between 70$^\circ$ and 110$^\circ$.
		Right: 
		the NLO corrected integrated asymmetry from $a$=$10^\circ$ to $b$=$170^\circ$.
		%JMR added
		The points are the results obtained from running the \KK~Monte Carlo generator as described in the text, where the error bars
		represent the statistical errors from the number of Monte Carlo events generated.
	}
	\label{12}
	\vspace{5mm}
\end{figure}

\end{document}